\documentclass[10pt,twocolumn,english,aps,manuscript,superscriptaddress,showpacs]{revtex4}

\usepackage{color}
\usepackage{amsmath}
\usepackage{amssymb}
\usepackage{graphicx}
\usepackage{esint}
\usepackage{url}
\usepackage{hyperref}

\usepackage{calc}       
\usepackage{rotating}
\usepackage{comment}
\usepackage{multirow}

\makeatletter

\@ifundefined{textcolor}{}
{%
 \definecolor{BLACK}{gray}{0}
 \definecolor{WHITE}{gray}{1}
 \definecolor{RED}{rgb}{1,0,0}
 \definecolor{GREEN}{rgb}{0,1,0}
 \definecolor{BLUE}{rgb}{0,0,1}
 \definecolor{CYAN}{cmyk}{1,0,0,0}
 \definecolor{MAGENTA}{cmyk}{0,1,0,0}
 \definecolor{YELLOW}{cmyk}{0,0,1,0}
}

\makeatother

\usepackage{babel}

\newif\ifTEXTFIG \TEXTFIGtrue

\newif\ifNOSUP \NOSUPfalse

\newif\ifNOFIG \NOFIGtrue

\newif\ifSFIG \SFIGfalse

\newif\ifXLFIG \XLFIGfalse

\newif\ifTtoD \TtoDfalse

\def\thefigure{\arabic{figure}}

\begin{document}

\setcounter{page}{1} 

\title{%
Rogue wave generation by inelastic quasi-soliton collisions in optical fibres
}

\author{M. Eberhard}
\email{m.eberhard@aston.ac.uk}
\homepage{www-users.aston.ac.uk/~eberhama}
\selectlanguage{english}%
\affiliation{Electrical, Electronic and Power Engineering, School of Engineering
and Applied Science, Aston University, Aston Triangle, Birmingham
B4 7ET, United Kingdom}

\author{A. Savojardo}
\email{A.Savojardo@warwick.ac.uk}
\selectlanguage{english}%
\affiliation{Department of Physics and Centre for Scientific Computing, The University
of Warwick, Coventry CV4 7AL, United Kingdom}

\author{A. Maruta}
\email{maruta@comm.eng.osaka-u.ac.jp}
\homepage{wwwpn.comm.eng.osaka-u.ac.jp/~maruta}
\selectlanguage{english}%
\affiliation{Division of Electrical, Electronic and Information Engineering, Graduate School of Engineering, Osaka University, Suita Campus, Japan}

\author{R.A. R\"{o}mer}
\email{r.roemer@warwick.ac.uk}
\homepage{www.warwick.ac.uk/rudoroemer}
\selectlanguage{english}%
\affiliation{Department of Physics and Centre for Scientific Computing, The University
of Warwick, Coventry CV4 7AL, United Kingdom}

\date{$Revision: 1.201 $, compiled \today}

\begin{abstract}
\vspace*{3ex}
We demonstrate a simple cascade mechanism that drives the formation and emergence of rogue waves in the generalized non-linear
Schr\"{o}dinger equation with third-order dispersion. This conceptually novel generation mechanism is based on \emph{inelastic collisions} of quasi-solitons and is well described by a resonant-like scattering behaviour for the energy transfer in pair-wise quasi-soliton collisions. Our results demonstrate a threshold for rogue wave emergence and the existence of a period of reduced amplitudes --- a "calm before the storm" --- preceding the arrival of a rogue wave event. Comparing with ultra-long time window simulations of $3.865\times 10^{6}$ps we observe the statistics of rogue waves in optical fibres with an unprecedented level of detail and accuracy, unambiguously establishing the long-ranged character of the rogue wave power-distribution function over seven orders of magnitude. 
\end{abstract}

\maketitle

\section{Introduction}
\label{sec-intro}

Historically, reports of ``monster" or ``freak" waves \cite{Dra64,Dra71,Mal74} on the earth's oceans have
been seen largely as sea men's tales \cite{Per06,Erk15}.  However, the recent availability of
reliable experimental observations \cite{Hop04,Per06} has proved their existence and
shown that these "rogues" are indeed rare events \cite{KhaP03}, governed by long tails in their probability distribution function
(PDF) \cite{OnoOSC04}, and hence concurrent with very large wave amplitudes \cite{AdcTD15,OnoRBM13}. 
As both deep water waves in the oceans and optical waves
in fibres can be described by similar generalized non-linear Schr\"{o}dinger
equations (gNLSE) they both show rogue waves (RW) and long-tail statistics \cite{DudDEG14,ChaHOG13,OnoOSC04}. 
The case of RW generation in optical fibres
during super-continuum generation has been observed experimentally
\cite{SolRKJ07,SolRJ08,ErkGD09,KibFD09}.  Recently, experimental data of long tails in the PDF have been collected \cite{RanWOS14,WalRS15}, as well as time correlations in various wave phenomena with RW occurrence studied \cite{BirBDS15}. 
RWs and long-tailed PDFs have also been found during high power femtosecond pulse filamentation in air \cite{KasBWD09}, in non-linear optical cavities \cite{MonBRA09} and in the output intensity of optically injected semiconductors laser \cite{BonFBG11}, mode-locked fiber lasers \cite{LecGSA12}, Raman fiber lasers \cite{RanS12} and fiber Raman amplifiers \cite{HamFDM08}.
%
However, it still remains largely unknown how RWs
emerge \cite{AkhP10,RubKRD10,AkhKBB16} and theoretical explanations range from high-lighting the importance of the non-linearity \cite{AdcTD15,KibFFM10,ChaHA11,KibFFM12} to  those based on short-lived linear superpositions of quasi-solitons during collisions \cite{OnoOSC05,HohKSK10}. 

Here we will show that there is also a process to generate 
rogue waves through an \emph{energy-exchange mechanism}  when collisions
become \emph{inelastic} in the presence of a third-order dispersion (TOD) term \cite{AkhSA10,WeeTTE15}.  Energy-exchange in NLSEs has indeed been observed experimentally \cite{LuaSYK06,MusKKL09} albeit not yet for TOD.
Recent studies confirmed experimentally and numerically that the presence of TOD in optical fibers turns the system convectively unstable and generates extraordinary optical intensities \cite{KhaP03,VorST08}. 
We derive a novel \emph{cascade model} that simulates the RW generation process directly without the need for a full numerical integration of the gNLSE. The model is validated using a massively parallel simulation \cite{EbeR16}, allowing us to achieve an unprecedented level of detail through the concurrent use of tens of thousands of CPU cores.
Based on statistics from more than $17 \times 10^{6}$ interacting quasi-solitons, 
find that the results of the full gNLSE integration and the cascade model exhibit the same quantitative, long-tail PDF.  This agreement highlights the importance of (i) a resonance-like two-soliton scattering coupled with (ii) quasi-soliton energy exchange in giving rise to RWs.
We furthermore find that the cascade model and the full gNLSE integration exhibit a "calm-before-the-storm" effect of reduced amplitudes prior to the arrival of the RW, hence hinting towards the possibility of RW prediction. Last, we demonstrate that there is a sharp threshold for TOD to be strong enough to lead to the emergence of RWs.

\section{Rogue waves as cascades of interacting solitons}
\label{sec-CM}

Analytical soliton solutions for the generalized non-linear Schr\"{o}dinger equation 
\begin{equation}
\partial_{z}u+\frac{i\beta_{2}}{2}\partial_{t}^{2}u-\frac{\beta_{3}}{6}\partial_{t}^{3}u - i\gamma| u|^{2}u = 0
\label{eq-gNLSE}
\end{equation}
are only known for $\beta_3=0$. Here $u(z,t)$ describes a slowly
varying pulse envelope, $\gamma$ the non-linear coefficient, $\beta_2$ ($<0$) the
normal group velocity dispersion and $\beta_3$ the TOD \cite{Agr13}.  
Due to the short distances of only a few kilometers in
a super-continuum experiment the effects of absorption, shock term and
delayed Raman response can be neglected \cite{DudDEG14}.  Our aim hence is not the most
accurate microscopic description possible, but the most simplistic numerical model that
is still capable of generating RWs.

Without TOD $\beta_3$, the model \eqref{eq-gNLSE} can be solved
analytically and a $u(z,t)$ describing the celebrated \emph{soliton} solutions can be found
\cite{ZakS72}.  Depending on the phase difference between two such solitons, attracting
or repulsing forces exist between them.  This then leads to them either
moving through each other unchanged or swapping their positions. The solitons emerge unchanged
with the same energy as they had before the collision.  This type of
collision is elastic and it is the only known type of collision for true
analytic (integrable) solitons \cite{AkhP10,Zak09}.
When we introduce $\beta_3$, the system can no longer be solved analytically
and no closed-form solutions are known in general.  Numerical
integration of \eqref{eq-gNLSE} shows that stable pulses still exist and
these propagate individually much like solitons in the $\beta_3=0$ case \cite{TakMKL10}.  These
\emph{quasi-solitons}, and collisions between them, are in many cases as elastic as
they are for integrable solitons.  
However, when two quasi-solitons have a matching phase, energy can be transferred between them leading to
one gaining and the other losing energy \cite{AkhSA10}.  In addition, the emerging pulses have to shed energy through
dispersive waves until they have relaxed back to a stable quasi-soliton
state \cite{AkhK95}.  


Based on this observation we can now replace the full numerical integration
of the gNLSE with a \emph{phenomenological cascade model} that tracks the collisions between
quasi-solitons.  Our starting point are quasi-solitons with exponential power tails as found in a super-continuum system after the
modulation instability has broken up the initial CW pump laser input into a
train of pulses \cite{Agr13,TaiHT86,Has84}.  We generate a list of such
quasi-solitons as our initial condition by randomly choosing the power
levels $P_q$, phases $\phi_q$ and frequency shifts
$\Omega_q$ in accordance with the statistics found in a real system at that
point of the integration. The shape of a quasi-soliton is approximated by
\begin{equation}
u_q(z,t)=\sqrt{P_q}\ \mathrm{sech}\!\left[\frac{(t-t_q)+z/v_q}{T_q}\right]\exp\!\left[i \phi_q \right]\label{eq-quasi-soliton}
\end{equation}
with effective period
\begin{equation}
T_q=\sqrt{\frac{|\beta_{2}+\beta_{3}\Omega_q |}{\gamma P_q}}
\label{eq-period}
\end{equation}
and inverse velocity
\begin{equation}
v_q^{-1}=\beta_{2}\Omega_q+\frac{\beta_{3}}{2}\Omega_q^{2}+\frac{\beta_{3}}{6T_q^{2}}
\label{eq-vp}
\end{equation}
for each quasi-soliton labelled by $q$.
Note that due to \eqref{eq-period}, the velocity $v_q$ of a quasi-soliton depends on its power in
addition to the frequency shift for $\beta_3\neq 0$.  We calculate which
quasi-solitons will collide first, based on their known initial times
$t_q$ and velocities $v_q$.  The phase difference $\phi_q - \phi_p$ between two quasi-solitons is drawn
randomly from a uniform distribution in the interval $[0, 2\pi[$.  Then we calculate the energy transferred from the smaller quasi-soliton to the larger via
\begin{equation}
\frac{\Delta E_{1}}{E_{2}}=\frac{\epsilon_{\mathrm{eff}}}{| v^{-1}_{1}-v^{-1}_{2}
|}\sin^{2}\left(\frac{\phi_1-\phi_2}{2}\right) ,
\label{eq-energy-transfer-eff}
\end{equation}
where $\Delta E_1$ is the energy gain for the higher energy
quasi-soliton and $E_2$ is the energy of the second quasi-soliton involved
in the collision. A detailed justification of \eqref{eq-energy-transfer-eff} and a discussion of the cross-section coefficient $\epsilon_\text{eff}$ will be given below. 
At this point we estimate the next collision to occur from the set of updated $v_q$ and
continue as described above until the simulation has reached the desired distance $z$. Obviously, this procedure is much simpler than a  numerical integration of the gNLSE \eqref{eq-gNLSE}. The main strength of the model is a new qualitative and quantitative understanding of RW emergence and dynamics.

\section{Rare events with ultra-long tails in the PDF}
\label{sec-pdf}

To compare the results with the full numerical integration we generate a power-distribution function (PDF) of the power levels at fixed distances $\Delta z$ from $u(z,t) = \sum_q{u_q(z,t)}$ 
using the full quasi-soliton waveform (\ref{eq-quasi-soliton}). 
The PDF for the complete set of $\gtrsim 17 \times 10^{6}$ pulses propagating over $1500$m is shown in Fig.\ \ref{fig-PDF} (a) and (b) for selected distances for cascade model and gNLSE, respectively, using, for the gNLSE, a highly-optimised, massively parallel and linearly-scaling numerical procedure (see Methods section). After $100$m, the PDF exhibits a roughly exponential distribution as seen in Fig.\ \ref{fig-PDF}(a).  With $\beta_3=0$ this exponential PDF remains stable from this point onwards (cp.\ inset).  However, with $\beta_3\neq 0$ the inelastic collision of the quasi-solitons leads to an ever increasing number of high-energy RWs. After $500$m, a clear deviation from the exponential distribution of the $\beta_3=0$ case has emerged and beyond $1000$m, the characteristic $L$-shape of a fully-developed RW PDF has formed. 
\ifTEXTFIG
\begin{figure*}[tbh]
\begin{centering}
(a)\includegraphics[width=0.95\columnwidth,clip]{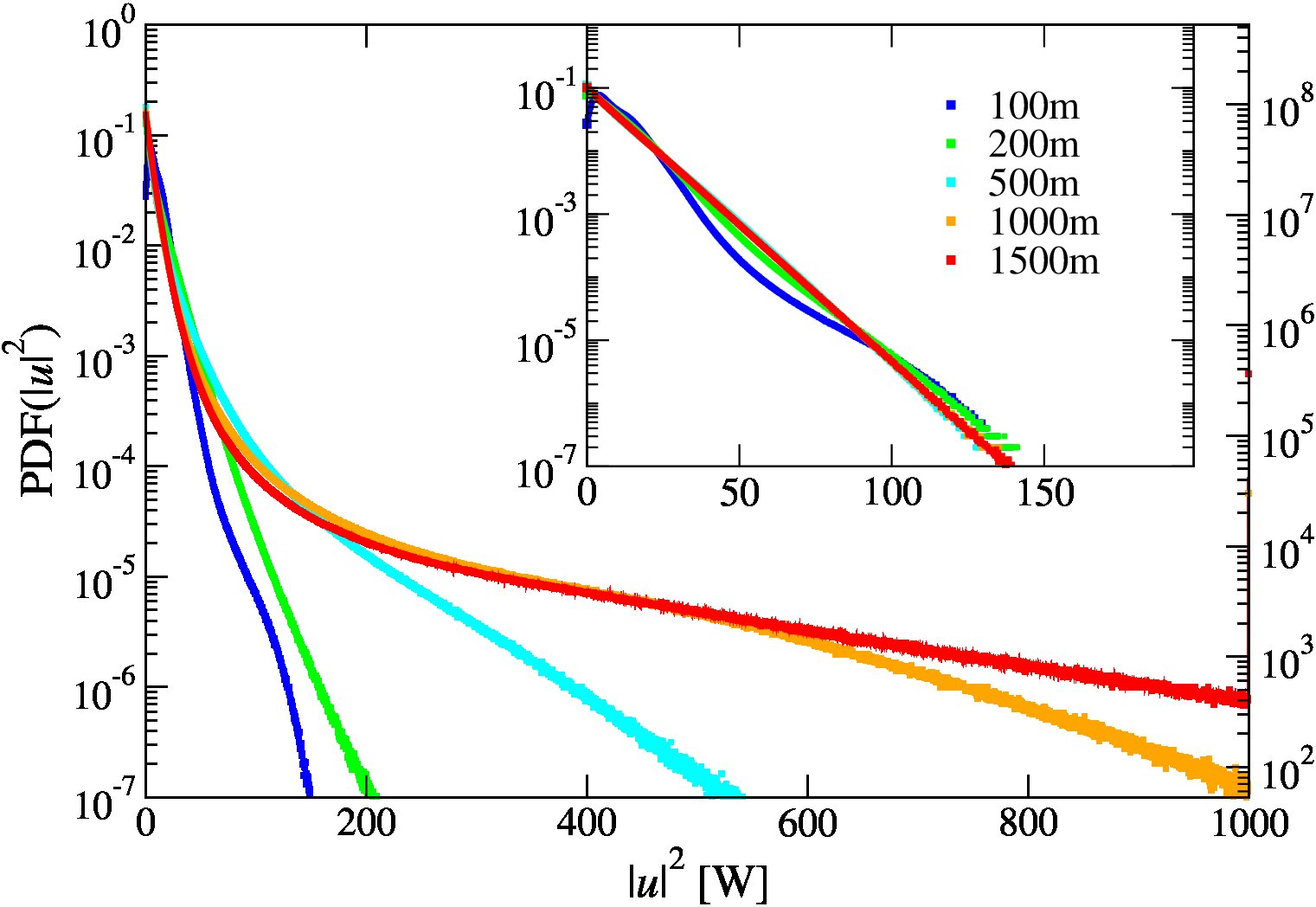}
%
%
(b)\includegraphics[width=0.95\columnwidth, clip]{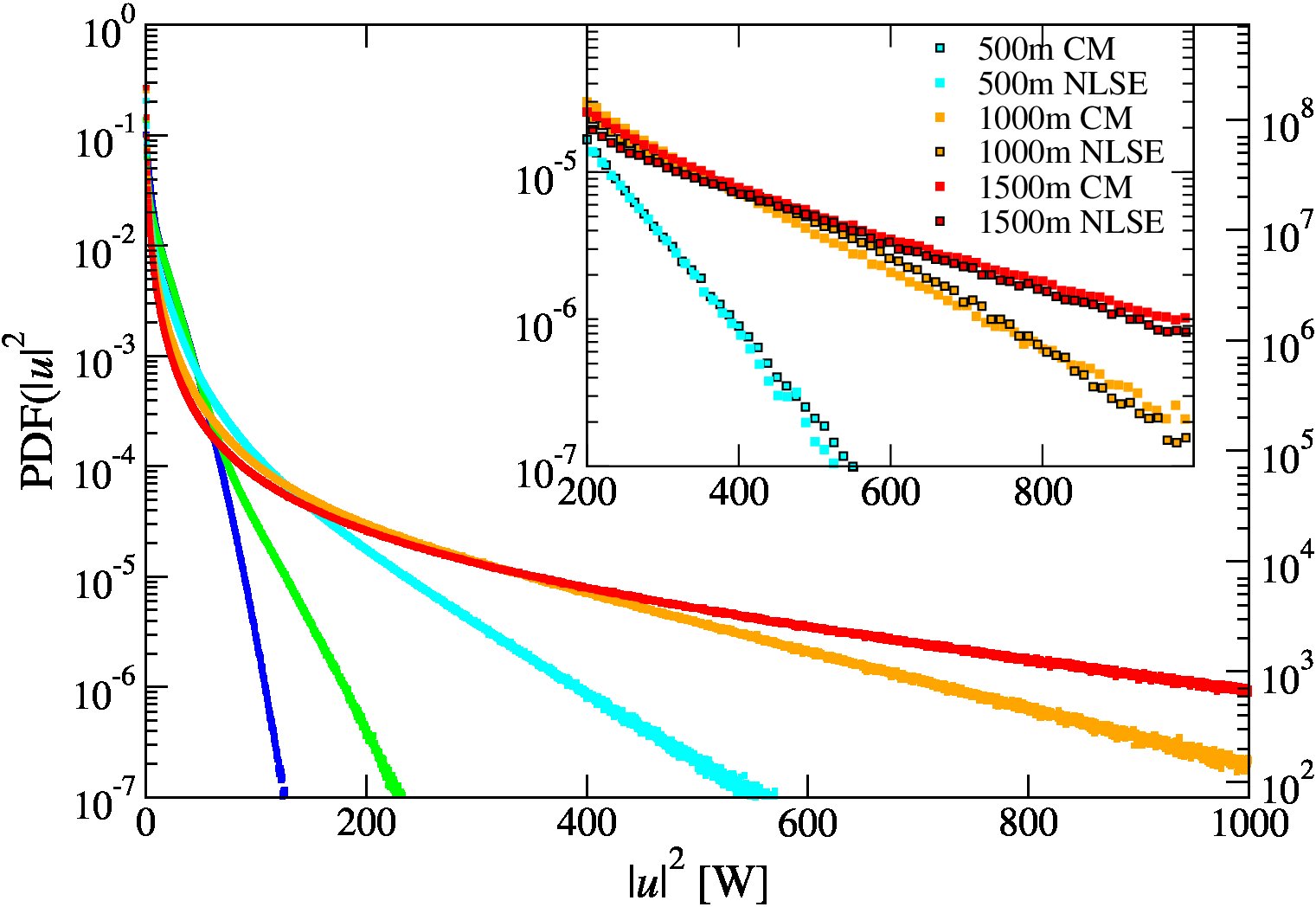}
\end{centering}

\protect\caption{(Color online) %
(a) PDFs of the intensity $|u|^{2}$ from the gNLSE \eqref{eq-gNLSE} at $\beta_{3}=2.64\times10^{-42}$s$^{3}$m$^{-1}$ using a large
time window of $\Delta t=3.865\times10^6$ps. The PDFs have been computed at distances $z= 100$m, $200$m, $500$m, $1000$m and $1500$m. The left vertical axis denotes the values of the normalized PDF while the right vertical axis gives the event count per bin.
The inset shows results for $\beta_{3}=0$.
%
%
%
(b) PDFs of the intensity $|u|^{2}$ from the cascade model for the same distances as in (a), using the same symbol and axes conventions. The inset shows a comparison between the results from the gNLSE (colored lines) and the cascade model (black lines and symbol outlines) for $z=500$m, $1000$m and $1500$m. Only every 50th symbol is shown.}
\label{fig-PDF}
\end{figure*}
\fi
The PDFs for both gNLSE and the cascade model in Figs.\ \ref{fig-PDF}(a) and (b) then continue to evolve
towards higher peak powers with some quasi-solitons becoming larger and larger. In the inset of Fig.\ \ref{fig-PDF}(b), we compare the long tail behaviour of both PDFs directly. We see that the agreement for PDFs is excellent taking into account that we have
reduced the full integration of the gNLSE to only discrete collision events
between quasi-solitons. Thus, we find that the essence of the emergence of RWs
in this system is very well captured by a process due to inelastic collisions.

\section{Mechanisms of the cascade}
\label{sec-pdf}

In Fig.\ \ref{fig-traces}, we show a representative example for the propagation of $u(z,t)$, in the full gNLSE and the cascade model, in a short $15$ps time range out of the full $3.865\times 10^{6}$ ps with $\beta_2=-2.6\times 10^{-28}$s$^2$m$^{-1}$ and $\gamma= 0.01$ W$^{-1}$m$^{-1}$.
A small initial noise leads to differences in the pulse powers and velocities and hence to eventual collisions of neighbouring pulses. In the enlarged trajectory plots Figs.\ \ref{fig-traces}(b) and (c), we see that for $\beta_3=0$ the solitons interact elastically and propagate on average with the group velocity of the frame. However, for finite $\beta_3$ in Fig.\ \ref{fig-traces}(c), most collisions are \emph{inelastic} and one quasi-soliton,
with higher energy, moves through the frame from left to right due to its
higher energy and group velocity mismatch compared to the frame. It
collides in rapid succession with the other quasi-solitons travelling, on average, at the
frame velocity. 
\ifTEXTFIG
\begin{figure*}[tbh]
\begin{centering}
(a)\includegraphics[width=0.37\textwidth,clip]{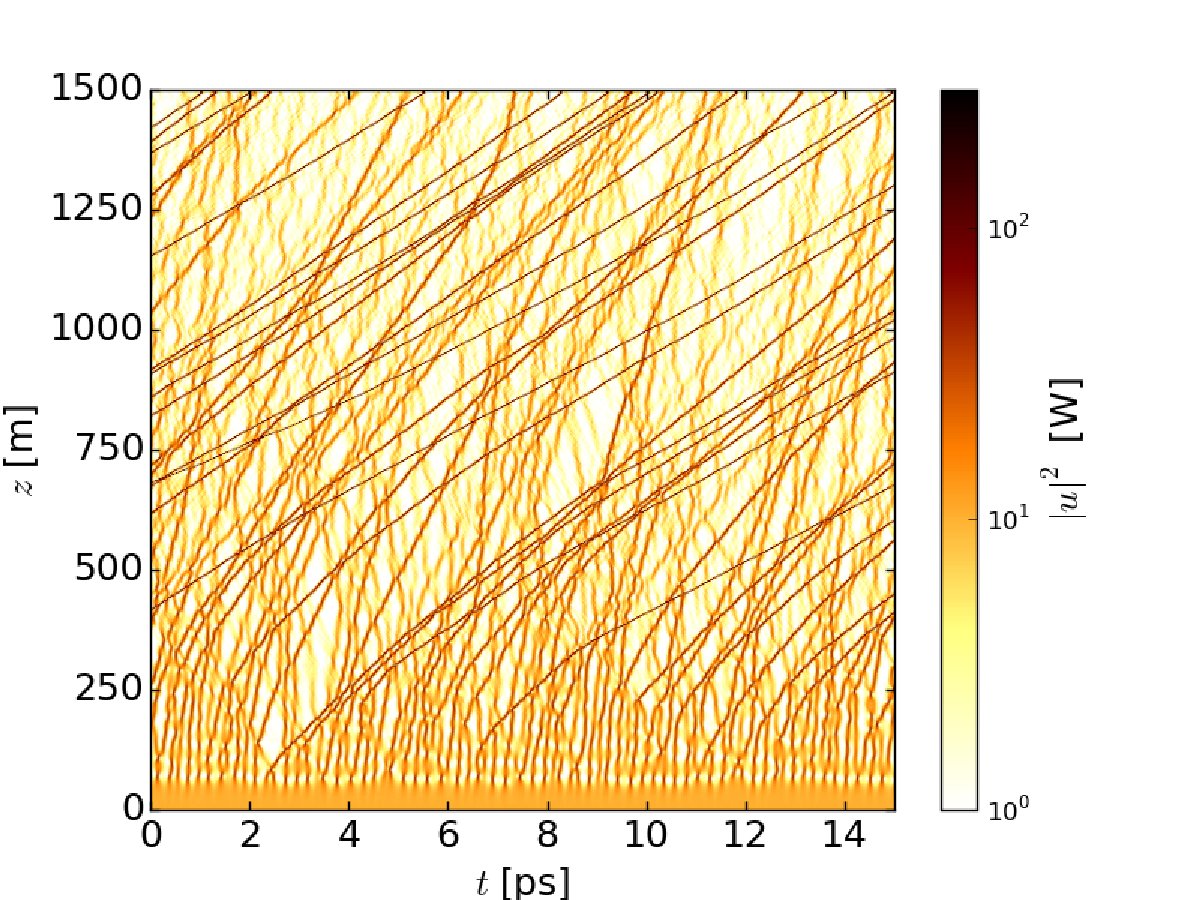}
\begin{minipage}{0.2\textwidth}\vspace*{-40ex}
(b)\includegraphics[width=\textwidth,clip]{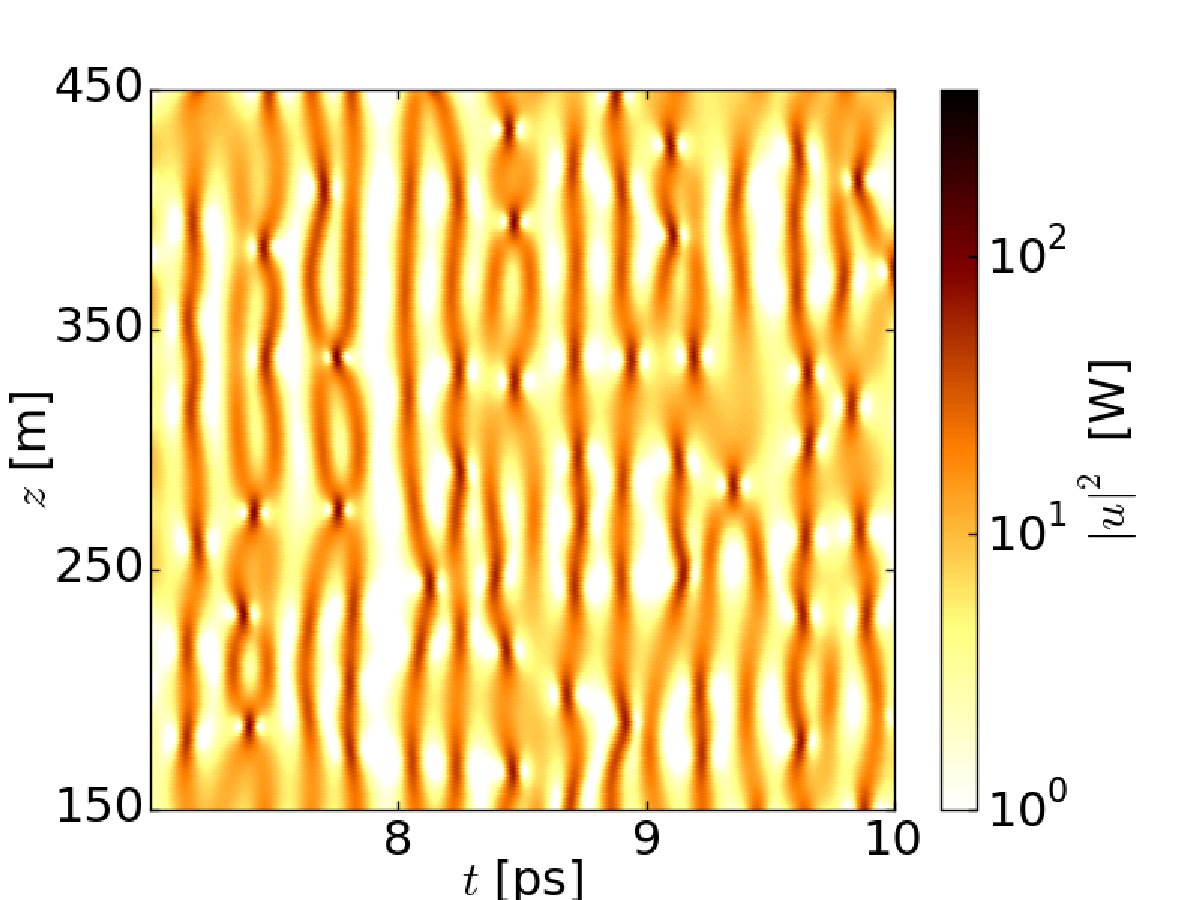}\\
(c)\includegraphics[width=\textwidth,clip]{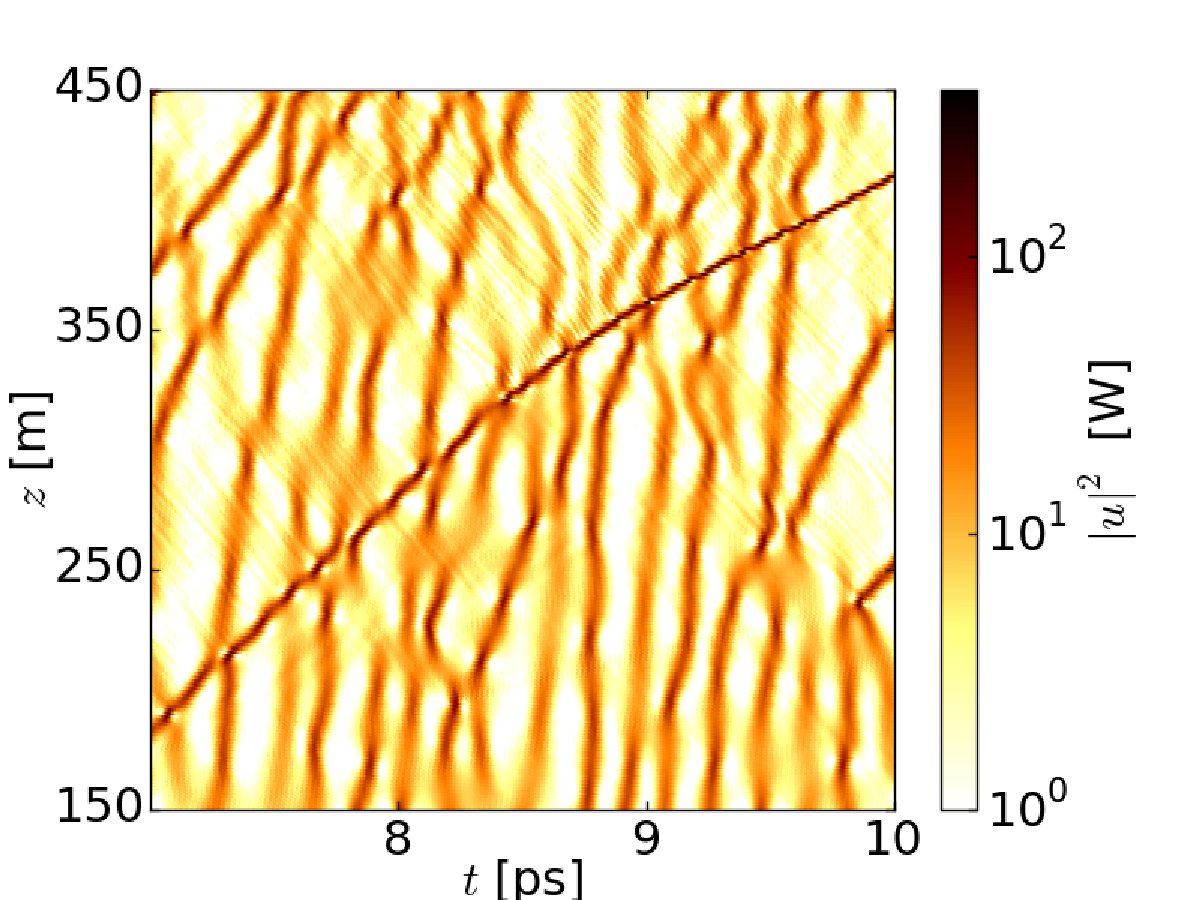}
\end{minipage}
(d)\includegraphics[width=0.37\textwidth,clip]{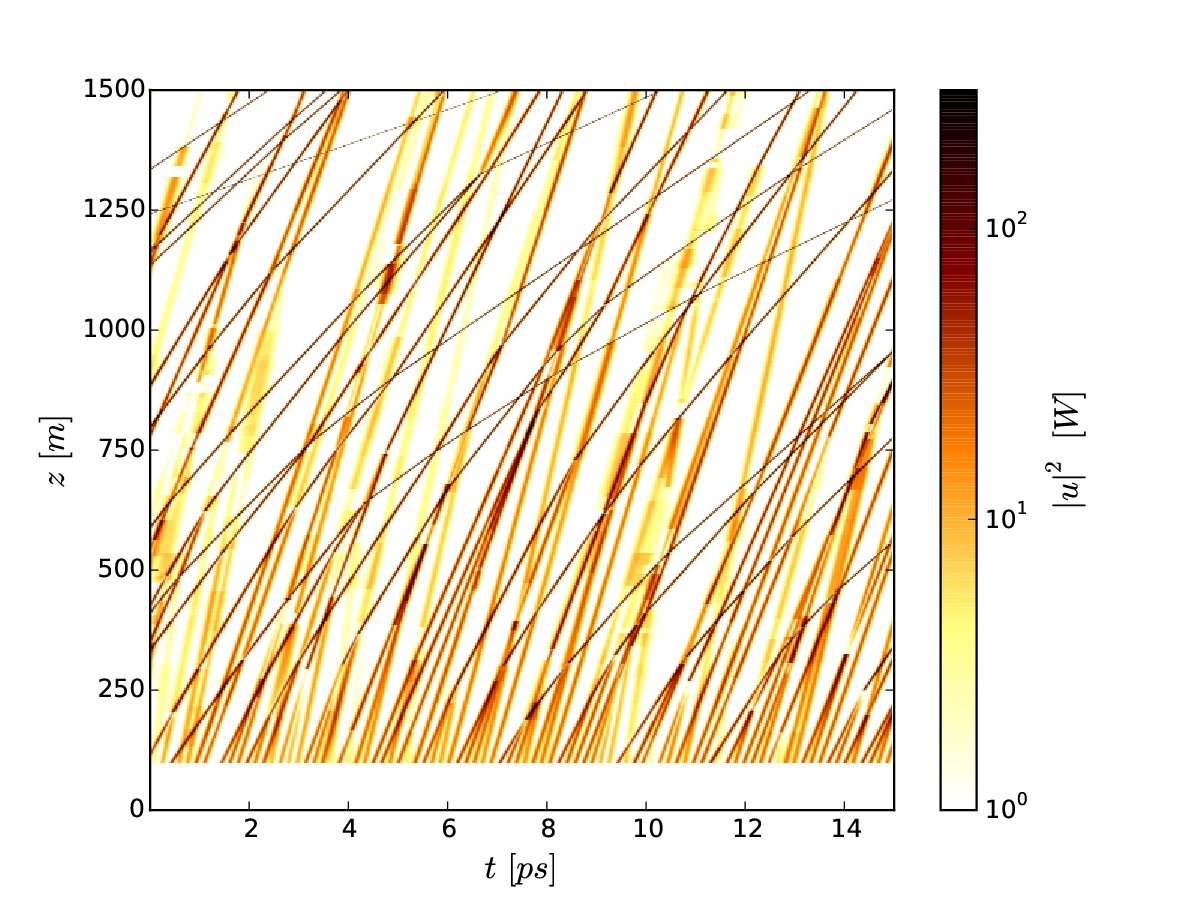}
\end{centering}

\protect\caption{(Color online)
(a) Intensity $|u(z,t)|^{2}$ for $\beta_{3}=2.64\times10^{-42}$s$^{3}$m$^{-1}$ of the gNLSE Eq.\ \eqref{eq-gNLSE} as function of the time
$t$ and distance $z$ in a selected time frame of $\Delta t= 15$ps and distance range $\Delta z=1.5$km. 
(b) $|u|^{2}$ with $\beta_{3}=0$ for a zoomed-in distance and time region,
(c) $|u|^{2}$  with $\beta_{3}$ value as in (a) for a region of (a) with $\Delta t$ and $\Delta z$ chosen identical to (b).
(d) Intensities $|u|^{2}$ as computed from the effective cascade model using the same shading/color scale as in (a). Note that we start the effective model at $z_0=100$m to mimic the effects of the modulation instability in (a).}
\label{fig-traces}
\end{figure*}
\fi
From Fig.\ \ref{fig-traces}(b), we see that our cascade model, in which we have replaced the intricate dynamics of the collission by an effective process, similarly shows quasi-solitons starting to collide inelastically, some emerging with higher energies and exhibiting a reduced group velocity. Note that in the cascade model we use an initial pulse power distribution that mimics the PDF of the gNLSE at $100$m, and thus only generate data from $100$m onwards (cp.\ Supplement).

The main ingredient of the cascade model, the energy exchange, is obvious in the gNLSE results: in almost all cases, energy is transferred from the
quasi-soliton with less energy to the one with more energy leading to the
cascade of incremental gains for the more powerful quasi-soliton. 
This pattern is visible throughout Fig.\ \ref{fig-traces}(a) where initial
 differences in energy of quasi-solitons become exacerbated over time
and larger and larger quasi-solitons emerge. These accumulate the
energy of the smaller ones to the point that the smaller ones eventually vanish into the background.
In addition, the group velocity of a quasi-soliton with TOD is dependent on the power of the quasi-soliton \cite{Agr13}.  Thus,
the emerging powerful quasi-solitons feature a growing group velocity
difference to their peers and this increases their collision rate leading to
even stronger growth.  This can clearly be seen from Fig.\ \ref{fig-traces}(a) where
larger-energy quasi-solitons start to move sideways as their velocity no longer
matches the group velocity of the frame after they have acquired energy from
other quasi-solitons due to inelastic collisions.
Indeed, the
relatively few remaining, soliton-like pulses at $1500$m can have peak
powers exceeding $1000$W. They are truly self-sustaining rogues that have
increased their power values by successive interactions and energy exchange
with less powerful pulses.

\section{Calm before the storm}
\label{sec-collisions}

Looking more closely at the temporal vicinity of waves with particularly large power values, we find that these tend to be preceded by a time period of reduced power values. This ``calm before the storm'' phenomenon can be observed in Fig.\ \ref{fig-calmb4storm}. In Fig.\ \ref{fig-calmb4storm}(a) we can clearly see an asymmetry in the normalized power $|u(\Delta t)|^2/ \langle |u(\Delta t)|^2 \rangle$ relative to the RW event at $\Delta t=0$ ($\Delta t<0$ denotes events before the RW). The average includes all RWs, defined here as large power events above a threshold  of $150$W (thresholds $200$ and $300$W show similar behaviour) and also two independent simulations of the gNLSE, both with parameters as in Fig.\ \ref{fig-traces}. 
The period of calm in power before the RW occurs lasts about $\Delta t=1.5$ps at $z= 200$m. It broadens for larger distances, but an asymmetry is retained even at $500$m. 
\ifTEXTFIG
\begin{figure*}[tbh]
%
%
\centering
(a)\includegraphics[width=0.95\columnwidth,clip]{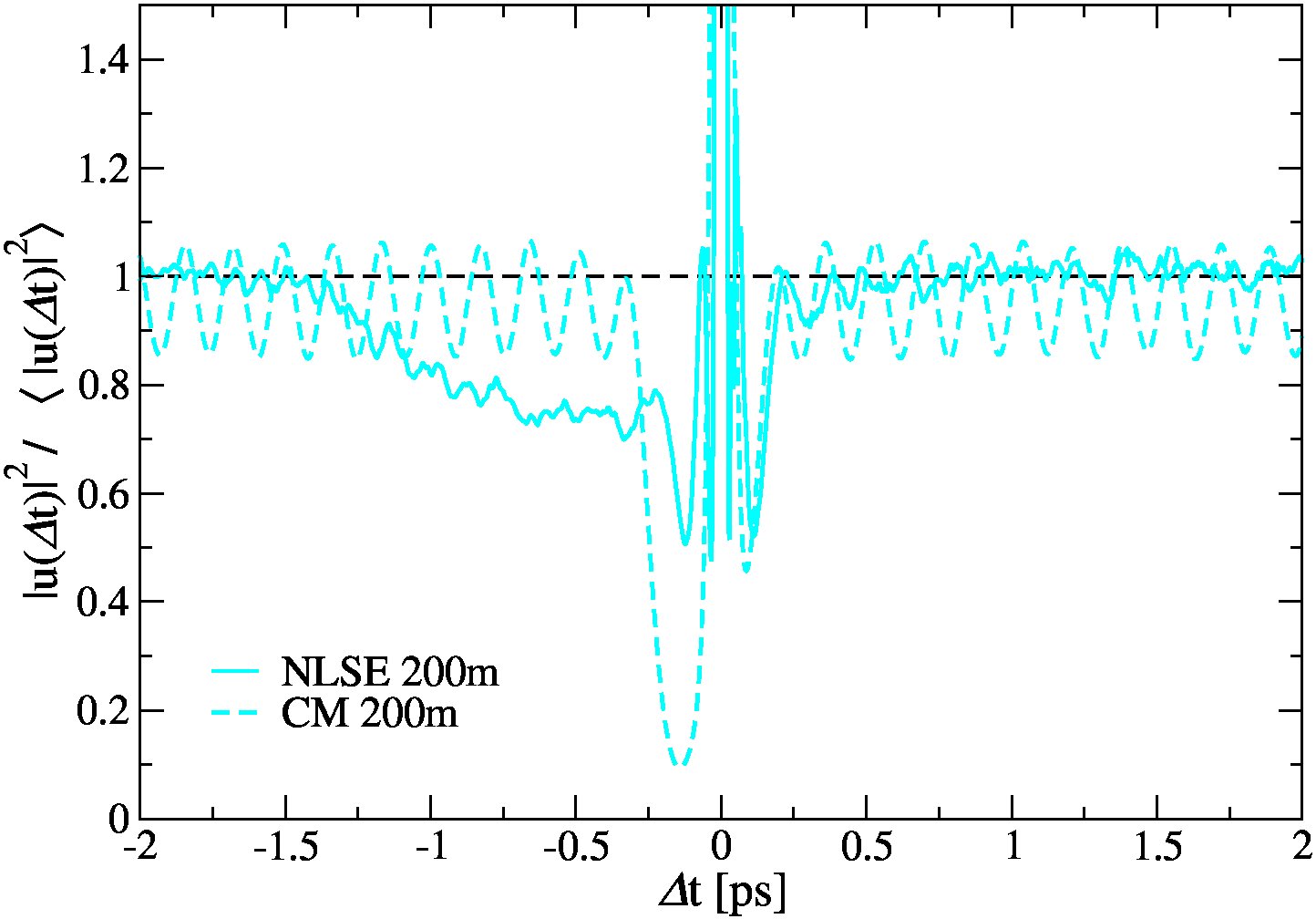}
%
%
\centering
(b)\includegraphics[width=0.95\columnwidth,clip]{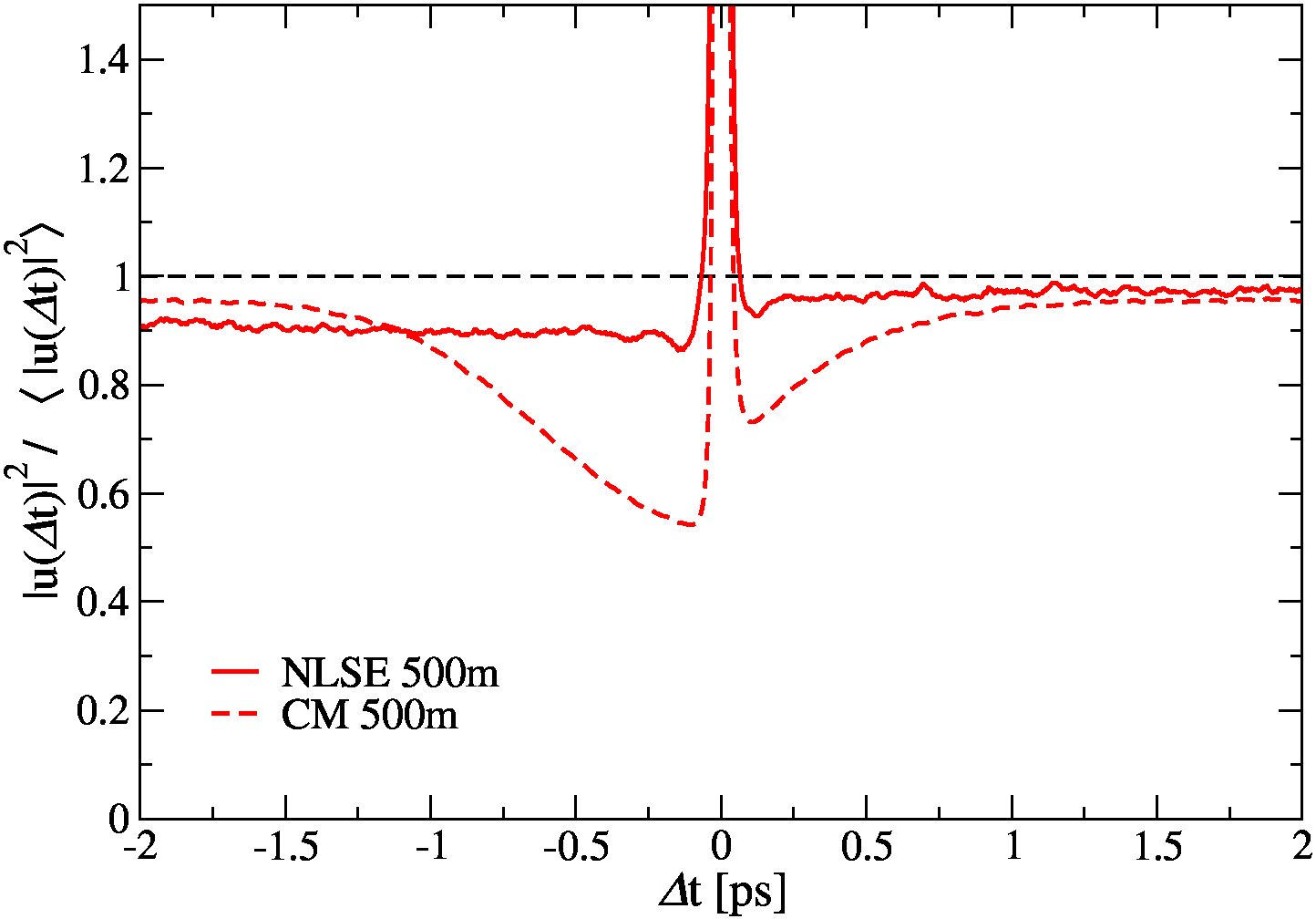}
\caption{\label{fig-calmb4storm} (Color online) 
Normalized averaged powers $|u(\Delta t)|^2/\langle |u(\Delta t)|^2 \rangle$ for times $\Delta t$ in the vicinity of a RW event at $\Delta t=0$. Panel (a), (b) corresponds to $200$ and $500$m, respectively. Solid lines in both panels
indicate averaged results for two gNLSE runs (with parameters as in Fig.\ \ref{fig-traces}), while dashed lines show the corresponding results for the cascade model. 
In both panels, we identify RWs as corresponding to powers equal to or larger than $150$W. The colours are chosen to indicate distances compatible with a full set of results $z= 150, \ldots, 1500$m given in the supplement.
Note that $|u(0)|^2/\langle |u(0)|^2 \rangle> 10$ in both panels.
}
\end{figure*}
\fi

This finding is further supported by Fig.\ \ref{fig-calmb4storm}(b), where we note that the ``calm before the storm'', already observed for the gNLSE, is even clearer and more pronounced for the cascade model. We observe strong oscillations away from $\Delta t =0$. These describe the simple quasi-soliton pulses which we used to model $u(t)$ in the cascade model. For the gNLSE, these oscillations are much less regular, although still visible. The time interval of the period of calm appears shorter in the cascade model while the amplitude reduction is stronger. We stress that the $z=100$m starting position for the cascade model remains a convenient choice. 

\section{A threshold for RW emergence}
\label{sec-collisions}

We find that the emergence of RW behaviour relies on the presence of ``enough'' $\beta_3\neq 0$ TOD --- or similar additional terms in Eq.\ \eqref{eq-gNLSE}. Even then, the conditions for the cascade to start are subtle as we show in Fig.\ \ref{fig-twosoliton} for two quasi-soliton collisions with $\beta_{3}=2.64\times10^{-42}$s$^{3}$m$^{-1}$.
\ifTEXTFIG
\begin{figure*}[tbh]
\vspace*{4ex}
\begin{centering}
\begin{minipage}{0.2\textwidth}\vspace*{-40ex}
(a)\includegraphics[width=\textwidth,clip]{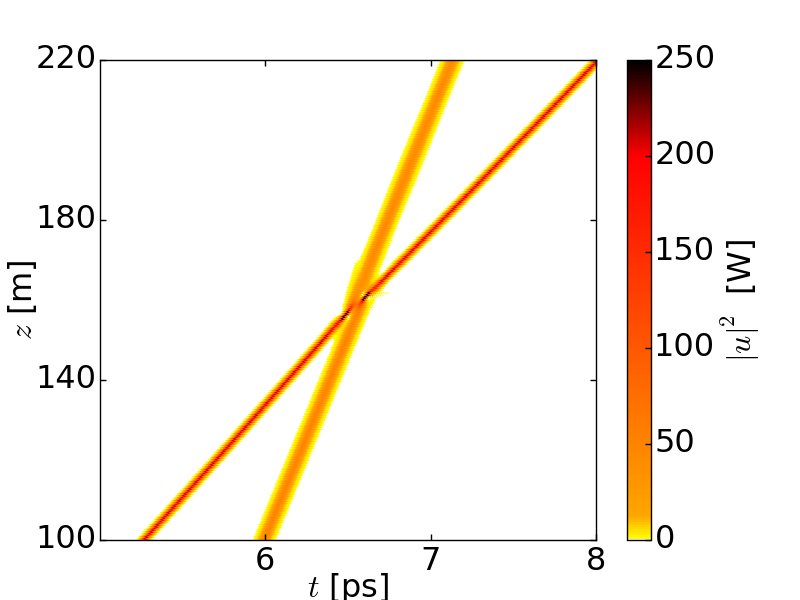}
(b)\includegraphics[width=\textwidth,clip]{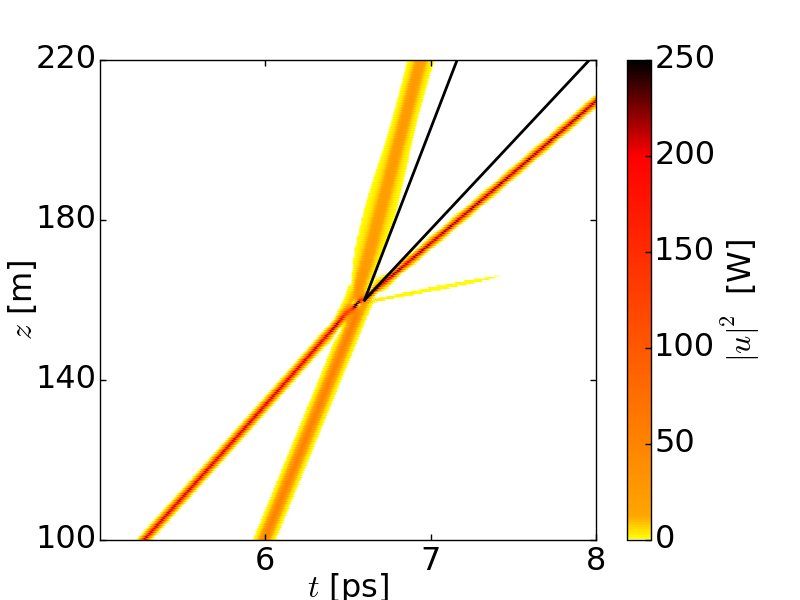}
\end{minipage}
(c)\includegraphics[width=0.37\textwidth,clip]{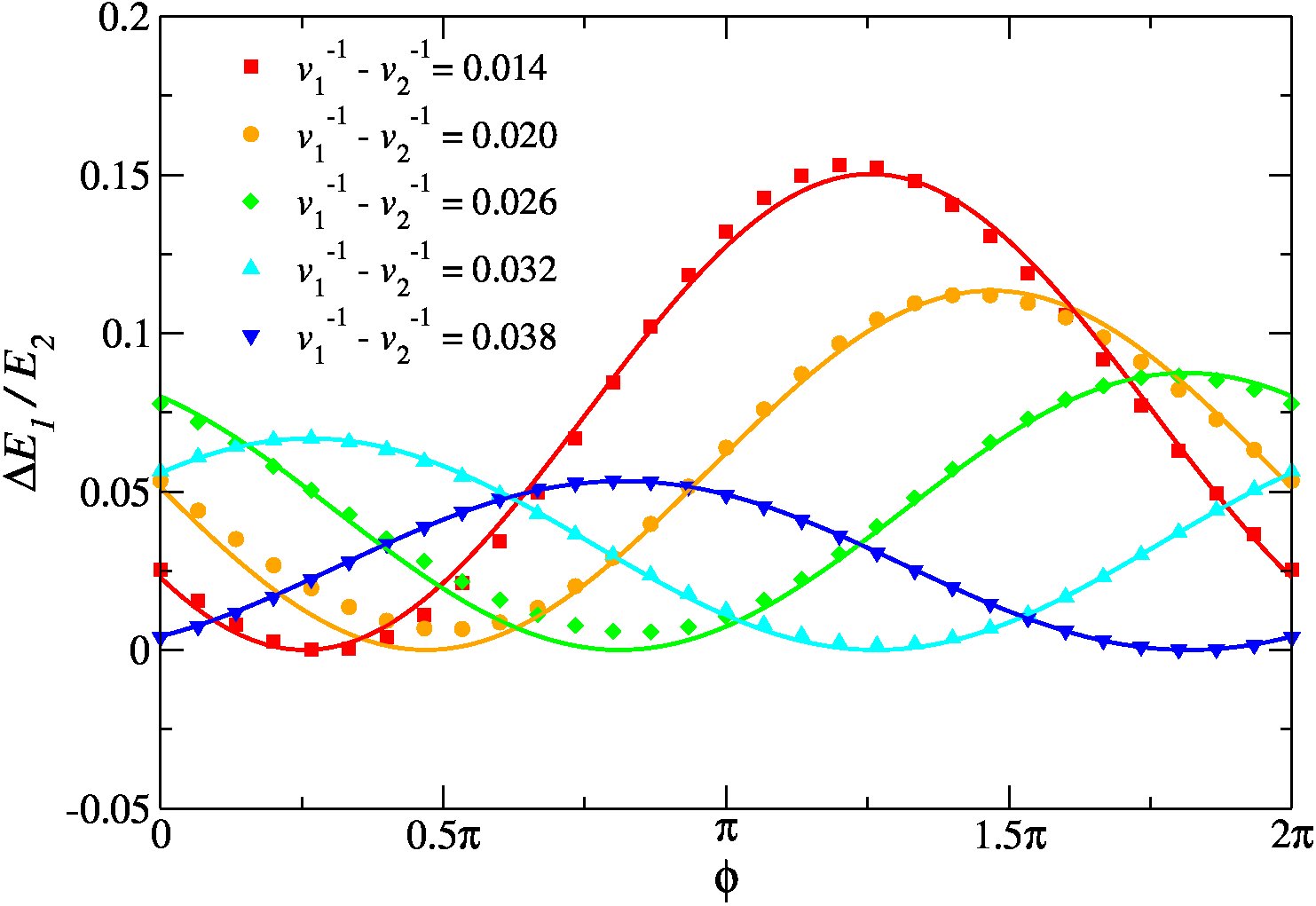}
(d)\includegraphics[width=0.365\textwidth,clip]{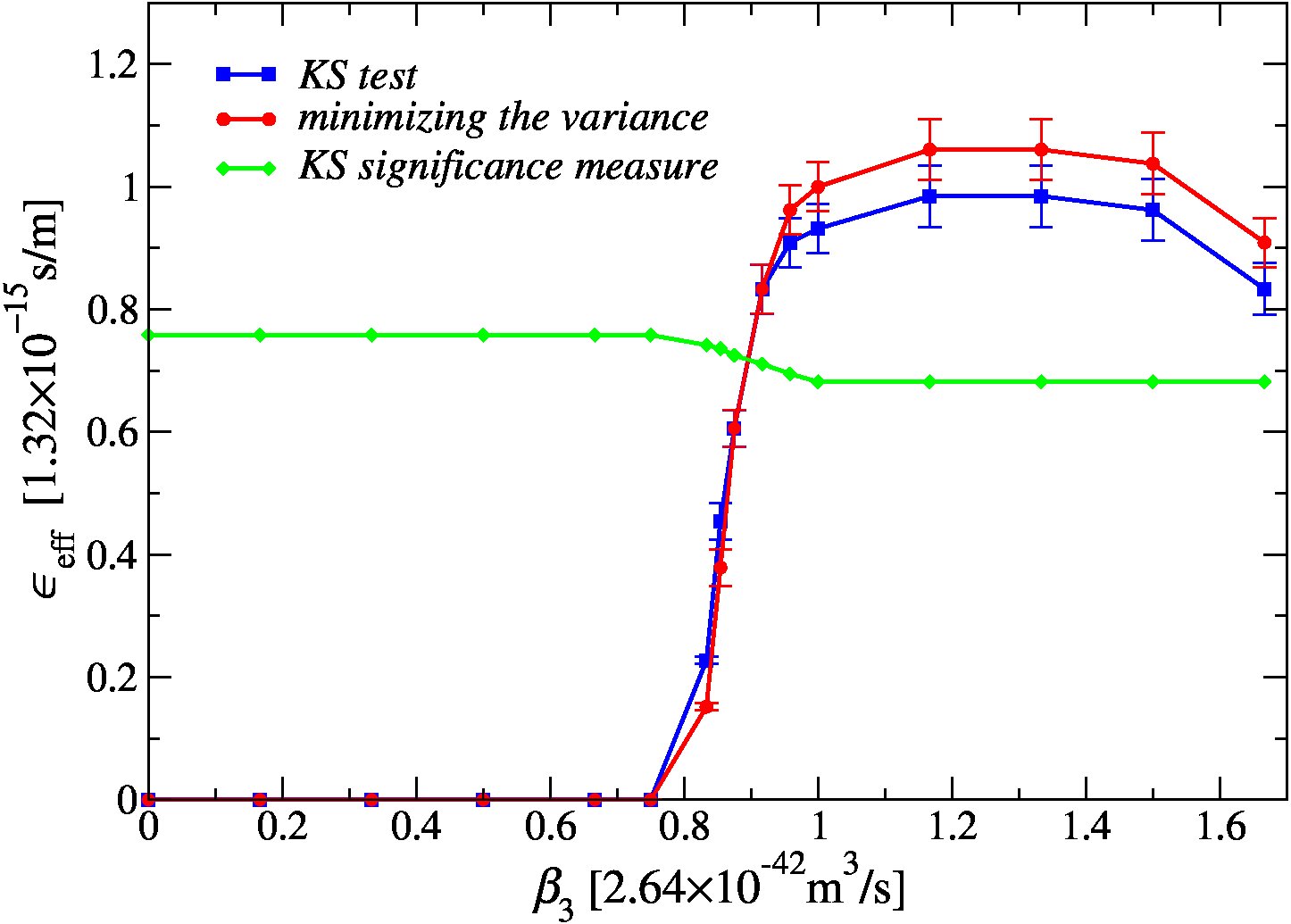}
\end{centering}

\protect\caption{(Color online) %
(a+b) Intensities $|u(z,t)|^2$ of the scattering between two quasi-solitons. The phase difference  $\phi$ was chosen to correspond to (a) the minimum and (b) the maximum of $\Delta E_1/E_2(\phi)$.
(c) $\Delta E_1/E_2(\phi)$ for various choices of initial speeds. The data points represents results of the gNLSE \eqref{eq-gNLSE} while the lines denote the fit \eqref{eq-energy-transfer-eff}.
(d) $\epsilon$ values have been obtained comparing the $\mathrm{PDF}$ from the gNLSE and the cascade model at short distance. 
}
\label{fig-twosoliton}
\end{figure*}
\fi
%
%
%
The initial conditions for both collisions have been chosen to be identical apart from the relative phase between the two quasi-solitons: an earlier quasi-soliton with initially large power ($200$W) is met by a later, initially weak pulse ($50$W).
Clearly, the collision of these two quasi-solitons as shown in (a) is (nearly) elastic, they simply exchange their velocities,
while retaining their individual power.  This situation retains much of the dynamics from the $\beta_3=0$ case.
In contrast, in (b), we see that after collision three pulses emerge: an 
early, much weaker quasi-soliton ($\sim 24$W), a later very high power
quasi-soliton ($\sim 245$W), and a final, very weak and dispersive wave
($\sim 0.002$W) --- the collision is highly inelastic.
We note that this process is similar to what was described in Refs.\ \onlinecite{KarS81,BurA94} for other NLSE variants.
Systematically studying many such collisions, we find that 
the outcome can be modelled quite accurately
using Eq.\ \eqref{eq-energy-transfer-eff} which is similar as in a two-body resonance process. 
In Fig.\ \ref{fig-twosoliton}(c), we show that agreement between Eq.\ \eqref{eq-energy-transfer-eff} and the numerical simulations is indeed remarkably good. Indeed, \eqref{eq-energy-transfer-eff} was used to generate the cascade model results for Figs.\ \ref{fig-PDF} and \ref{fig-traces}.
We have further verified the accuracy of \eqref{eq-energy-transfer-eff} by simulating a large number of
individual collision processes with varying relative
phases and varying initial quasi-soliton parameters. 
%
In Eq.\ \eqref{eq-energy-transfer-eff}, 
$\epsilon_{\mathrm{eff}}$ is an empirical cross-section
coefficient (see the supplement for an analytical  justification assuming two-particle scattering). 
It depends on $\beta_3$ as shown in Fig.\
\ref{fig-twosoliton}(d).  
The transition from a regime without RWs to a regime with well pronounced RWs appears rather abrupt. This indicates that already small, perhaps only local changes in $\beta_3$, and hence in the local composition of the optical fibre itself \cite{AkhKBB16,DegMGF16}, can lead to dramatic changes for the emergence of RWs. Indeed, we find that once a RW has established itself in the large-$\beta_3$ region, it continues to retain much of its amplitude when entering a $\beta_3=0$ region.

\section{Discussion and Conclusions}
\label{sec-conclusions}

Our results emphasize the crucial role played by quasi-soliton interactions in the energy exchange underlying the formation of RWs via the proposed cascade mechanism. While interactions are known to play an important role in RW generation \cite{MusKKL09,SluP16,Sun16}, the elucidation of the full cascade mechanism including its resonance-like quasi-soliton pair scattering and details such as the "calm before the storm", will be essential ingredients of any attempt at RW predictions. In addition, these features are quite different from linear focusing of wave superpositions \cite{OnoOSC05,HohKSK10} and allow the experimental and observational distinction of both mechanisms.

RWs emerge when $\beta_3$ is large enough as shown in Fig.\ \ref{fig-twosoliton}. Their appearance is very rapid in a short range $0.8 \lesssim \beta_3/2.64\times 10^{-42} \mathrm{s}^3\mathrm{m}^{-1}\lesssim 1$. This can be understood as follows: the dispersion relation, in the moving frame, is $\beta(\omega)= \beta_2 (\omega - \omega_0)^2/2 + \beta_3 (\omega - \omega_0)^3 /6$. The anomalous dispersion region of $\beta(\omega)<0$, with soliton-like excitations, ends at $\omega_c - \omega_0 \geq - 3 \beta_2 / \beta_3$ beyond which $\beta(\omega)\geq 0$ and dispersive waves emerge. From \eqref{eq-period}, we can estimate the spectral width as $2 (\omega_c - \omega_0) = 2 \pi \sqrt{\gamma P/|\beta_2|}$. This leads to the condition
\begin{equation}
\beta_3 \geq \frac{3 |\beta_2|}{\pi} \sqrt{\frac{|\beta_2|}{\gamma P}} \approx 0.9 \times (2.64 \times 10^{-42}) \mathrm{s}^3\mathrm{m}^{-1},
\label{eq-beta3c}
\end{equation}
which is in very good agreement with the numerical result of Fig.\ \ref{fig-twosoliton}(c). In Eq.\ \eqref{eq-beta3c} the $\beta_3$ threshold that leads to fibres supporting RWs depends on the peak power $P$. In deriving the numerical estimate in \eqref{eq-beta3c} we have used a typical $P \sim 50$W as appropriate after about $\sim 100$m (cp.\ Fig.\ \ref{fig-PDF} and also the movies in \cite{EbeSMR16b}). Once such initial, and still relatively weak RWs have emerged, the condition \eqref{eq-beta3c} will remain fulfilled upon further increases in $P$ due to quasi-soliton collisions, indicating the stability of large-peak-power RWs.
%
Our estimation of the effective energy-transfer cross-section parameter $\epsilon_\mathrm{eff}$ can of course be improved. However, we believe that \eqref{eq-energy-transfer-eff} captures the essential aspects of the quasi-soliton collisions already very well. 
Thus far, we have ignored fibre attenuation. Clearly, this would cause energy dissipation and eventually lead to a reduction in the growth of RWs and hence give rise to finite RW lifetimes. But as long as the fibre contains colliding quasi-solitons of enough power, RWs will still be generated. 

Up to now, we have used the term RW only loosely to denote high-energy quasi-solitons as shown in Figs.\ \ref{fig-traces} and \ref{fig-PDF}. Indeed, a strict definition of a RW is still an open question and qualitative definitions such as \emph{a pulse whose amplitude} (or energy or power) \emph{is much higher than surrounding pulses} are common \cite{AkhKBB16}. Our results now suggest a quantifiable operational definition at least for \emph{normal} waves in optical fibres described by the gNLSE: a large amplitude wave is \emph{not} a RW if it occurs as frequently as expected for the PDF at $\beta_3=0$ (cp.\ Fig.\ \ref{fig-PDF}).  
We emphasise that both high spatial and temporal resolution are required to obtain reliable statistics for RWs in optical fibres for reliable predictions of the PDF. A small time window in the simulation can severely distort the tails of the PDF, and hence their correct interpretation (see supplement).

We find that RWs are preceded by short periods of reduced wave amplitues. This ``calm before the storm'' has been observed previously \cite{BirBDS15} in ocean and in optical multifilament RWs, but not yet in studies of optical fibres. We remark that we first noticed the effect in our cascade model, before investigating it in the gNLSE as well. This highlights the usefulness of the cascade model for qualitatively new insights into RW dynamics. More results are needed to ascertain if the periods of calm can be used as reliable predictors for RW occurrence, i.e., reducing false positives.

\section{Methods}
\label{sec-methods}


The numerical simulations of \eqref{eq-gNLSE} were performed using the
split-step Fourier method \cite{Agr13} in the co-moving frame of reference. 
A massively parallel implementation based on the discard-overlap/save method
\cite{PreFTV92C} was implemented to allow for simulations with $2^{31}$ intervals of $\Delta t = 1.8$fs and hence long time windows
up to $3.865\times10^6$ps with several kilometres in propagation distance.  We assume periodic boundary
conditions in time and, as usual, a coordinate frame moving with the group velocity.  
The code was shown to scale linearly up to 98k cores (see Supplement).

We start the simulations with a continuous wave of $P_0=10$W power at
$\lambda_{0}=1064$nm.  For the fibre, we assume the parameters
$\beta_2=-2.6\times10^{-28}$s$^{2}$m$^{-1}$, $\gamma=0.01$
W$^{-1}$m$^{-1}$, and varying
$\beta_{3}$ up to $1.7 \times (2.64\times10^{-42}$s$^{3}$m$^{-1})$, see Fig.\ \ref{fig-twosoliton}.  Due to the modulation
instability, we observe, after seeding with a small $10^{-3}$W Gaussian noise, a break-up into individual pulses within the first
$100$m of the simulation with a density of $\sim 5.88$ pulses/ps.  Throughout
the simulation, we check that the energy remains conserved.  The PDF of
$|u|^2$ is computed as the simulation progresses.
For the two-quasi-soliton interaction study, we use the massively-parallel
code as well as a simpler serial implementation.  The collision runs are
started using pulses of the quasi-soliton shapes \eqref{eq-quasi-soliton}
with added phase difference $\exp(i \phi)$ in the advanced pulse.

The effective cascade model assumes an initial condition of quasi-solitons of the
same density of $5.88$ quasi-solitons/ps.  Their starting times are evenly distributed with separation $\Delta t\sim 0.17$ps while their
initial powers are chosen to mimic the distribution observed
in the gNLSE at $z\sim 100$m. The time resolution is $2 \times 10^{-3}$ps and we simulate the propagation in $1500$ replicas of time windows of $4000$ps duration. This gives an effective duration of $6\times 10^{6}$ps. For all such $30\times 10^{6}$ quasi-soliton pulses, we
compute their speeds, find the distance at which the next two-quasi-soliton
collision will occur and compute the quasi-soliton energy exchange via
\eqref{eq-energy-transfer-eff}.  The power of the emerging two pulses is
calculated from $E_q=2 P_q T_q$ and the algorithm 
proceeds to find the next collision.  The PDF of $|u|^2$ is computed 
assuming that the shape of each quasi-soliton is given by
\eqref{eq-quasi-soliton}.

\subsection*{Acknowledgements}

We thank George Rowlands for stimulating discussions in the early stages of the project. We are grateful to the EPSRC for provision of computing resources through
the MidPlus Regional HPC Centre (EP/K000128/1), and the national facilities
HECToR (e236, ge236) and ARCHER (e292).  We thank the Hartree Centre for use
of its facilities via BG/Q access projects HCBG055, HCBG092, HCBG109. 

\subsection*{Contributions}

ME, AM and RAR planned the study. ME and RAR computed the high-precision PDFs;
ME, RAR and AS studied the two-soliton collisions.  ME and RAR devised the cascade model; AS ran the
cascade simulations.  All authors discussed the results and worked on the
paper. Data created during this research are openly available from the Aston University data archive \cite{EbeSMR16b}.

\subsection*{Competing financial interests}

The authors declare no competing financial interests.

\subsection*{Supporting Citations} References \cite{EbeSMR16b,Nag10,PreFTV92C,CheTE10,SanA03} appear in the Supporting Material.




\ifNOFIG
\clearpage
\fi 

\ifSFIG
\pagestyle{empty}

\clearpage

\setcounter{figure}{0}

\newpage

\begin{figure}[h]
\begin{centering}
(a)\includegraphics[width=\columnwidth,clip]{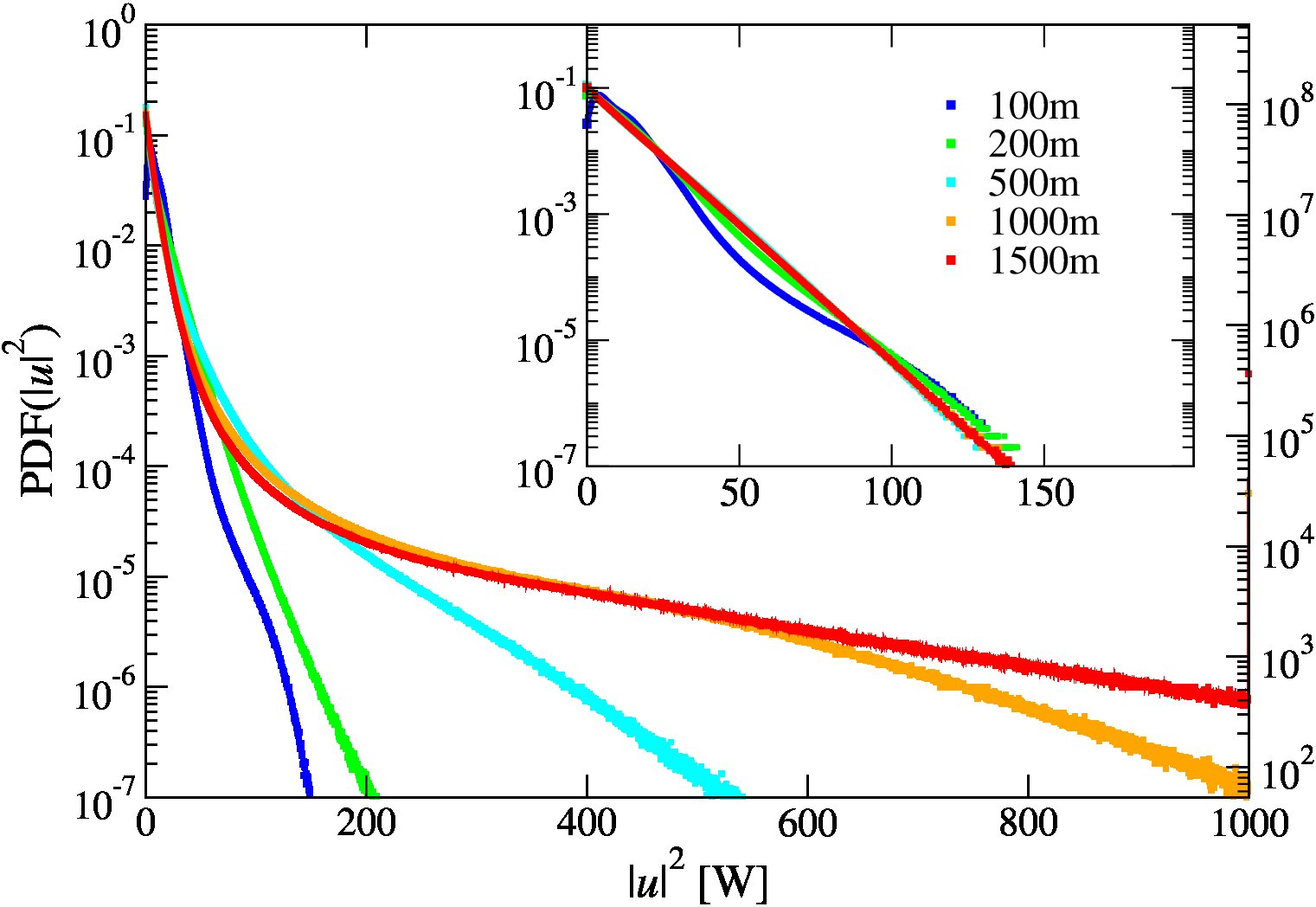}
\par\end{centering}


\begin{centering}
(b)\includegraphics[width=\columnwidth, clip]{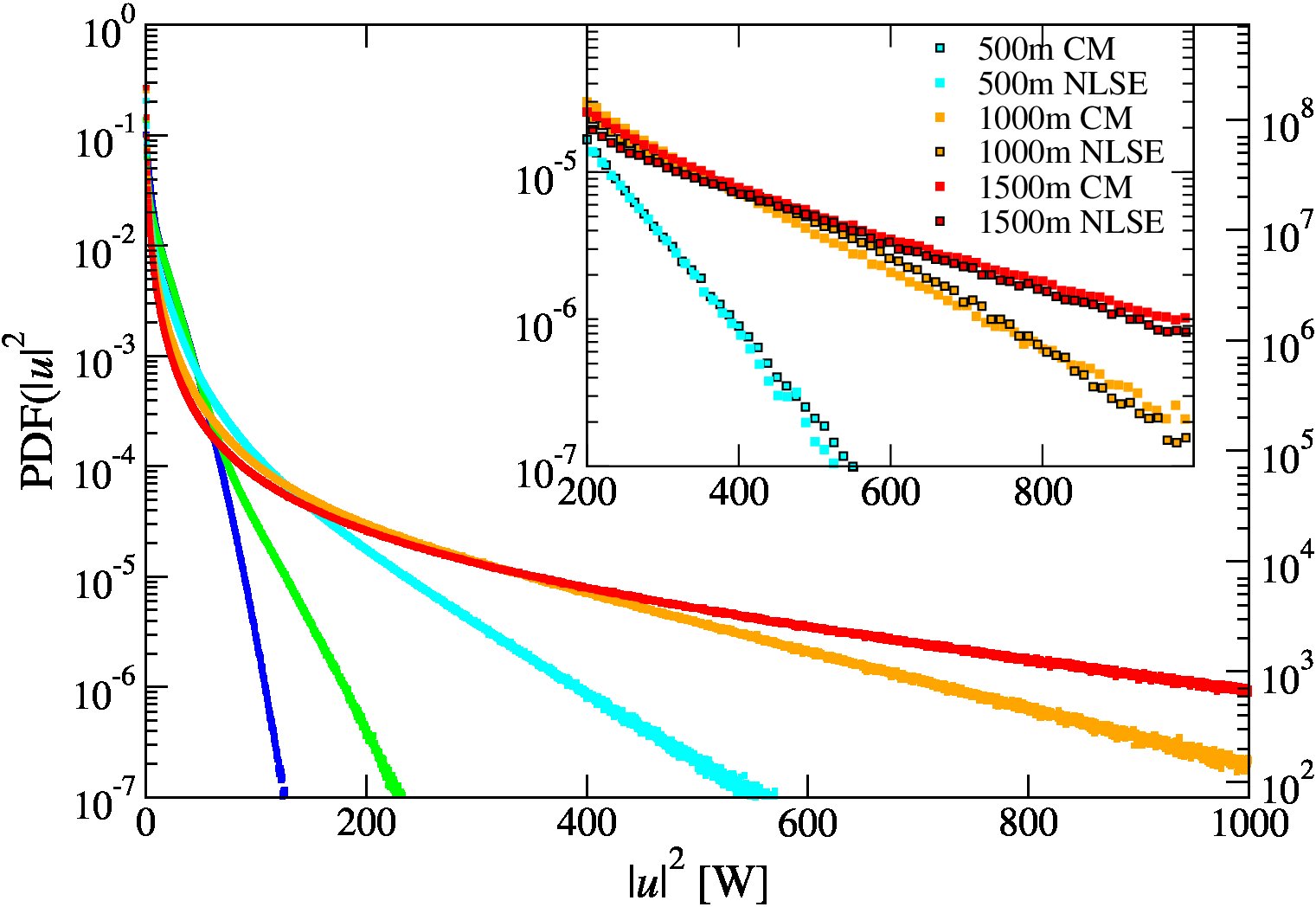}
\par\end{centering}

\protect\caption{(Color online) %
(a) PDFs of the intensity $|u|^{2}$ from the gNLSE \eqref{eq-gNLSE} at $\beta_{3}=2.64\times10^{-42}$s$^{3}$m$^{-1}$ using a large
time window of $\Delta t=3.865\times10^6$ps. The PDFs have been computed at distances $z= 100$m, $200$m, $500$m, $1000$m and $1500$m. The left vertical axis denotes the values of the normalized PDF while the right vertical axis gives the event count per bin.
The inset shows results for $\beta_{3}=0$.
%
%
%
(b) PDFs of the intensity $|u|^{2}$ from the cascade model for the same distances as in (a), using the same symbol and axes conventions. The inset shows a comparison between the results from the gNLSE (colored lines) and the cascade model (black lines and symbol outlines) for $z=500$m, $1000$m and $1500$m. Only every 50th symbol is shown.}
\label{fig-PDF}
\end{figure}

\newpage

\begin{figure}[h]
\begin{centering}
(a)\includegraphics[width=\columnwidth,clip]{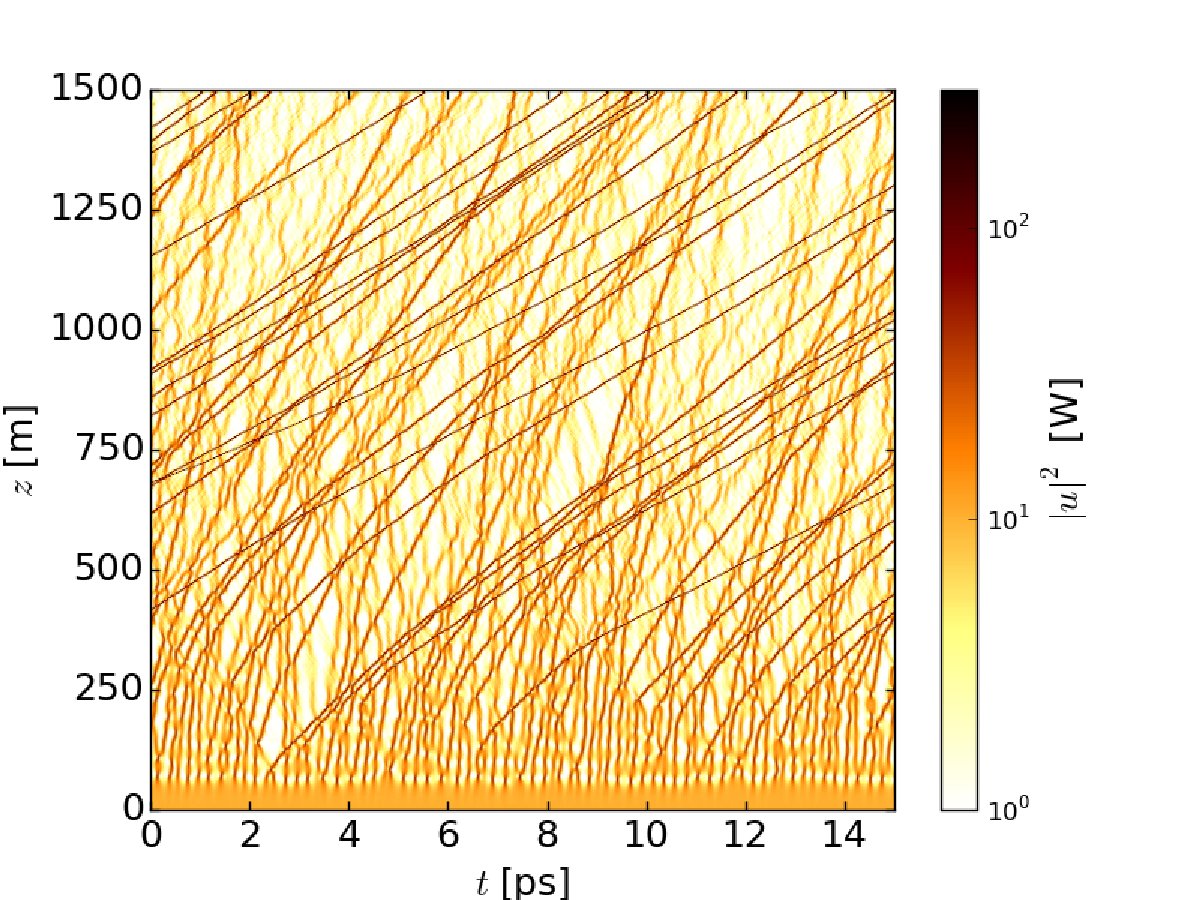}
\end{centering}

\begin{centering}
(b)\includegraphics[width=0.44\columnwidth,clip]{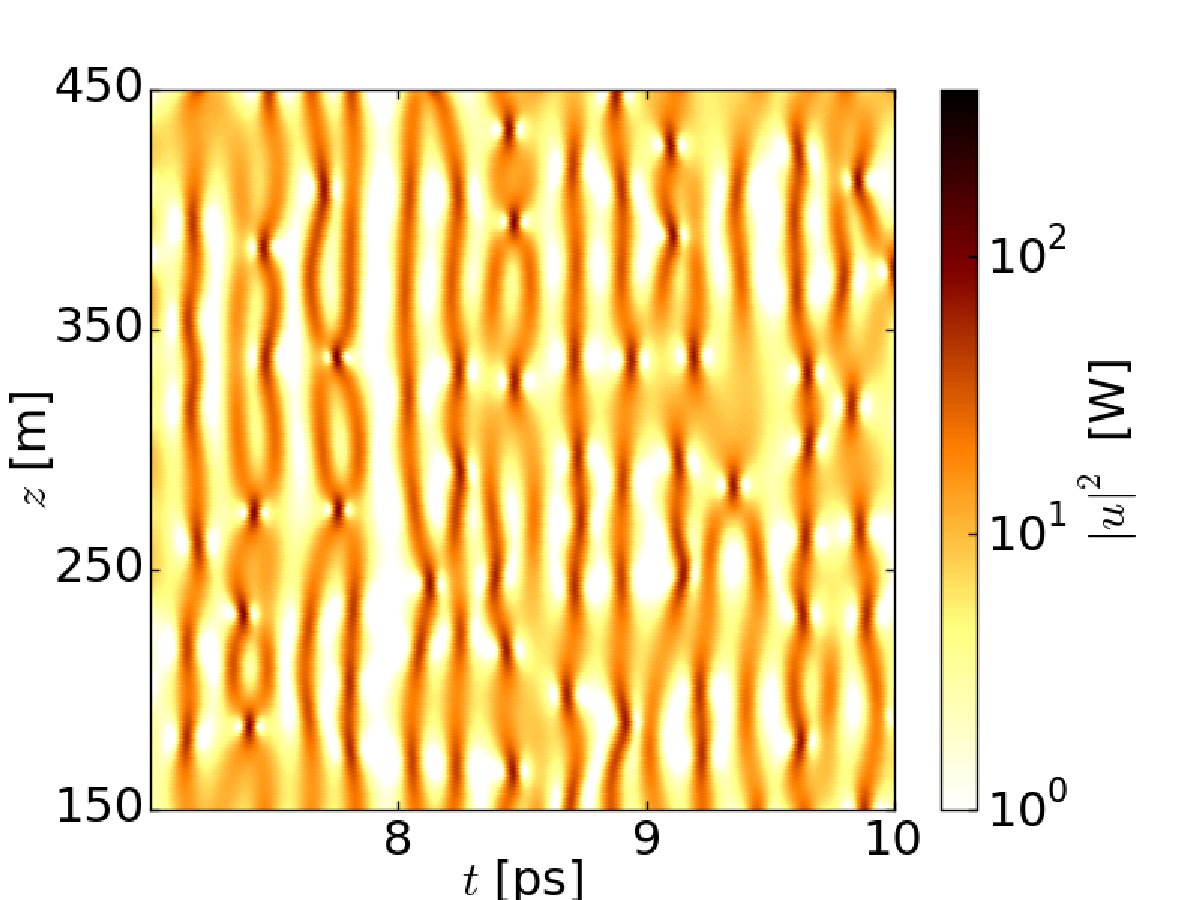}
(c)\includegraphics[width=0.44\columnwidth,clip]{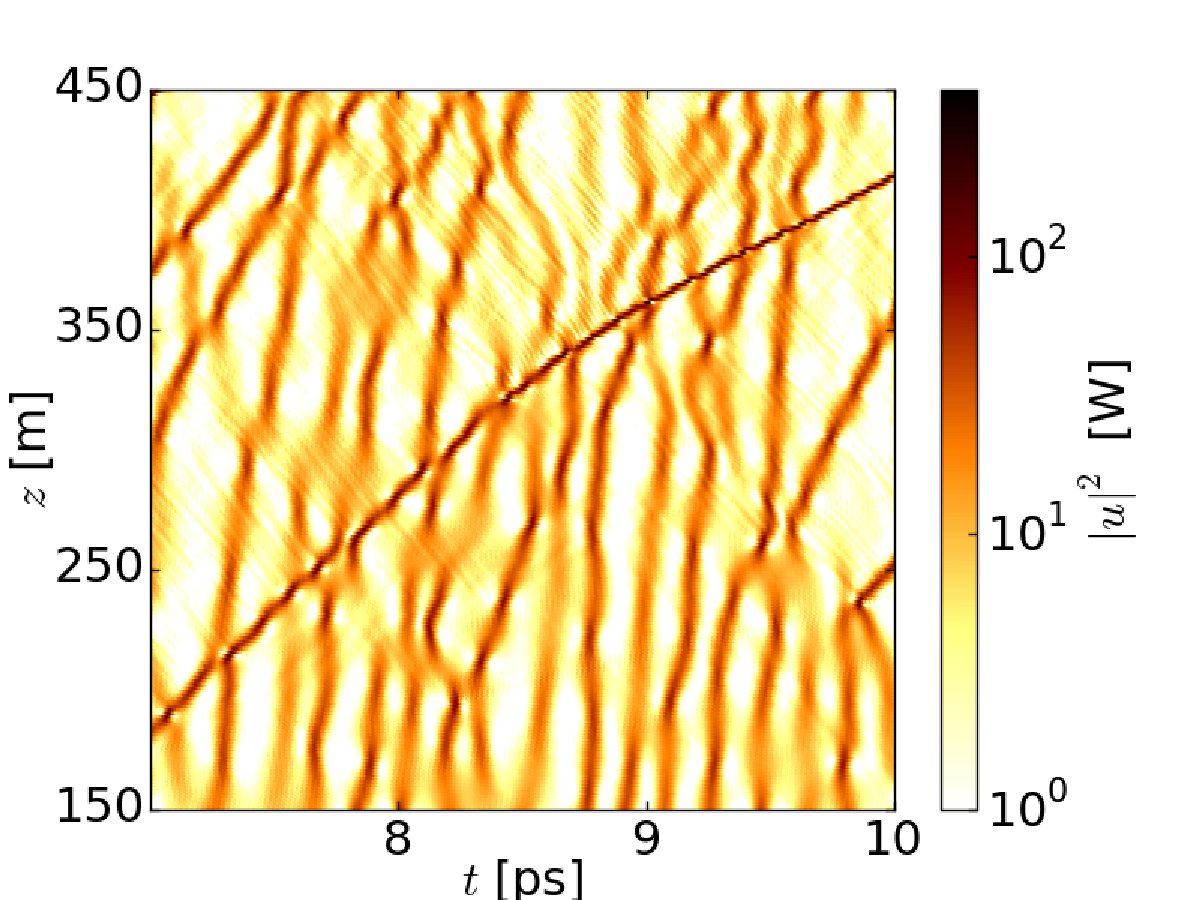}
\end{centering}

\begin{centering}
(d)\includegraphics[width=\columnwidth,clip]{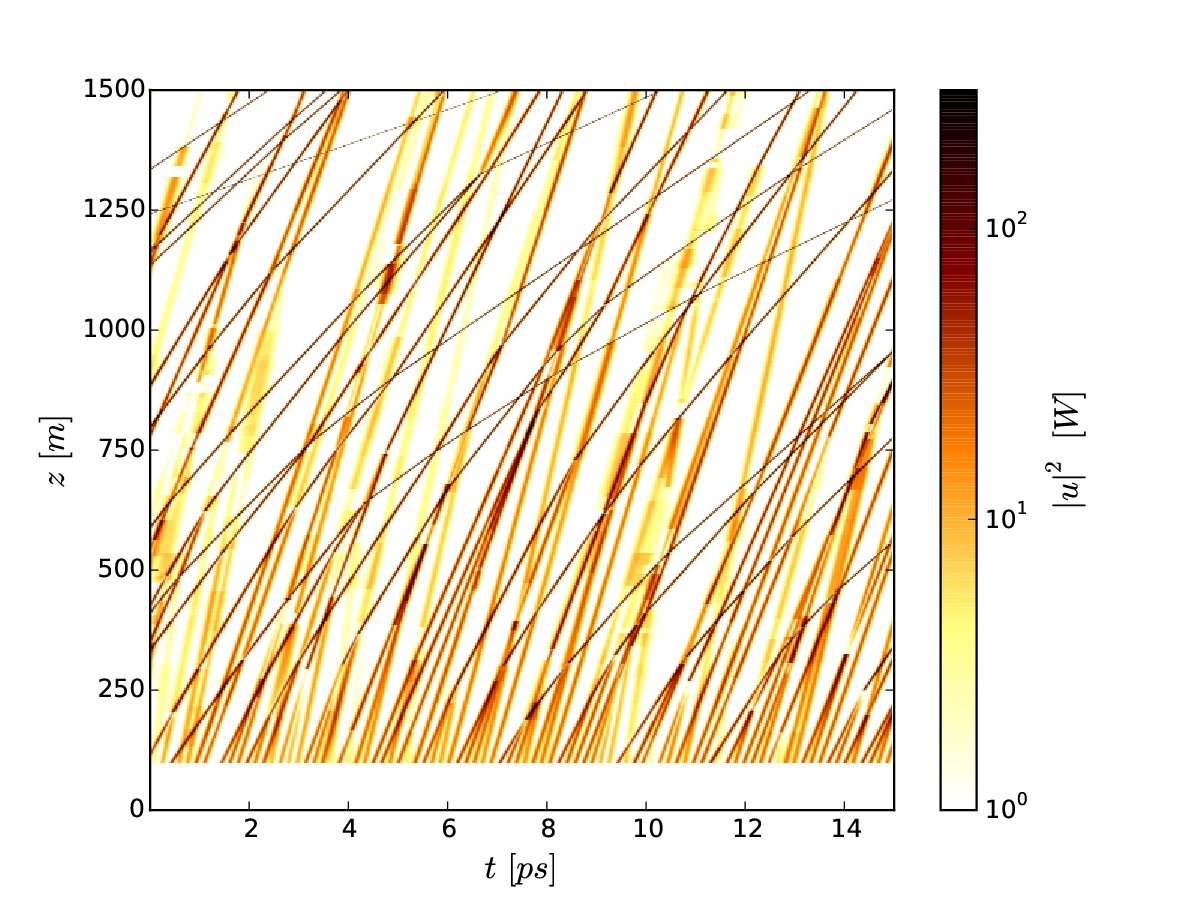}
\end{centering}

\protect\caption{(Color online)
(a) Intensity $|u(z,t)|^{2}$ for $\beta_{3}=2.64\times10^{-42}$s$^{3}$m$^{-1}$ of the gNLSE Eq.\ \eqref{eq-gNLSE} as function of the time
$t$ and distance $z$ in a selected time frame of $\Delta t= 15$ps and distance range $\Delta z=1.5$km. 
(b) $|u|^{2}$ with $\beta_{3}=0$ for a zoomed-in distance and time region,
(c) $|u|^{2}$  with $\beta_{3}$ value as in (a) for a region of (a) with $\Delta t$ and $\Delta z$ chosen identical to (b).
(d) Intensities $|u|^{2}$ as computed from the effective cascade model using the same shading/color scale as in (a). Note that we start the effective model at $z_0=100$m to mimic the effects of the modulation instability in (a).}
\label{fig-traces}
\end{figure}

\newpage

\begin{figure}[h]

\begin{centering}
(a)\includegraphics[width=0.44\columnwidth,clip]{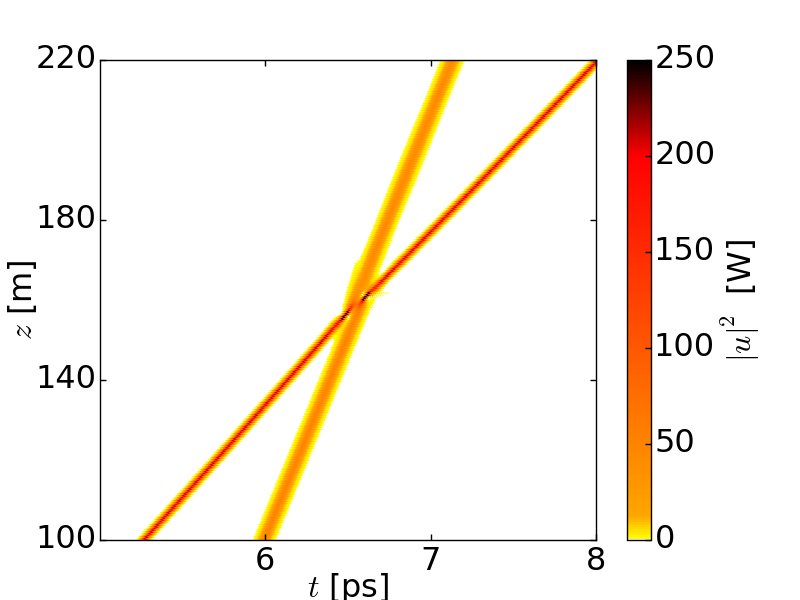}
(b)\includegraphics[width=0.44\columnwidth,clip]{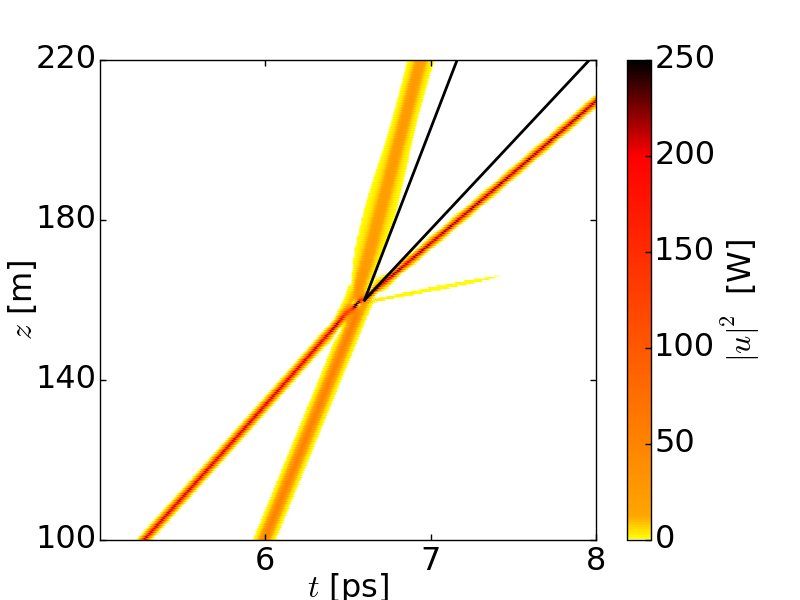}
\par\end{centering}

\begin{centering}
(c)\includegraphics[width=\columnwidth,clip]{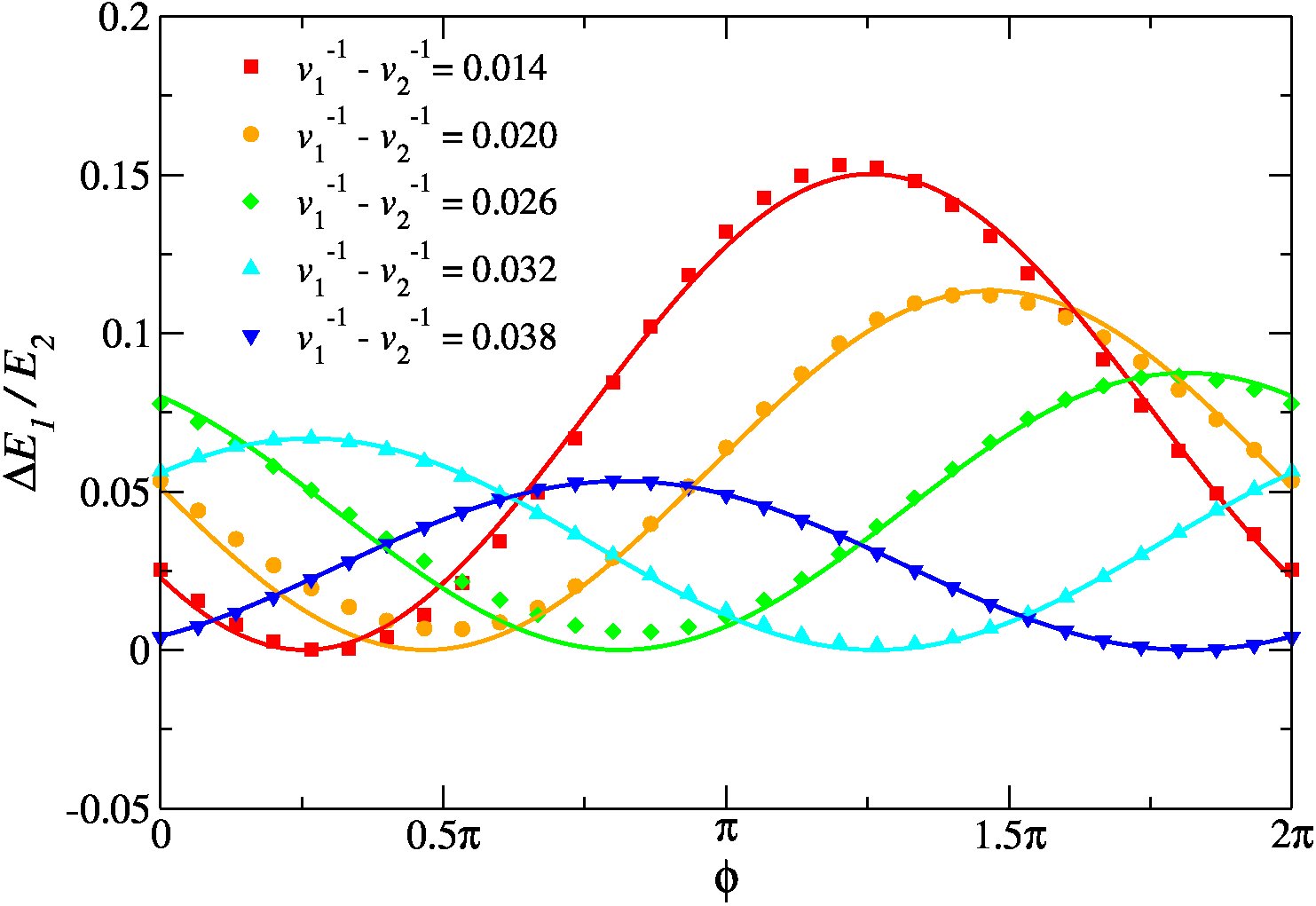}
\par\end{centering}

\begin{centering}
(d)\includegraphics[width=\columnwidth,clip]{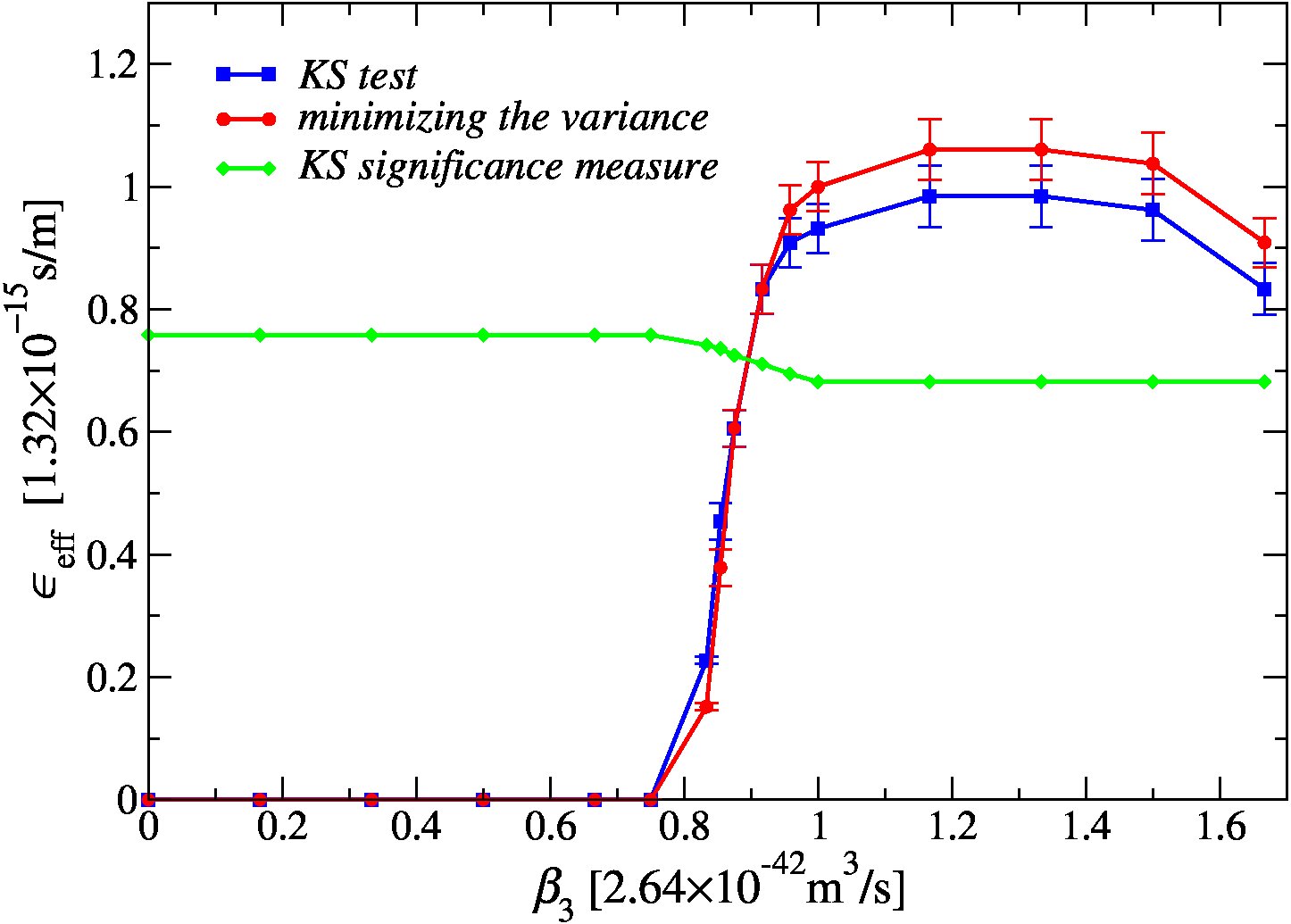}
\par\end{centering}

\protect\caption{(Color online) %
(a+b) Intensities $|u(z,t)|^2$ of the scattering between two quasi-solitons. The phase difference  $\phi$ was chosen to correspond to (a) the minimum and (b) the maximum of $\Delta E_1/E_2(\phi)$.
(c) $\Delta E_1/E_2(\phi)$ for various choices of initial speeds. The data points represents results of the gNLSE \eqref{eq-gNLSE} while the lines denote the fit \eqref{eq-energy-transfer-eff}.
(d) $\epsilon$ values have been obtained comparing the $\mathrm{PDF}$ from the gNLSE and the cascade model at short distance. 
}
\label{fig-twosoliton}
\end{figure}

\begin{figure}[h]


\centering
(a)\includegraphics[width=\columnwidth,clip]{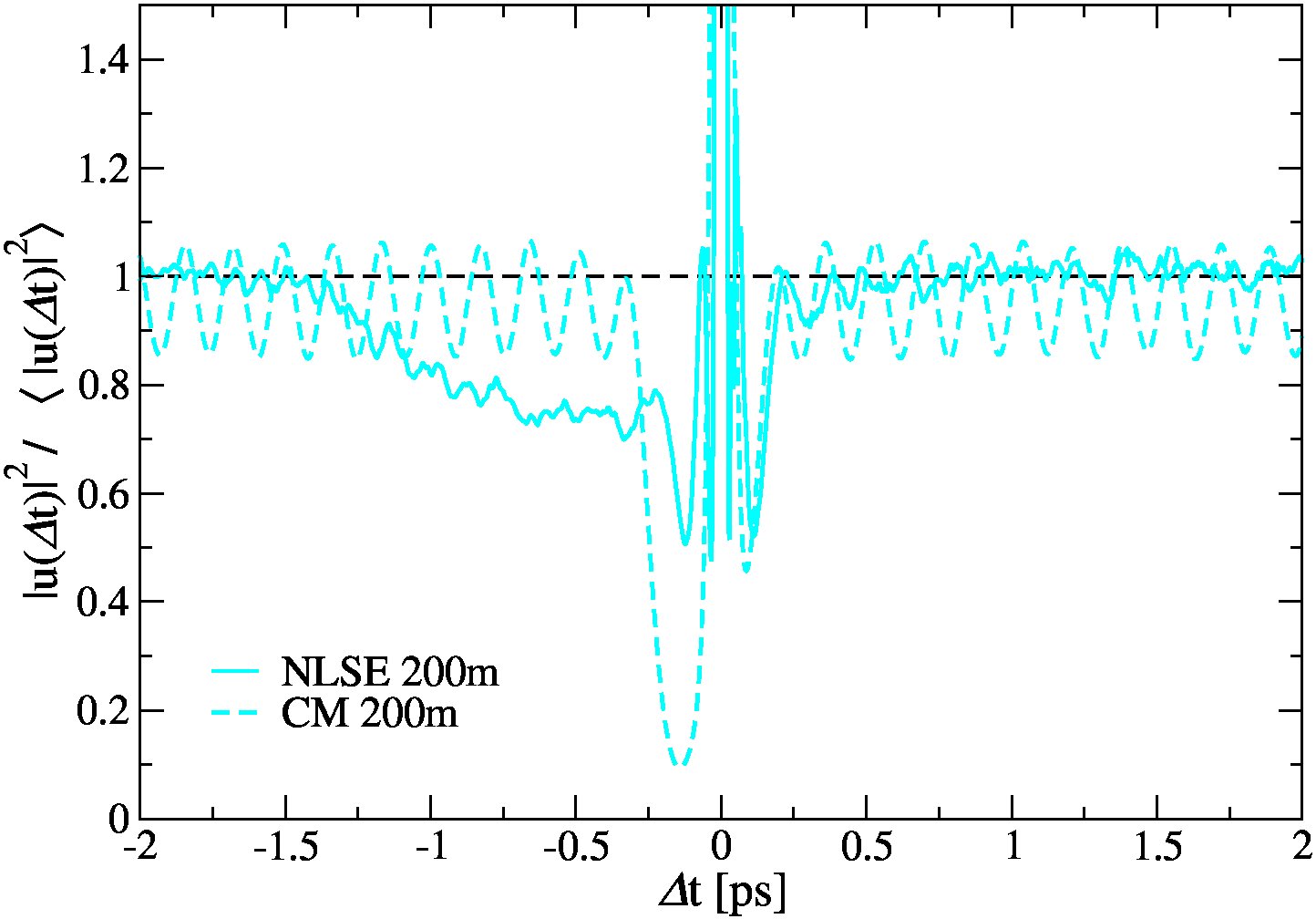}


\centering
(b)\includegraphics[width=\columnwidth,clip]{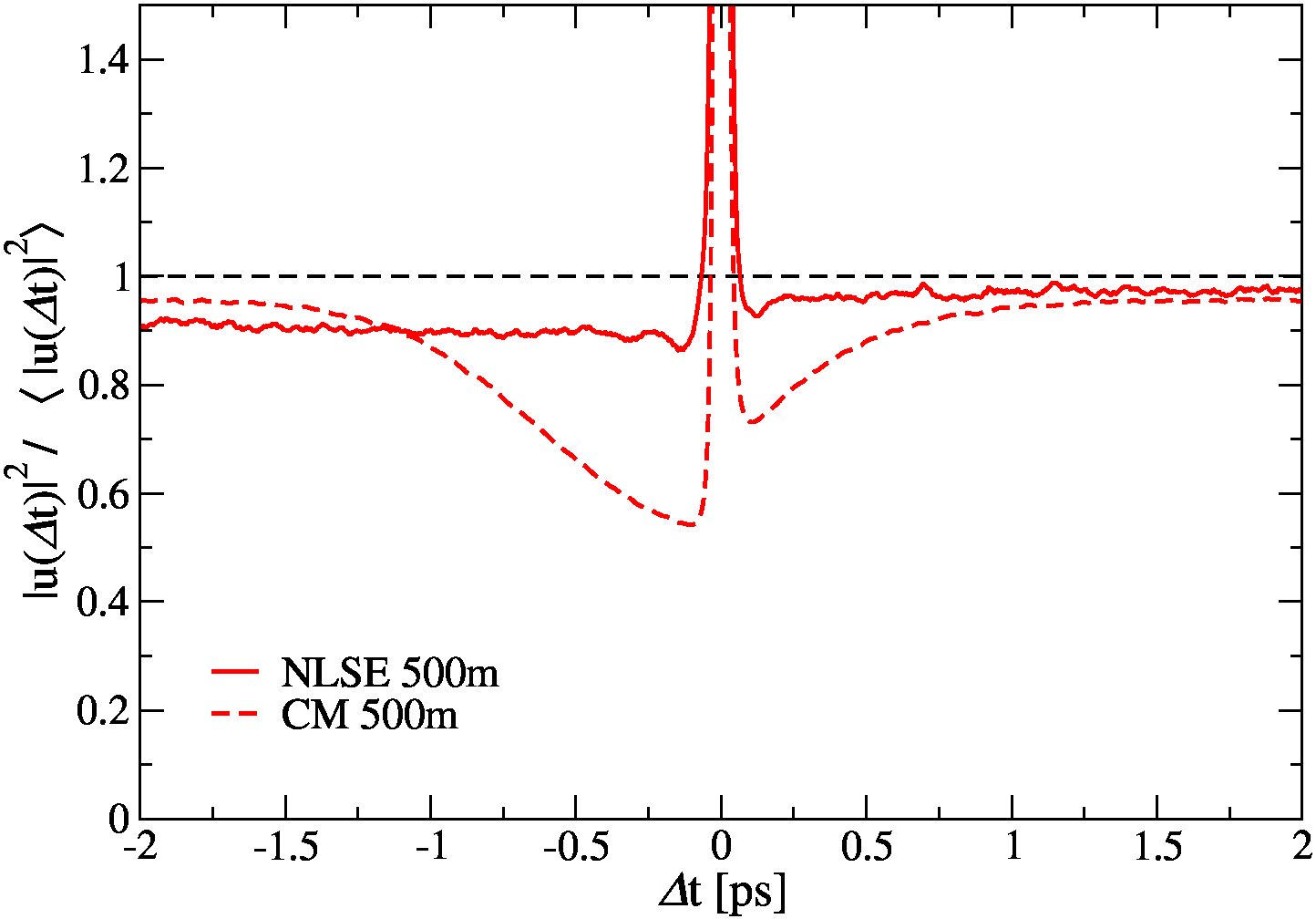}

\caption{\label{fig-calmb4storm} (Color online) 
Normalized averaged powers $|u(\Delta t)|^2/\langle |u(\Delta t)|^2 \rangle$ for times $\Delta t$ in the vicinity of a RW event at $\Delta t=0$. Panel (a), (b) corresponds to $200$ and $500$m, respectively. Solid lines in both panels
indicate averaged results for two gNLSE runs (with parameters as in Fig.\ \ref{fig-traces}), while dashed lines show the corresponding results for the cascade model. 
In both panels, we identify RWs as corresponding to powers equal to or larger than $150$W. The colours are chosen to indicate distances compatible with a full set of results $z= 150, \ldots, 1500$m given in the supplement.
Note that $|u(0)|^2/\langle |u(0)|^2 \rangle> 10$ in both panels.
}
\end{figure}

\newpage

\fi

\ifXLFIG
\pagestyle{empty}
\clearpage

\setcounter{figure}{0}

\newpage
\begin{figure*}[h]
\begin{centering}
(a)\includegraphics[width=\columnwidth]{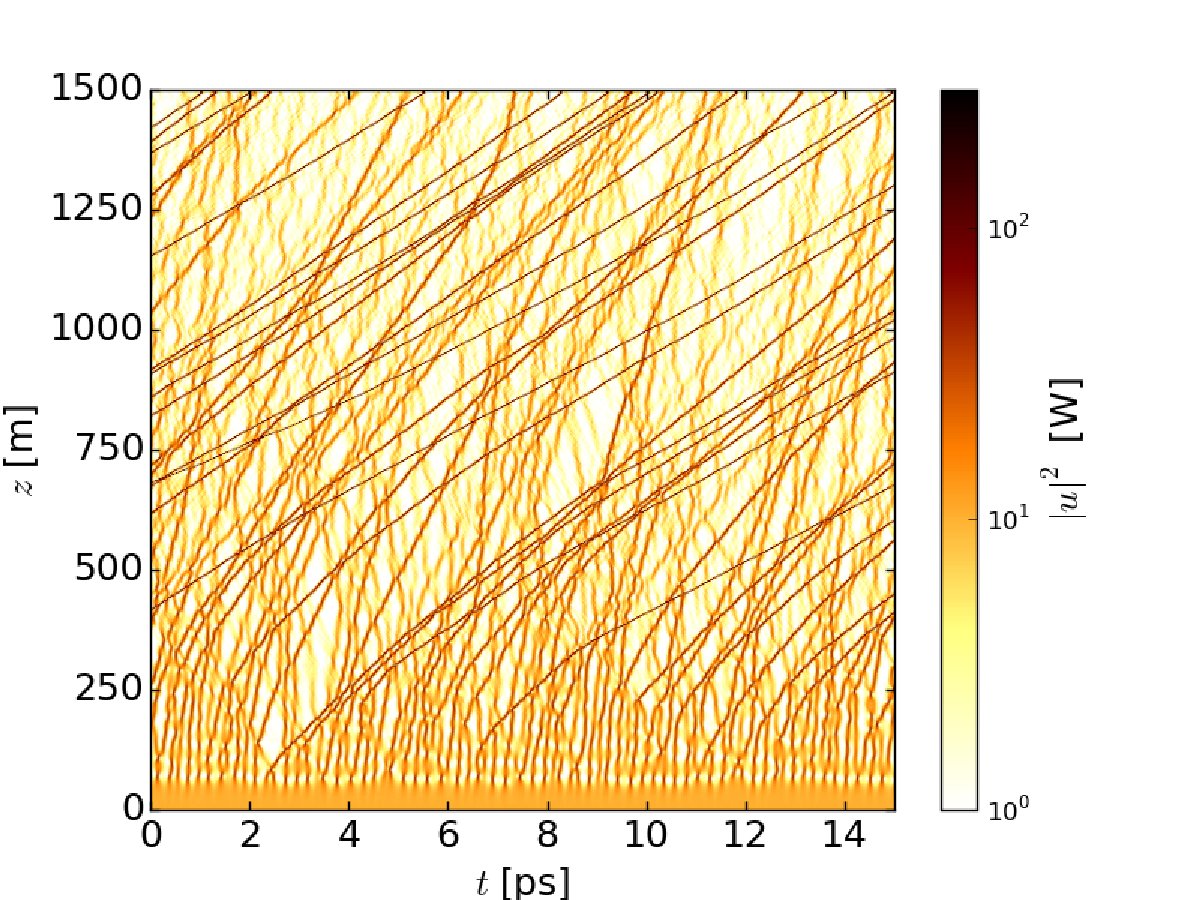}

(b)\includegraphics[width=0.44\columnwidth]{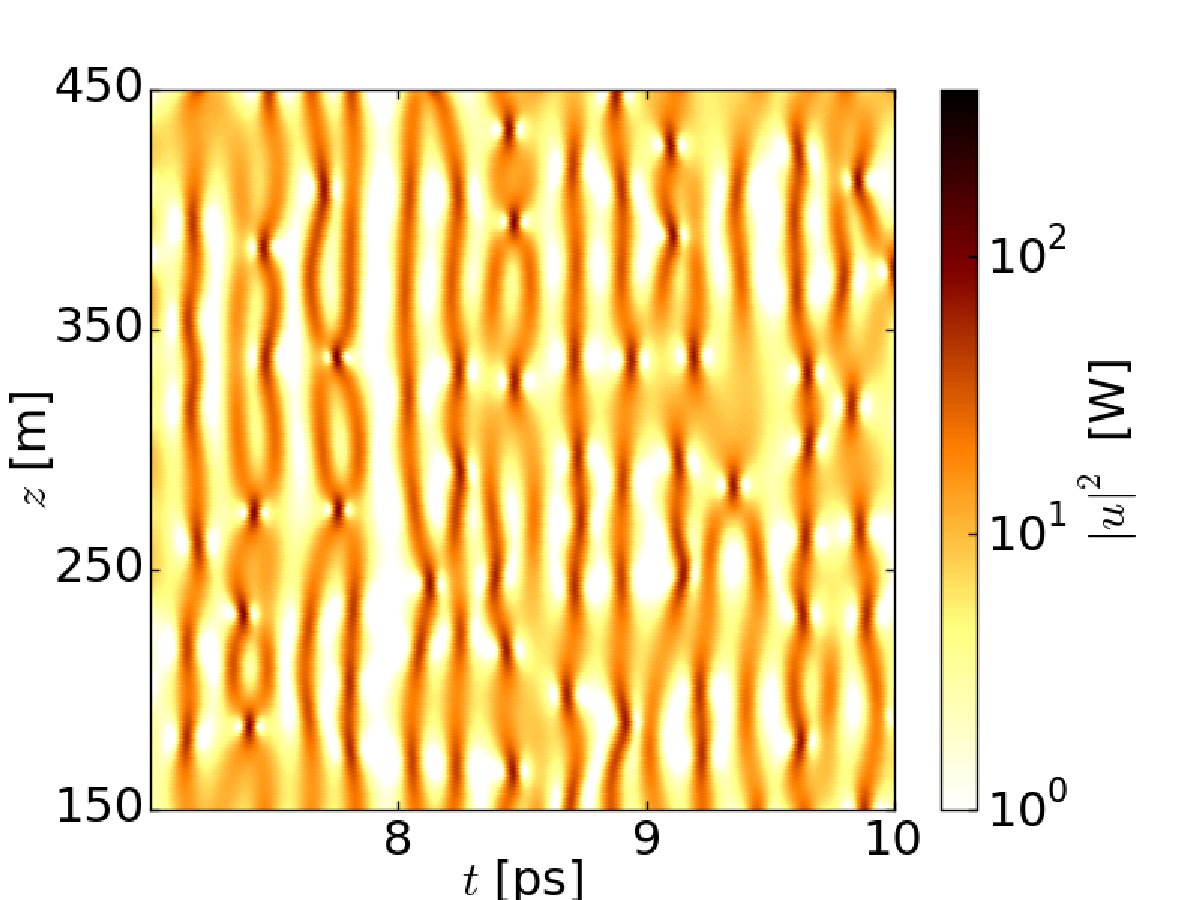}
(c)\includegraphics[width=0.44\columnwidth]{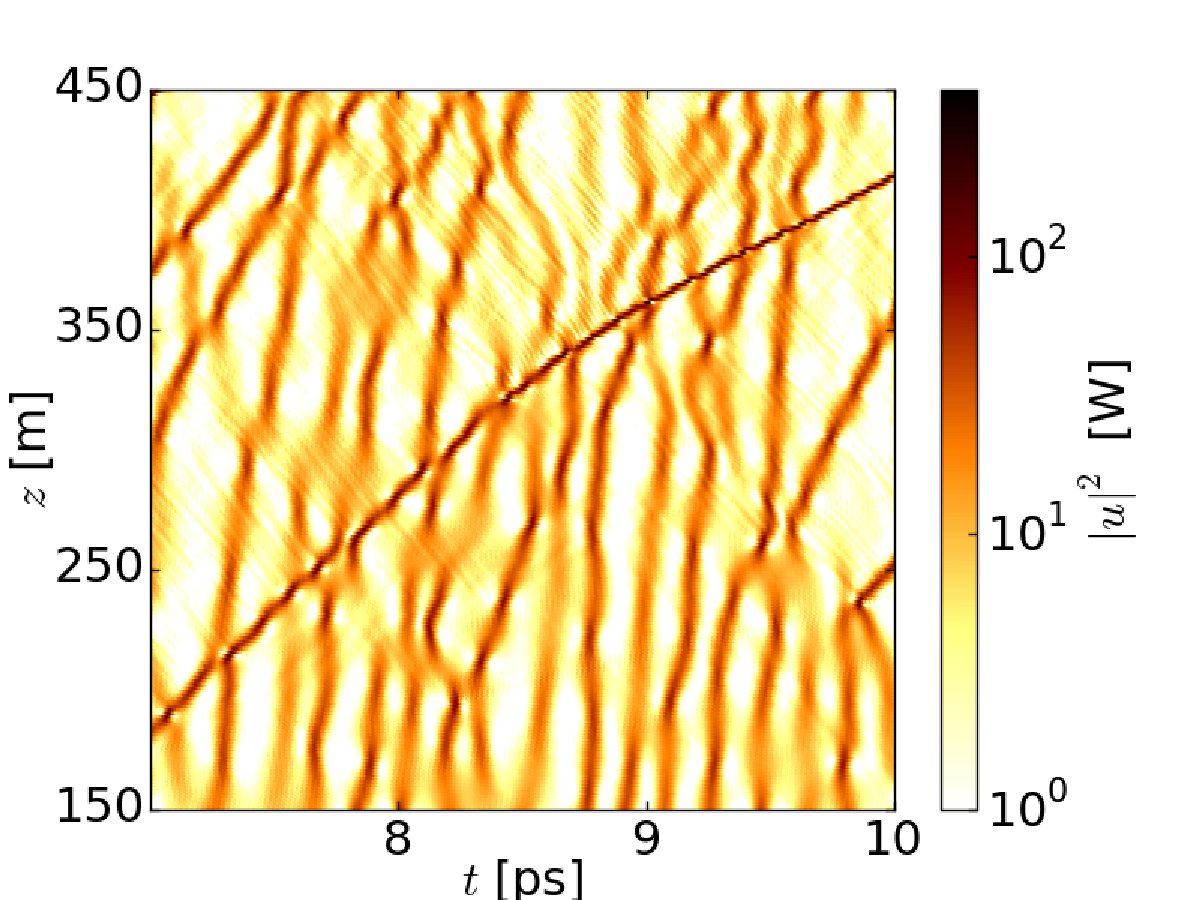}

(d)\includegraphics[width=\columnwidth]{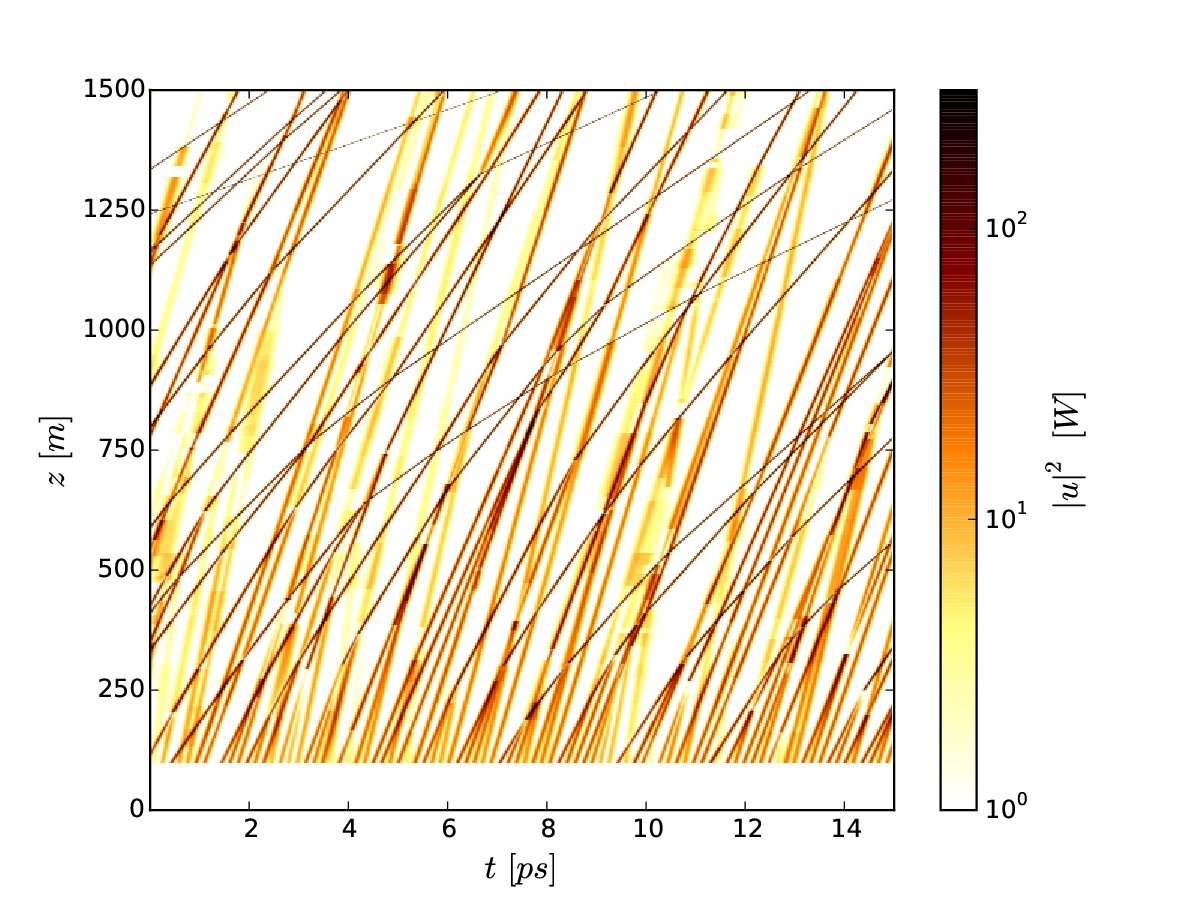}
\end{centering}

\end{figure*}

\newpage
\begin{figure*}[h]
\begin{centering}
(a)\includegraphics[width=\columnwidth,clip]{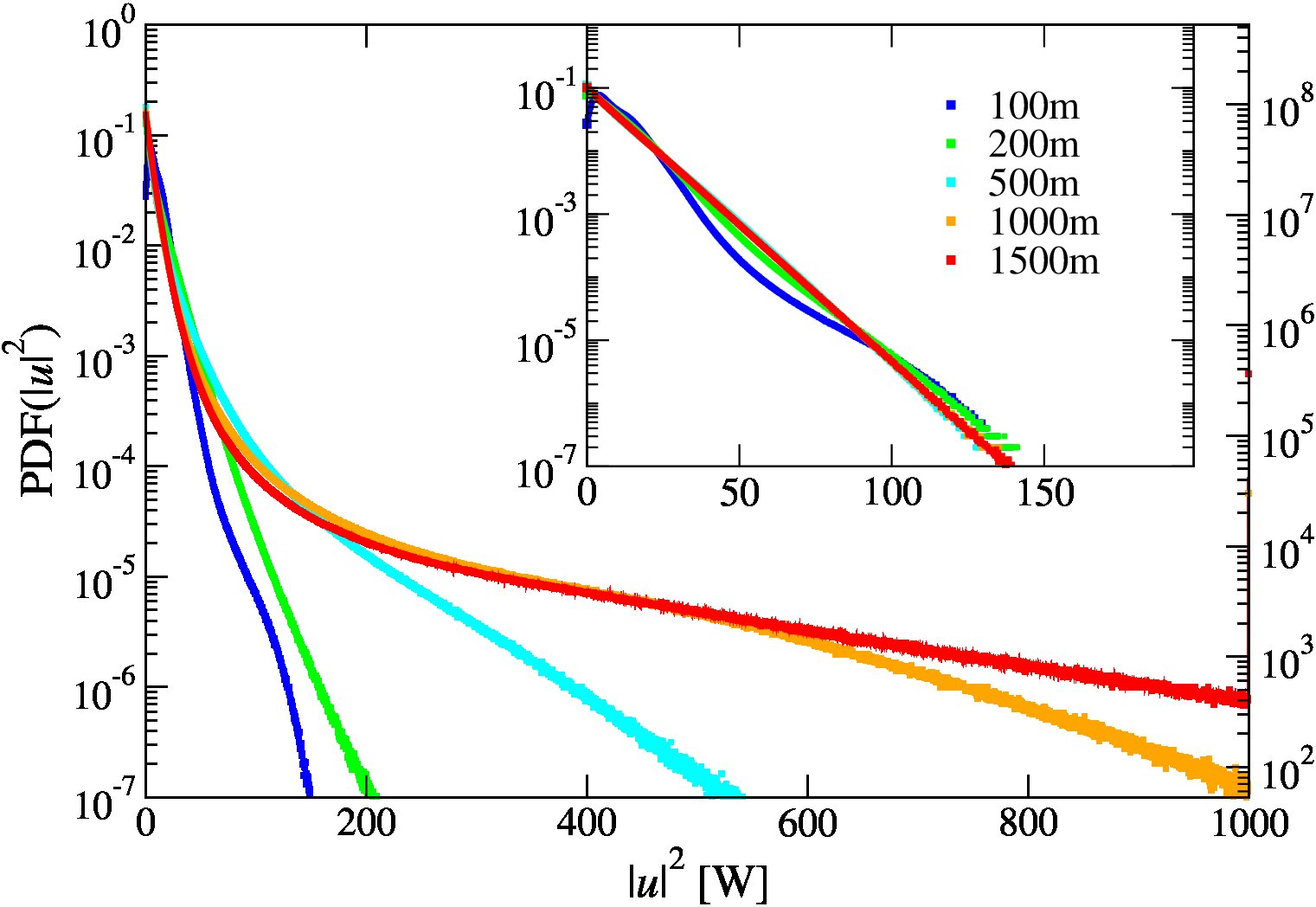}
\par\end{centering}


\begin{centering}
(b)\includegraphics[width=\columnwidth, clip]{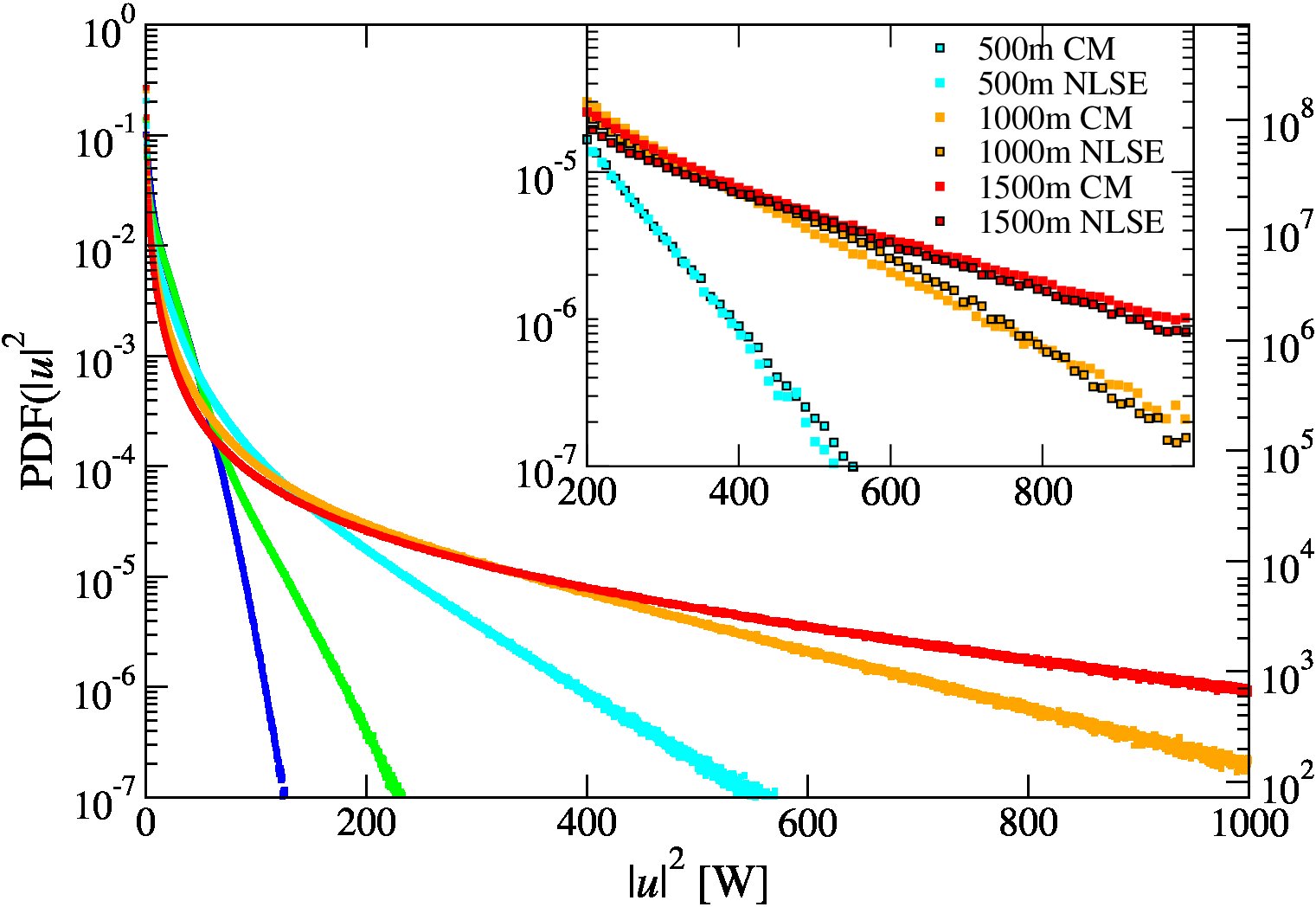}
\par\end{centering}

\end{figure*}

\newpage
\begin{figure*}[h]

\begin{centering}
(a)\includegraphics[width=0.44\columnwidth,clip]{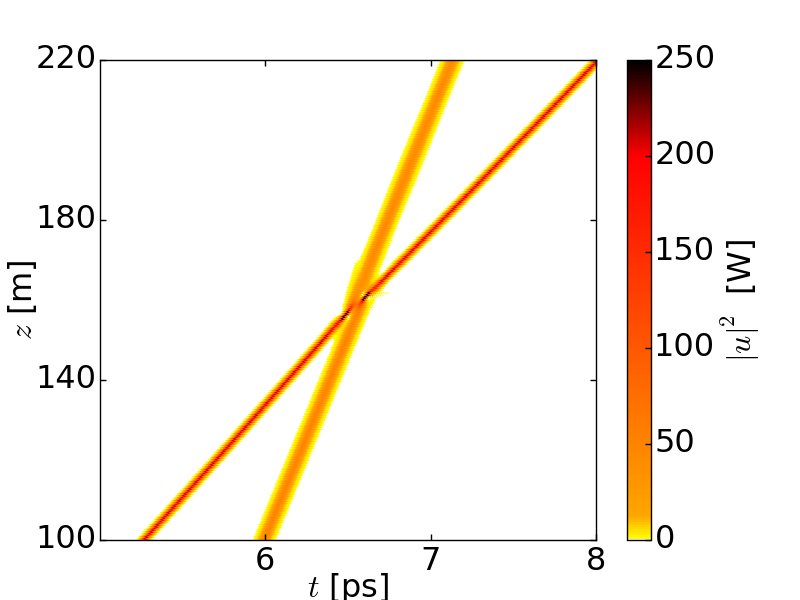}
(b)\includegraphics[width=0.44\columnwidth,clip]{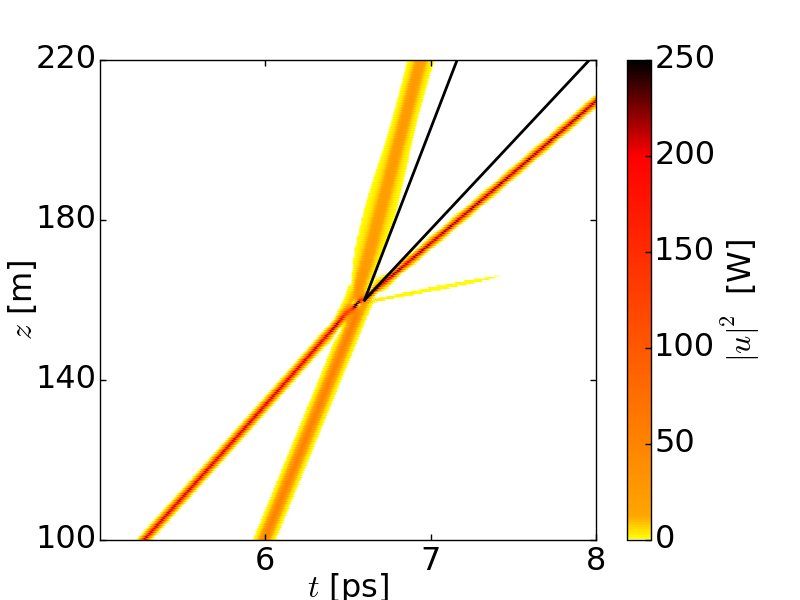}
\par\end{centering}

\begin{centering}
(c)\includegraphics[width=\columnwidth,clip]{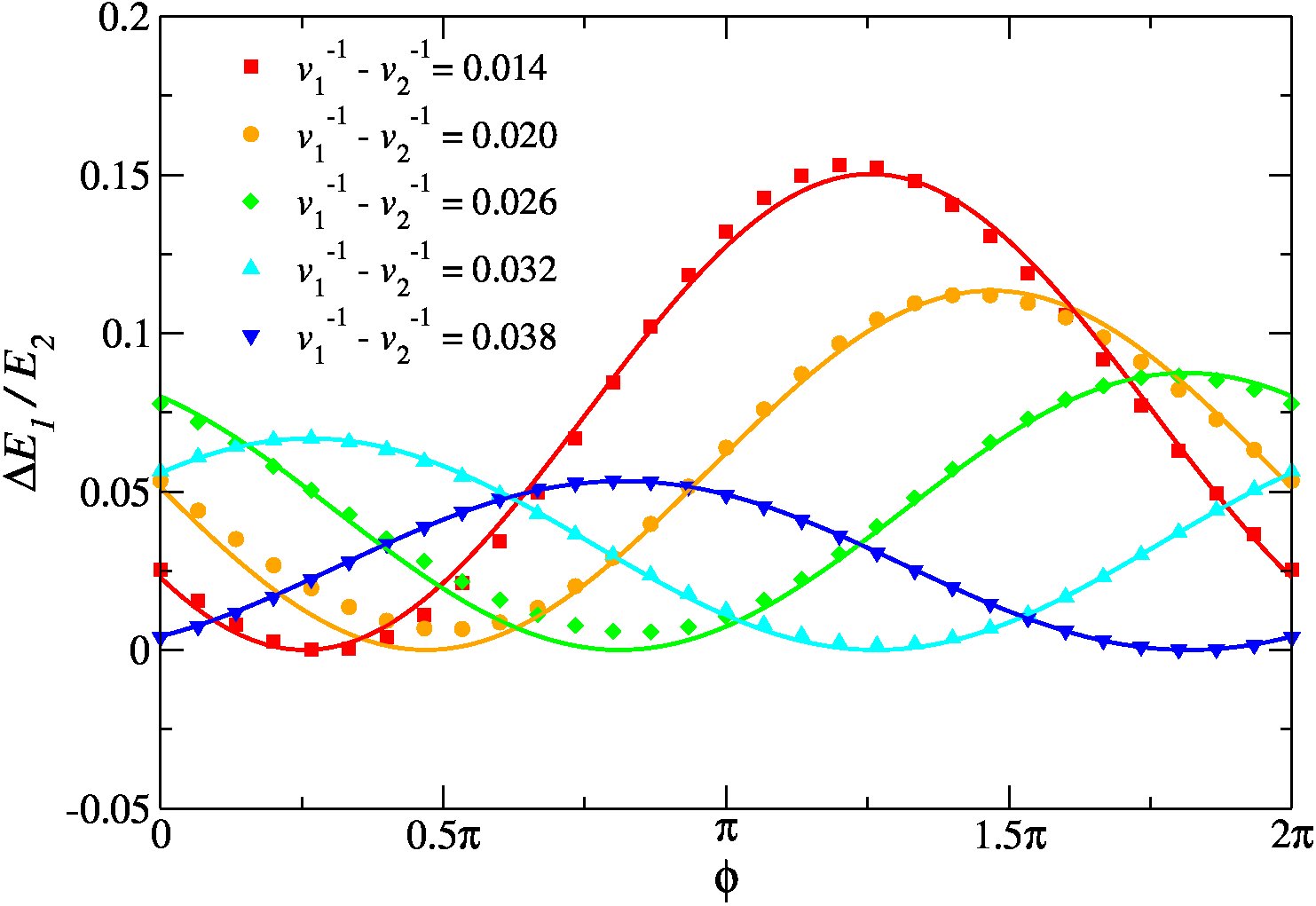}
\par\end{centering}

\begin{centering}
(d)\includegraphics[width=\columnwidth,clip]{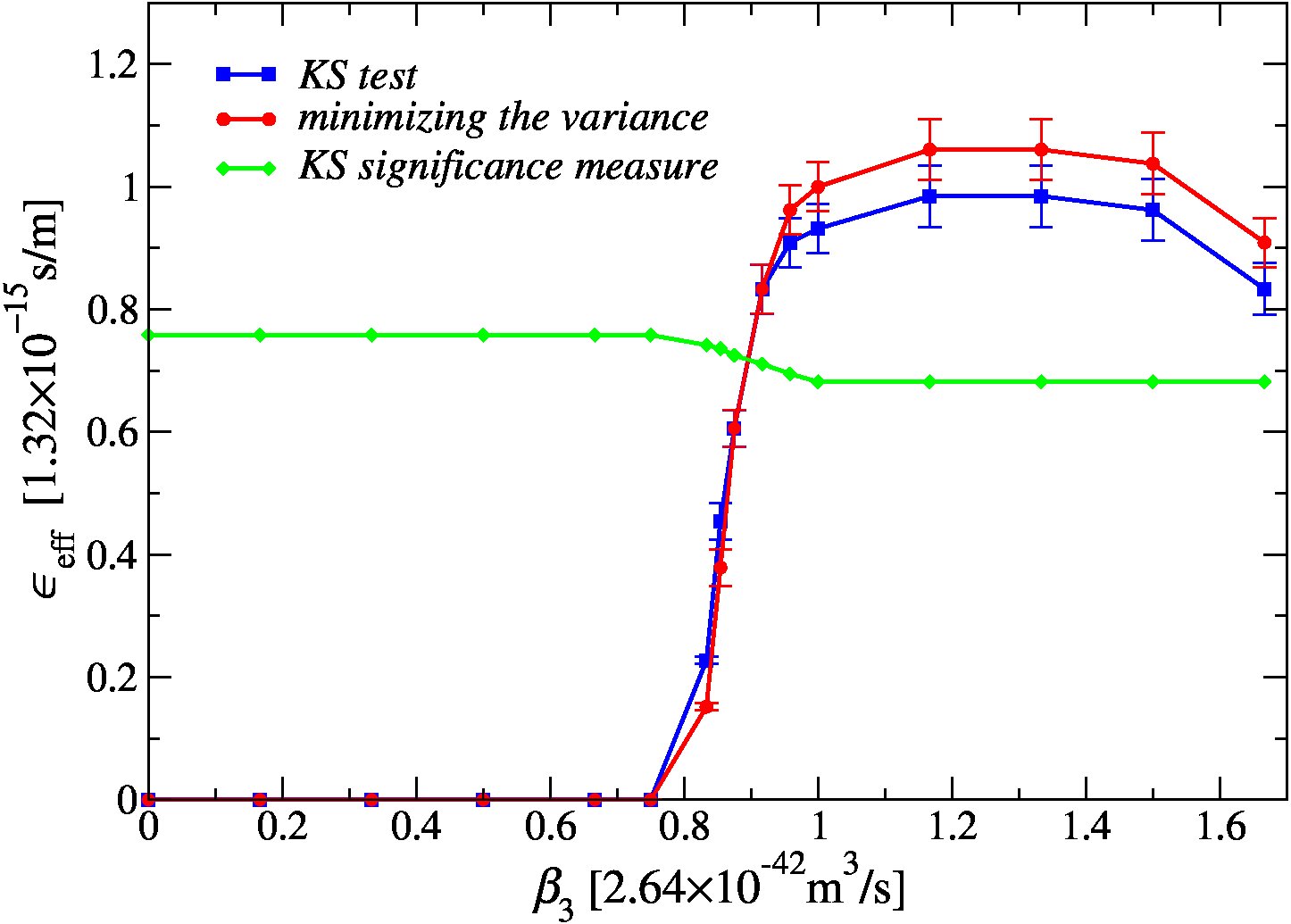}
\par\end{centering}

\end{figure*}
 
\begin{figure*}[h]


\centering
(a)\includegraphics[width=\columnwidth,clip]{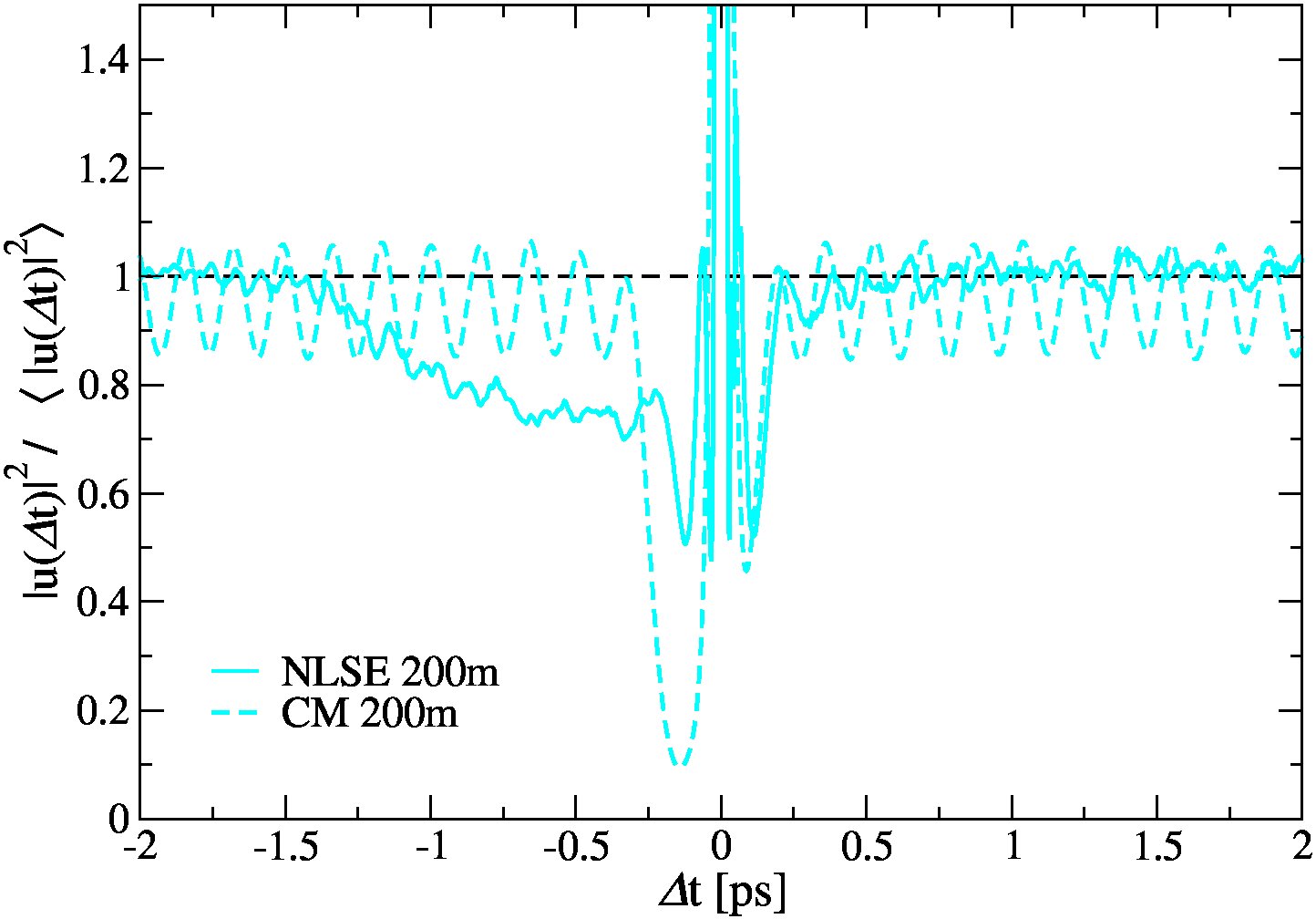}


\centering
(b)\includegraphics[width=\columnwidth,clip]{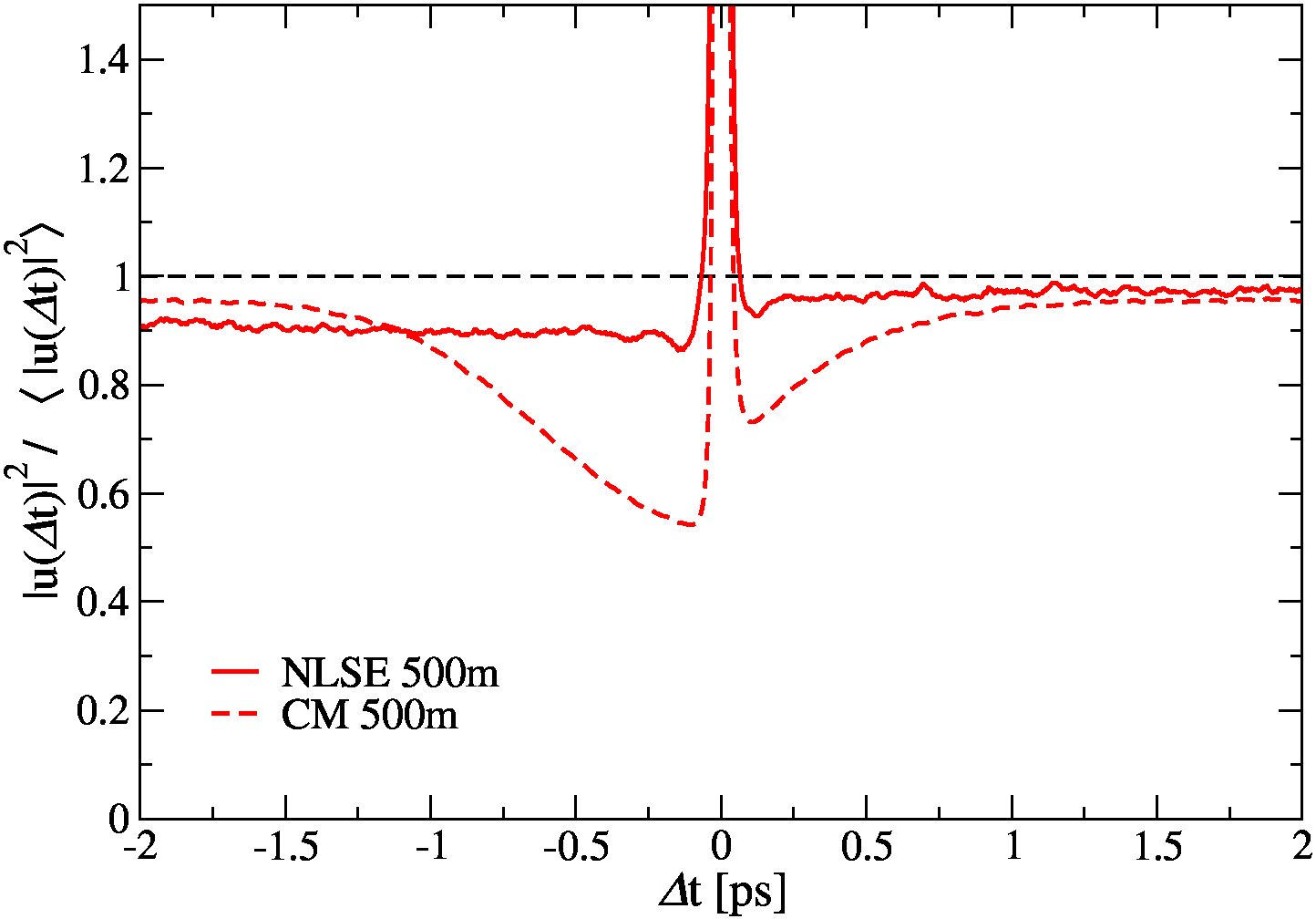}

\end{figure*}

\newpage

\fi

\ifNOSUP\end{document}\else%

\clearpage\newpage
\setcounter{figure}{0}
\setcounter{table}{0}
\setcounter{section}{0}
\setcounter{equation}{0}
\def\thefigure{S\arabic{figure}}
\def\thetable{S\arabic{table}}
\def\thesection{S\arabic{section}}
\def\theequation{S\arabic{equation}}
\setcounter{page}{1}
\pagestyle{plain}


\onecolumngrid
\begin{center}
\noindent\textbf{\large Supporting Material\\[2ex] Rogue wave generation due to inelastic quasi-soliton collisions in optical fibres}\\[2ex]

\noindent
Marc Eberhard, Antonino Savojardo, Akihiro Maruta and Rudolf A R\"{o}mer\\[2ex]

\end{center}
\twocolumngrid



\section{Spectra of RWs for $\beta_3\neq 0$}

\begin{figure}[b]
  \centering
  
  \includegraphics[width=\columnwidth,clip]{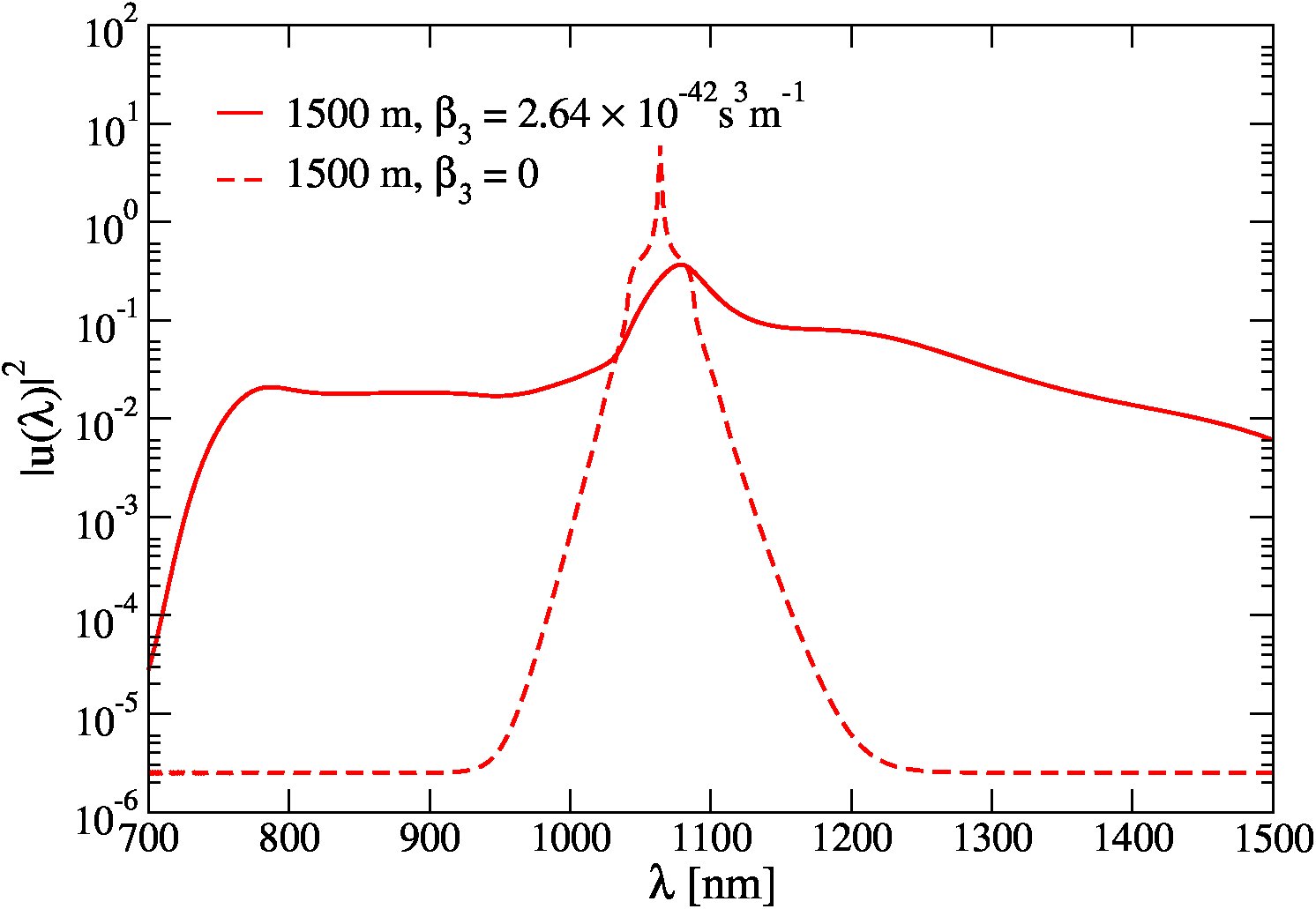}
    \caption{\label{Sfig-spectrum_comparison} Variation of $|u|^2$ with wave length $\lambda$ for $\beta_3 = 0$ (dashed line) and $\beta_3 = 2.64\times 10^{-42}$s$^{-4}$m$^{-1}$ (solid line) at $z= 1500$m.}

\end{figure}

In Fig.\ \ref{Sfig-spectrum_comparison}, we show the spectra $|u(\lambda)|^2$ for the gNLSE at long distance $z= 1500$m for $\beta_3=0 $ and $\beta_3 = 2.64\times 10^{-42}$s$^{-4}$m$^{-1}$. The much broader spectrum for $\beta_3 \neq 0$ shows how the TOD has led to wave excitation across a broad range of wave lengths.

\section{More on the characterization of PDFs for RWs}

\subsection{PDFs for small time window simulations}

\begin{figure}[tb]
  \centering
  \includegraphics[width=\columnwidth,clip]{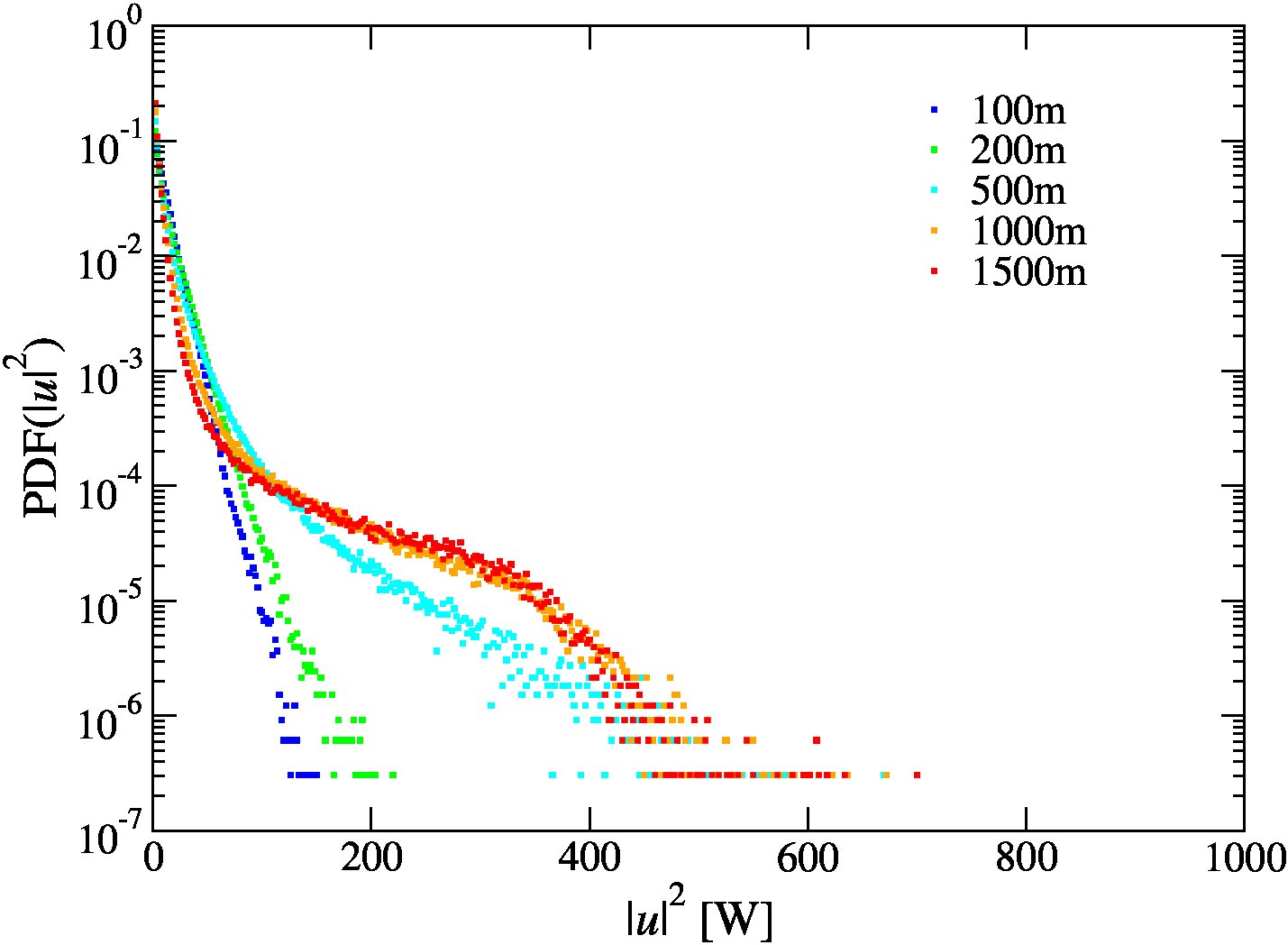}
  \caption{\label{Sfig-PDF-Small-t} PDFs of the intensity $|u|^{2}$ from the gNLSE \eqref{eq-gNLSE} at $\beta_3=1$ using a
time window of $\Delta t=200$ps. The PDFs have been computed at distances $z= 100$m, $200$m, $500$m, $1000$m and $1500$m. After $500$ m the tails 
saturate in comparison with the massively parallel simulation (Fig.\ \ref{fig-PDF} a). }
\end{figure}

The ultra-long time window simulations of the gNLSE \eqref{eq-gNLSE} unambiguously establish the long-range nature of RW PDFs. As shown in Fig.\ \ref{fig-PDF}, the tails for large powers $|u|^2$ are clearly visible.
In contrast, if the time window of a numerical simulation were chosen too small, it would
lead to only a few (and eventually only one) powerful quasi-soliton(s) to
extract all the energy from the system. Hence the resulting PDF would have a bump at some high
peak power given by the total available energy and reflecting the small numerical system size.  
This effect is
demonstrated in Fig.\ \ref{Sfig-PDF-Small-t} where we show PDFs
derived for a short, $200$ps time window simulation, a factor $\sim20000$ shorter than our high-precision parallel 
simulation.  We observe a PDF with an artificial  "knee" for high peak powers.
Only by increasing the length of the time window, such that even the most powerful quasi-solitons can not pass through the time window more than once throughout a simulation, does one recover the true PDF of Fig.\ \ref{fig-PDF}.


\subsection{Fitting distributions and their tails}

\begin{figure}[tb]
  \centering
  (a)\includegraphics[width=\columnwidth,clip]{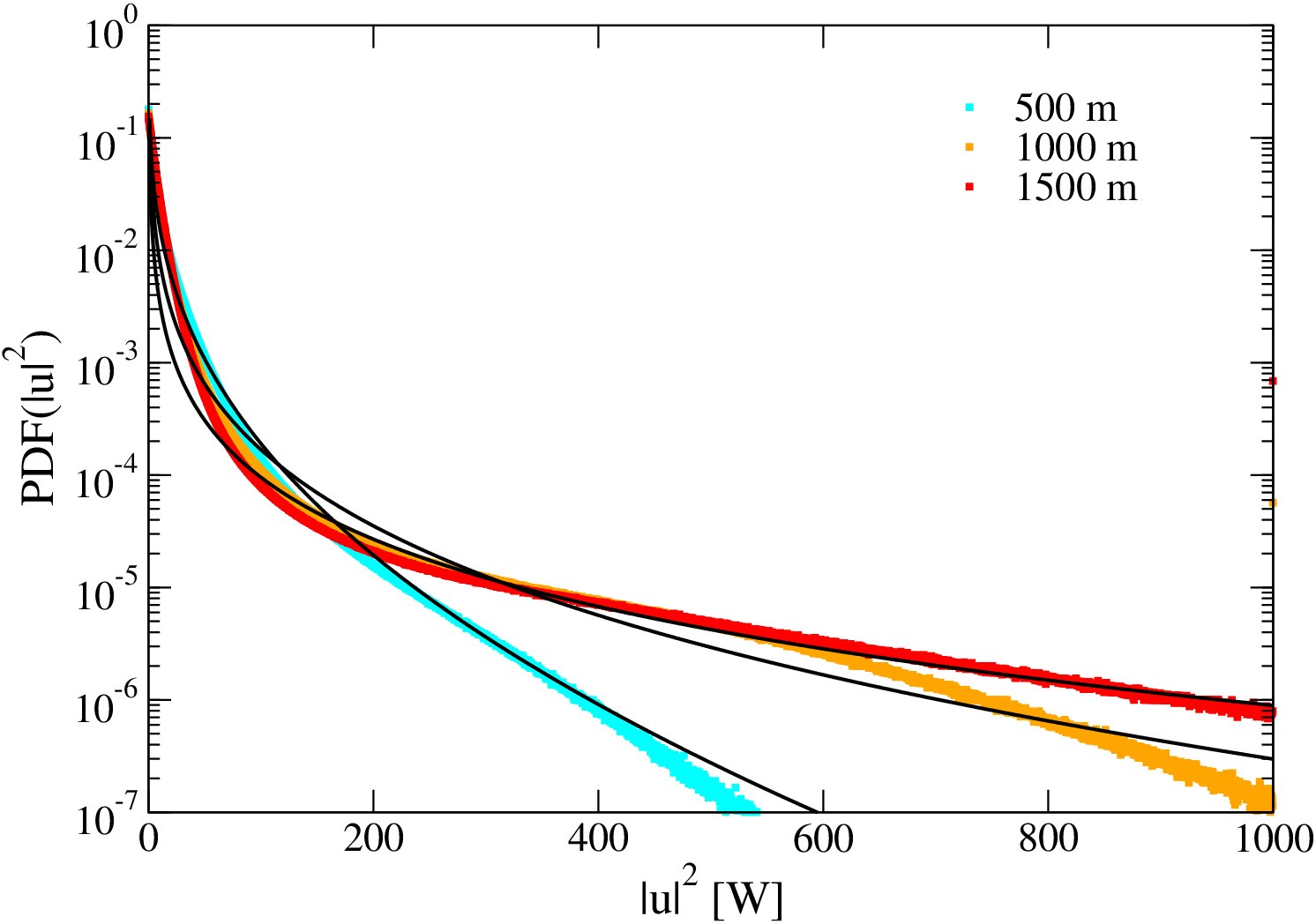}
  (b)\includegraphics[width=\columnwidth,clip]{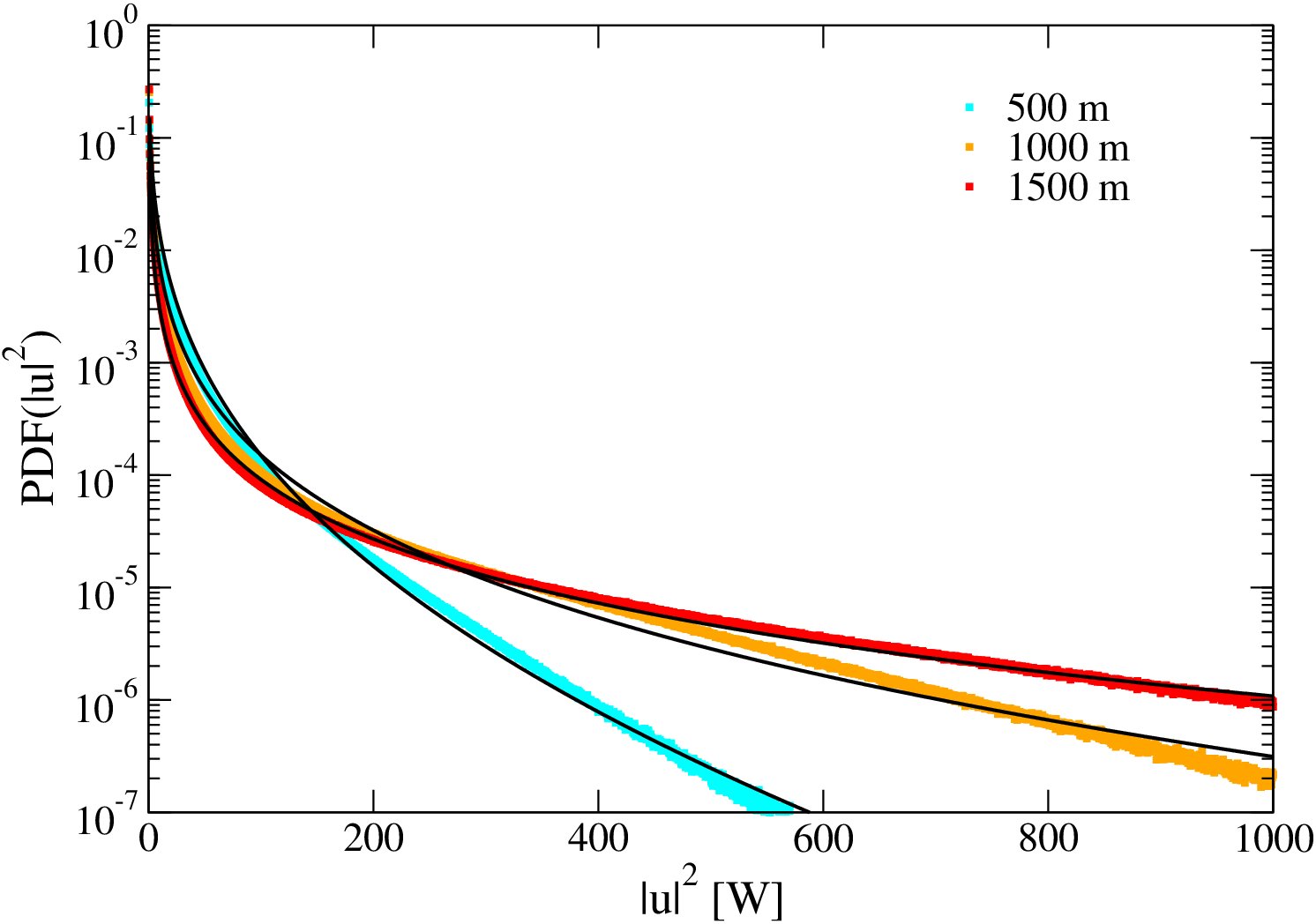}
  \caption{\label{Fit-PDFs-Weibull} (a) PDFs calculated for the gNLSE at $500$, $1000$ ad $1500$m for $\beta_3=2.64\times 10^{-42}$s$^3$m$^{-1}$, the PDFs have been fitted using the Weibull function 
  (full black lines). (b) Same calculation and fits for the cascade model.}
\end{figure}

The full PDFs have also been fitted using the Weibull function
\begin{equation}
W(|u|^2)={b}{a}^{-{b}}(|u|^2)^{{b}-1}\exp\left[-\left(\frac{|u|^2}{a}\right)^{b}\right]\label{eq:W(|u|^2)}
\end{equation}
the PDFs with the respective fits are shown in Fig.\ \ref{Fit-PDFs-Weibull} (a) for the gNLSE and Fig.\ \ref{Fit-PDFs-Weibull} (b) for the cascade model. Every fit has been performed taking the log of the PDF$(|u|^2)$ and $W(|u|^2)$, the resulting coefficients are in Tab.\ \ref{Stab-Fit-coeff}. Our results suggest that while a Weibull fit  \cite{BirBDS15} is indeed possible for the tails, a systematic and consistent variation of the fitting parameters with distance travelled is not obvious. While, e.g., the PDF for $200$, $500$ and $1500$m as shown in Fig.\ \ref{fig-PDF}(a) appears sub-exponential in the tails, we find that the tail of the PDF for $1000$m is super-exponential for $z \gtrsim 600$m.

\begin{figure}[tb]
  \centering
  \includegraphics[width=\columnwidth,clip]{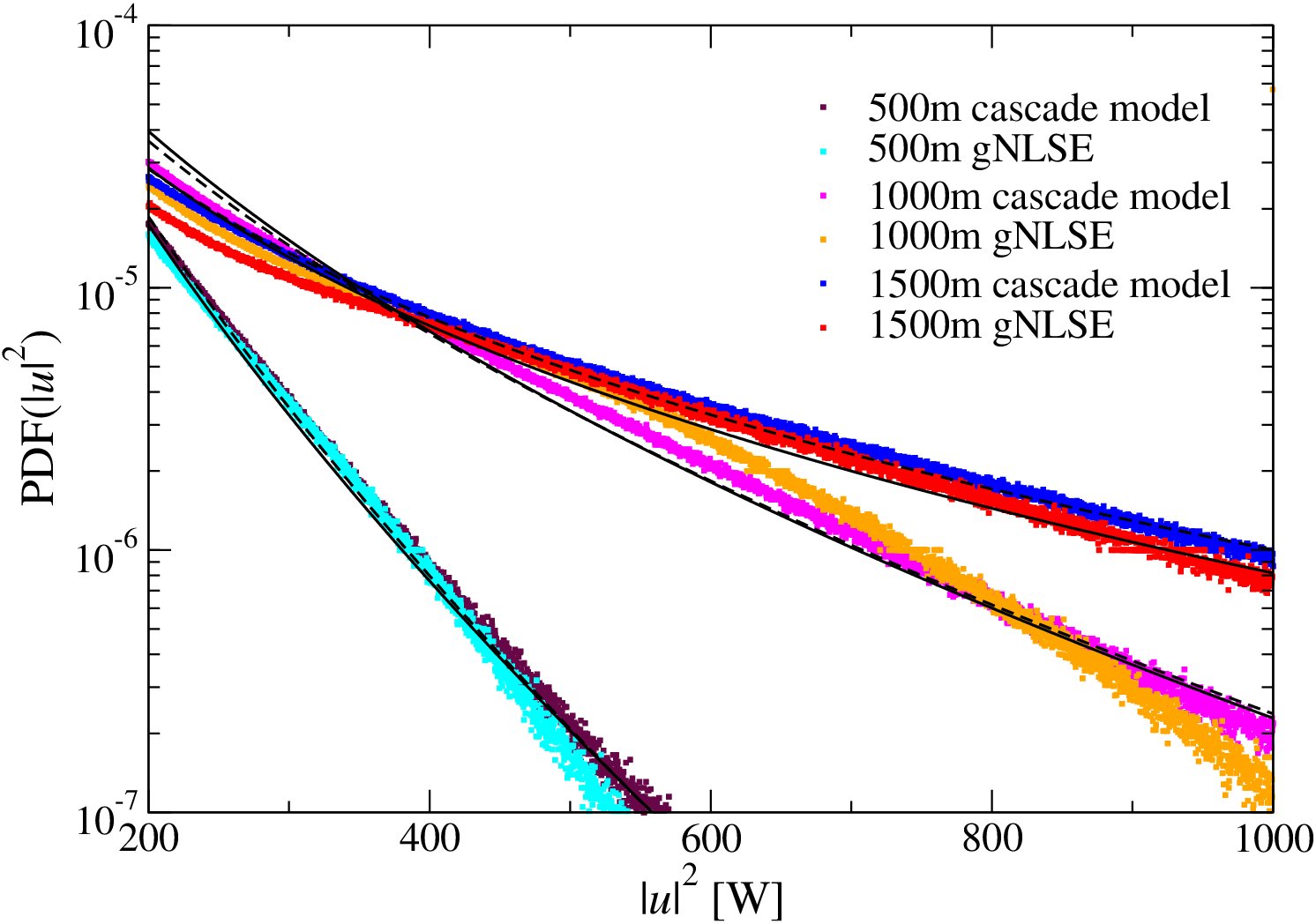}
  \caption{\label{Sfig-Fit-comparison-PDFs} Fits for the cascade model (dashed black lines) and for the gNLSE (full black lines) PDFs.}
\end{figure}

We also fitted the tails of PDF$(|u|^2)$ with
\begin{equation}
F(|u|^2)=F_{0}~\mathrm{exp}\left[-\left(\frac{|u|^2}{a}\right)^b\right],
\label{Seq-Fit-comparison-PDFs}
\end{equation}
Both the PDFs and the respective fits are shown in Fig. \ref{Sfig-Fit-comparison-PDFs} while the resulting coefficients are given in Tab.\ \ref{Stab-Fit-coeff}. As for the Weibull fit, the results are not convincing and hint towards a continuing development of the shape of the PDF as the propagation continues.

\begin{table*}[tb]
\begin{center}
\begin{tabular}{|c|c|c|c|c|c|c|}
\hline 
\multicolumn{3}{|c|}{Fit} & $a$ & $b/10^{-1}$ & $F_0/10^{-3}$ & $\chi^{2}$ \tabularnewline
\hline 
\hline 
\multirow{4}{*}{500 m} & 
  \multirow{2}{*}{gNLSE} 
    & W & $~$4.0 $\pm$ 0.5 & 4.94 $\pm$ 0.15 & - & 0.005\tabularnewline
    \cline{3-7} 
    &  & F & $~~$9 $\pm$ 21 & 6 $\pm$ 3 & 8 $\pm$ 30 & 0.2 \tabularnewline
    \cline{2-7} 
  & \multirow{2}{*}{cascasde model } 
    & W & $~$3.0 $\pm$ 0.5 & 4.60 $\pm$ 0.16 & - & 0.49\tabularnewline
    \cline{3-7} 
    &  & F &  12 $\pm$ 3 & 6.3 $\pm$ 3$~~$ & 6 $\pm$ 2 & 0.9\tabularnewline
\hline 
\multirow{4}{*}{1000 m} & 
  \multirow{2}{*}{gNLSE} 
    & W & $~$0.9 $\pm$ 0.4 & 3.2 $\pm$ 0.2 & - & 0.017\tabularnewline
    \cline{3-7} 
    &  & F & $~~~$8 $\pm$ 10$~$ & 47 $\pm$ 10 & 4 $\pm$ 5 & 0.05\tabularnewline
    \cline{2-7} 
  & \multirow{2}{*}{cascasde model } 
    & W & 0.7 $\pm$ 0.2 & 3.04 $\pm$ 0.14 & - & 0.005\tabularnewline
    \cline{3-7} 
    &  & F &  11 $\pm$ 5$~$ & 4.9 $\pm$ 0.4 & 2 $\pm$ 1 & 0.02\tabularnewline
\hline 
\multirow{4}{*}{1500 m} & 
  \multirow{2}{*}{gNLSE} 
    & W & 0.07 $\pm$ 0.03 & 2.1 $\pm$ 0.1 & - & 0.02\tabularnewline
    \cline{3-7} 
    &  & F & $~$0.006 $\pm$ 0.001$~$ & 2.1 $\pm$ 0.4 & 200 $\pm$ 70 & 0.03\tabularnewline
    \cline{2-7} 
  & \multirow{2}{*}{cascasde model } 
    & W & 0.04 $\pm$ 0.01 & 1.94 $\pm$ 0.05 & - & 0.003\tabularnewline
    \cline{3-7} 
    &  & F &  0.008 $\pm$ 0.016 & 2.1 $\pm$ 0.3 & 120 $\pm$ 140 & 0.006\tabularnewline
\hline 
\end{tabular}
\par\end{center}

\protect\caption{\label{Stab-Fit-coeff}Fitted coefficients using the Weibull function (W) and $F(|u|^2)$ (F).}

\end{table*}

\section{Collisions: Determining the energy gain}

\subsection{Deriving the energy gain formula}

In Fig.\ \ref{fig-twosoliton} we observe an energy transfer due to inelastic scattering. 
This energy \emph{gain} of quasi-soliton $1$ from quasi-soliton $2$, can be written as \cite{Nag10}
\begin{equation}
\Delta E_{1}=\int\mathcal{G}(P_{1},\Omega_{1},P_{2},\Omega_{2}; z,t,\phi)dzdt
\label{Seq-density-energy-transfer},
\end{equation}
where $\mathcal{G}$ is an energy density and $\phi$ is the phase difference between the two quasi-solitons. It is convenient to change variables
$s = t-z/v_{1}$, $w=t-z/v_{2}$. 
Then \eqref{Seq-density-energy-transfer} becomes 
\begin{equation}
\Delta E_{1}=\frac{1}{\mid v_{1}^{-1}-v_{2}^{-1}\mid}\int\mathcal{G}\left(P_{1},\Omega_{1},P_{2},\Omega_{2}; s,w,\phi\right)ds dw.
\label{Seq-transfer}
\end{equation}
Eq.\ \eqref{Seq-transfer} shows that a large difference between the (inverses of the) $v_q$ results in a reduced energy gain for the larger quasi-soliton in agreement with the results in the main paper.
In section \ref{sec-collisions}, we showed the importance of $\phi$ for the energy transfer. Fourier-expanding \eqref{Seq-transfer}, we can write
\begin{equation}
\frac{\Delta E_{1}}{E_{2}}=\frac{1}{\mid v_{1}^{-1}-v_{2}^{-1}\mid}{\sum_{n=0}^{\infty}}
\epsilon_{P_{1},\Omega_{1},P_{2},\Omega_{2}}(n) \cos\left[n(\phi-\phi_0)\right],
\end{equation}
where $\epsilon(n)$ are Fourier coefficients (we suppress the $P_q$ and $\Omega_q$ indices for a moment) and $\phi_0$ is the phase difference for which the gain has a maximum. 
We approximate the above formula with just the first two coefficients. These two coefficients are related; indeed the larger quasi-soliton always gains energy ($\Delta E_{1} > 0$), hence $\epsilon(0)\geq |\epsilon(1)|$ and because for a certain $\phi$ the energy gain is zero, we have $\epsilon(0)=-\epsilon(1)$. Thus we can write
\begin{equation}
\frac{\Delta E_{1}}{E_{2}}\simeq\frac{\epsilon_{P_{1},\Omega_{1},P_{2},\Omega_{2}}}{| v^{-1}_{1}-v^{-1}_{2}| }\sin^{2}\left(\frac{\phi-\phi_{o}}{2}\right),\label{Seq-energy-transfer}
\end{equation}
where we have used $1-\cos\left(\phi\right)=2\sin^{2}\left(\frac{\phi}{2}\right)$ and defined  $\epsilon=2\epsilon(0)$. 
The value of $\epsilon$ is yet undetermined while the dependence on the group velocities and the phase difference it is clear. As shown in Fig.\ \ref{fig-twosoliton}(c), \eqref{Seq-energy-transfer} provides an excellent description of the energy gain in pair-wise quasi-soliton collisions.

\subsection{Effective coupling constant $\epsilon_{\mathrm{eff}}$}

We want to model a system of many colliding quasi-solitons as shown in Fig.\ \ref{fig-traces}(a). The parameter $\epsilon_{P_{1},\Omega_{1},P_{2},\Omega_{2}}$ depends on the individual power and frequency shift of each quasi-soliton pair. In order to devise a tractable model, we have to find an effective $\epsilon_\mathrm{eff}$ that describes the average properties of the $u$-amplitudes well. We therefore choose a distance $z=500$, where from Fig.\ \ref{fig-traces}(a) we see that well-developed quasi-soliton pulses exist, while the situation is not yet RW dominated as shown in Fig.\ \ref{fig-PDF}(a). 
We then use a constant trial value for $\epsilon_\mathrm{eff}$ and apply \eqref{eq-energy-transfer-eff} to all quasi-soliton collisions in the cascade model computing data similar to Fig.\ \ref{fig-traces}(d) and Fig.\ \ref{fig-PDF}(b). We then repeat the calculation with another trial $\epsilon_\mathrm{eff}$. For different $\epsilon_\mathrm{eff}$, we compare the PDF created from the cascade model with the PDF obtained from the gNLSE and choose $\epsilon_\mathrm{eff}$ such that the agreement is best (see below). We note that this process was followed for the $\epsilon_\mathrm{eff}$ values shown  in Fig.\ \ref{fig-twosoliton}(d) for the different $\beta_3$ values. Furthermore, we have checked that similar results can be obtained by using $z=300$ as the starting point of the analysis.
Once $\epsilon_\mathrm{eff}$ is determined, we use it to compute the results for the cascade model, starting at $z=100$m and "propagating" all the way to $1500$m as described in section \ref{sec-methods}. We emphasize the good agreement of the PDFs for $z\neq 500$m. 

\subsection{Estimating $\epsilon_{\mathrm{eff}}$}

\begin{figure}[tb]
  \centering
  \includegraphics[width=\columnwidth,clip]{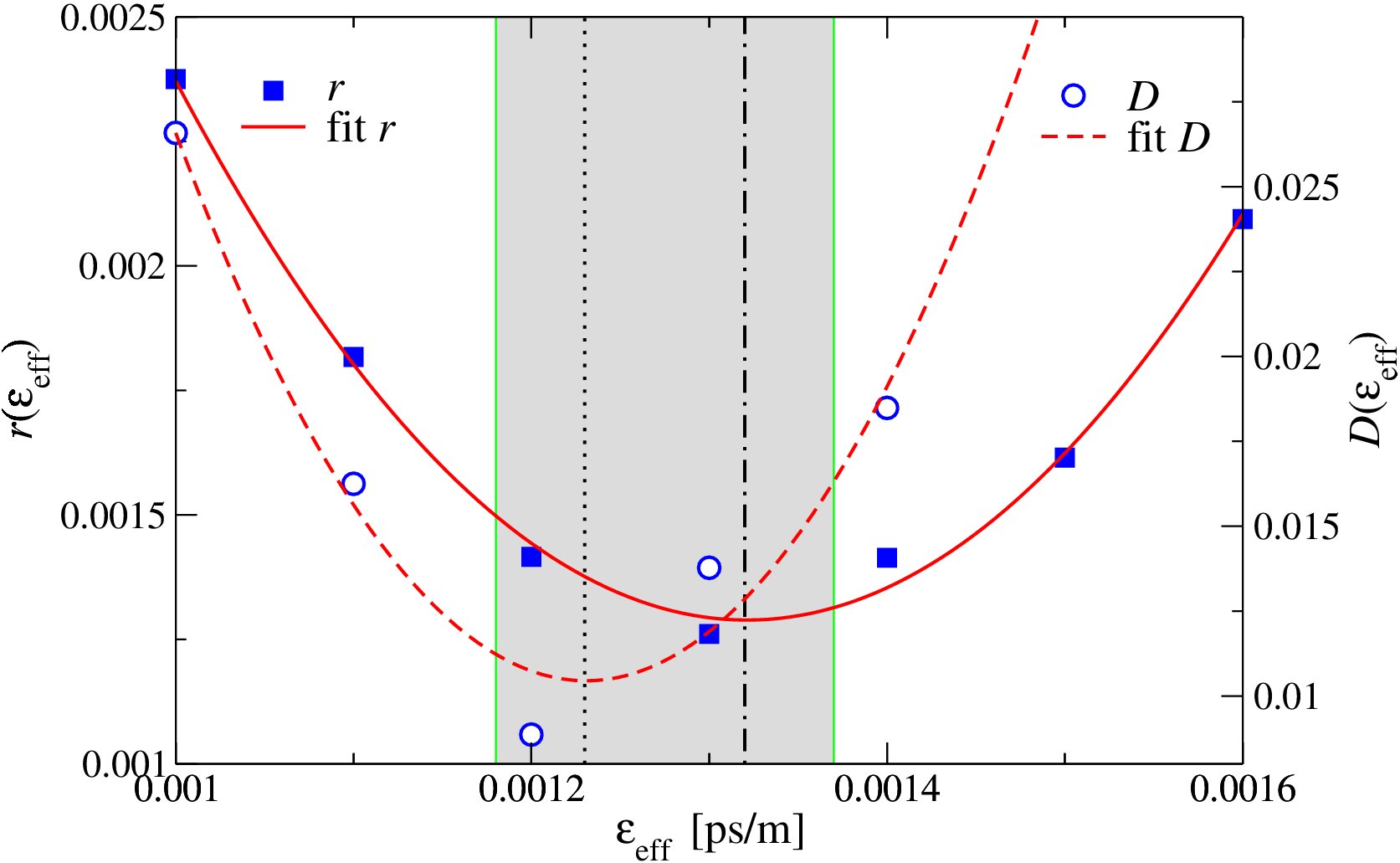}
  \caption{Relative variance $r$ (filled squares) and largest difference $D$ (open circles) calculated for different values of $\epsilon_{\mathrm{eff}}$ at $\beta_3=2.64\times 10^{-42}$s$^3$m$^{-1}$. Parabolic fits to the data are shown as lines. The vertical dotted line denotes the estimated $\epsilon_{\mathrm{eff}}=(1.23 \pm 0.05)$fs/m at which $D$ is minimal, the grey region indicates the error of that estimate. The vertical dashed-dotted line denotes the estimate $\epsilon_{\mathrm{eff}}=(1.32 \pm 0.05)$fs/m from $r$.}
  \label{Sfig-Minimum-on-RD}
\end{figure}

We are interested in determining $\epsilon_\mathrm{eff}$ such that the PDFs of the gNLSE and the cascade model (CM) agree at $z=500$m. Since we are interested in RWs we want that agreement to be good in the tail region of $|u|^2>150$W. We therefore define the relative variance 
 \begin{equation}
r(\epsilon_{\mathrm{eff}}) = \frac{\sum_i 
	\left[
	\log\mathrm{PDF}_{\mathrm{gNLSE}}(|u_i|^2) -
	\log\mathrm{PDF}_{\mathrm{CM}}(|u_i|^2,\epsilon_{\mathrm{eff}})
	\right]^2}
	{\sum_i \log\left[ \mathrm{PDF}_{\mathrm{gNLSE}}(|u_i|^2) \right]^2}
	\label{Seq-R}
\end{equation}
and minimize it with respect to $\epsilon_\mathrm{eff}$ as shown in Fig.\ \ref{Sfig-Minimum-on-RD}. The $\epsilon_\mathrm{eff}$ at minimum is our estimate with accuracy $\epsilon_\mathrm{eff} \sqrt{r(\epsilon_\mathrm{eff})}$.
The results for $\epsilon_{\mathrm{eff}}$ calculated using this variance minimization are shown in Fig. \ref{fig-twosoliton}(d) (red line).

\begin{figure}[tb]
  \centering
  \includegraphics[width=\columnwidth,clip]{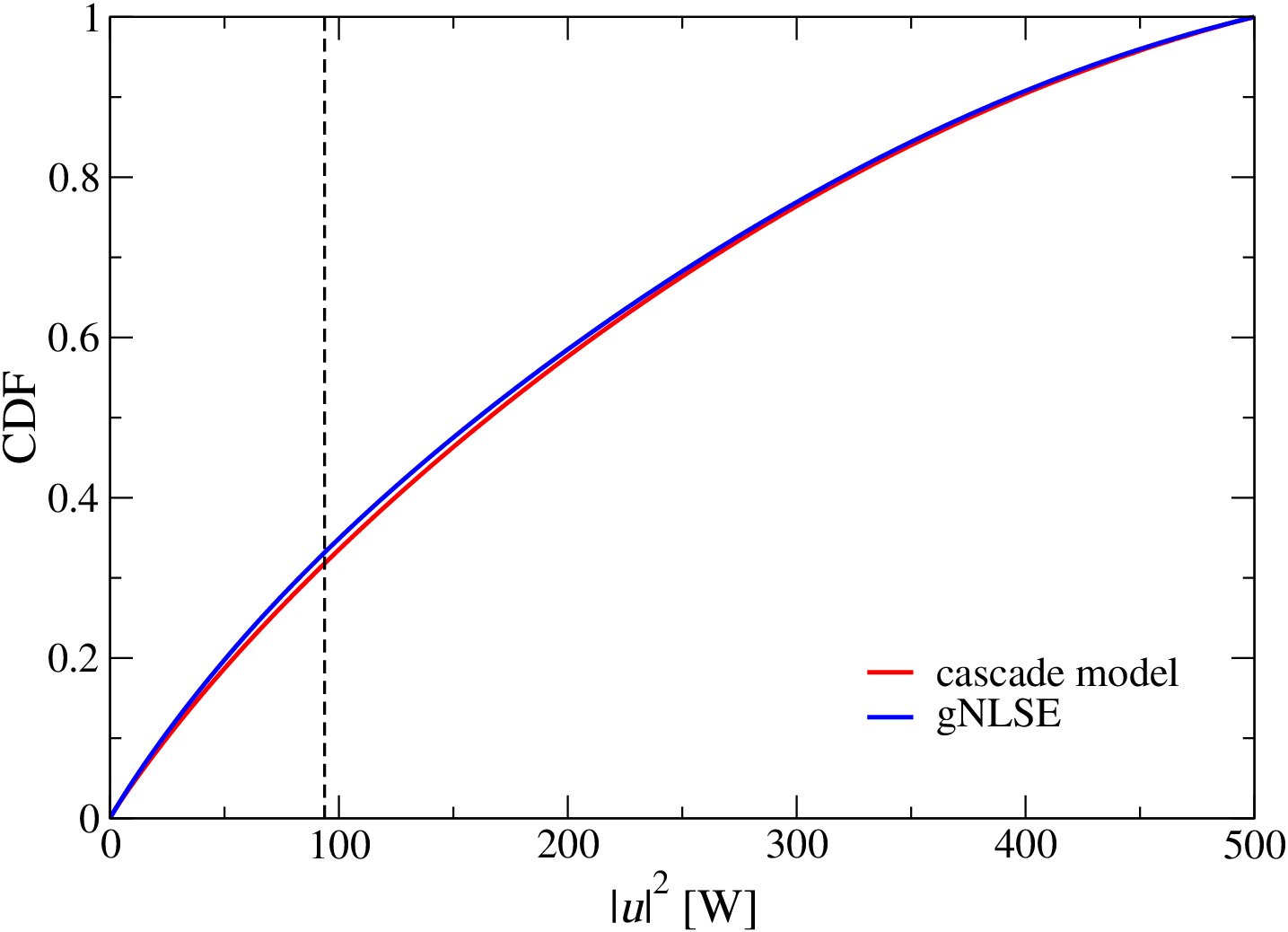}
  \caption{\label{Sfig-CDF} CDFs for the gNLSE (blue line) and for the cascade model (red line) calculated for $\beta_3=2.64\times 10^{-42}$s$^3$m$^{-1}$ using the modified version of the KS test.
  The dashed black line at $93.75$ W corresponds to the value $D$ .}
\end{figure}

As a second test, we perform a Kolmogorov-Smirnov (KS)-like two-sample test \cite{PreFTV92C} between $\log\mathrm{PDF}_{\mathrm{gNLSE}}(|u_i|^2)$ and $\log\mathrm{PDF}_{\mathrm{CM}}(|u_i|^2,\epsilon_{\mathrm{eff}})$. We need to renormalize the data count as $\widetilde{N}_{i}= \log(1+N_i)$ with each $j$ denoting a $|u_i|^2$ bin and overall $\widetilde{N}= \sum_{j=1}^{N_\mathrm{bins}}\log(1+N_i)$. Hence the effective number of data is given by 
\begin{equation}
\widetilde{N}_e = \frac{\widetilde{N}_\mathrm{gNLSE} \widetilde{N}_\mathrm{CM}}{\widetilde{N}_\mathrm{gNLSE}+\widetilde{N}_\mathrm{CM}} .
\end{equation}
Following the KS prescription, we then define the difference
\begin{equation}
 D = \underset{i\in [1,N_\mathrm{bins}]}{\mathrm{max}}\left|\mathrm{CDF}_{\mathrm{gNLSE}}(i)-\mathrm{CDF}_{\mathrm{CM}}(i)\right|.
\end{equation}
and minimize it with respect to $\epsilon_\mathrm{eff}$ as shown in Fig.\ \ref{Sfig-Minimum-on-RD}.
As usual, with $\lambda = (\sqrt{\widetilde{N}_e}+0.12+0.11/\sqrt{\widetilde{N}_e})D$, a KS-like accuracy can be given as 
\begin{equation}
Q_\mathrm{KS}(\lambda)=2\sum_{j=1}^{\infty}(-1)^{j-1}e^{-2j^{2}\lambda^2} ,
\end{equation}
although it should no longer be interpreted probabilistically.
The results for $\epsilon_{\mathrm{eff}}$ calculated using this KS-like test are also shown in Fig. \ref{fig-twosoliton}(d) with $Q_\mathrm{KS}$ given for each $\beta_3$ value. It is important to note that $Q_\mathrm{KS}$ remains roughly constant for all $\beta_3$ values, indicating a comparable level of similarity between $\mathrm{PDF}_{\mathrm{CM}}$ and $\mathrm{PDF}_{\mathrm{gNLSE}}$ across the full $\beta_3$ range.
In Fig.\ \ref{Sfig-Minimum-on-RD}, we show the $\epsilon_\mathrm{eff}$ dependence of the test while Fig.\ \ref{Sfig-CDF} displays the CDFs .

\section{Independent quasi-solitons}

\begin{figure}[tb]
  \centering
  \includegraphics[width=\columnwidth,clip]{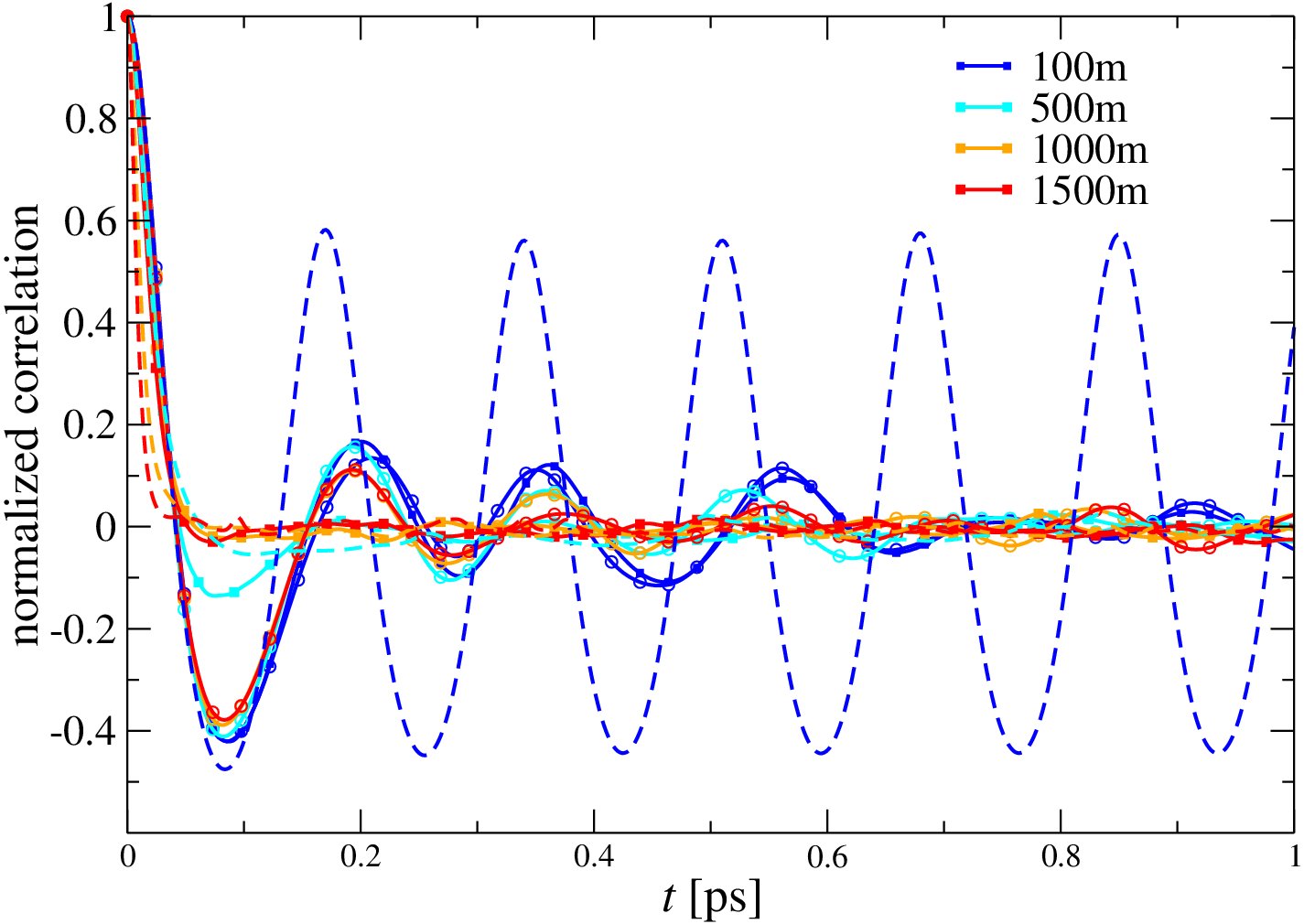}
  \caption{\label{Sfig-autocorrelation} Time autocorrelation ${\cal C}_z(t)/{\cal C}_z(0)$ for the gNLSE (squares and solid lines) and for the cascade model (dashed lines) at selected $z$ values as indicated ($\beta_3=2.64\times 10^{-42}$s$^3$m$^{-1}$). 
The circles with solid lines indicate the case $\beta_3=0$.
  }
\end{figure}

In Ref.\ \cite{BirBDS15}, the authors studied RW events in three data sets, one from ocean waves and two based on optical devices. They compute, e.g., the time-series autocorrelation function ${\cal C}_z(\tau)= \int dt' 
\left[ |u(z,t')|^2 - \langle |u(z,t)|^2\rangle_t \right] 
\left[ |u(z,\tau-t')|^2 - \langle |u(z,t)|^2\rangle_t \right] $ (see section \ref{sec-collisions}). 
In Fig.\ \ref{Sfig-autocorrelation}, we show ${\cal C}_z(\tau)$ for our gNLSE data as well as for the cascade model at various distances. 
At $100$m quasi-solitons are clearly correlated because of the initial conditions in both models, but from $200$m onwards ${\cal C}_z(\tau)$ is nearly zero after a small fraction of picoseconds supporting the notion of largely independently travelling quasi-solitons that interact only when in close spatial and temporal proximity. In agreement with Ref.\ \cite{BirBDS15}, we hence find  a quick decay of ${\cal C}_z(t)$ after $z>100$m which supports the notion of well-separated individual quasi-solitons (cp.\ Fig.\ \ref{Sfig-autocorrelation}). Furthermore, the agreement between gNLSE results and the cascade model is very good. Also, the correlation ${\cal C}_z(\tau)$ is close to what has been reported in Ref.\ \cite{BirBDS15}.

\section{Further details of the cascade model}

\subsection{Choice of initial PDF}

\begin{figure}[tb]
  \centering
  \includegraphics[width=\columnwidth,clip]{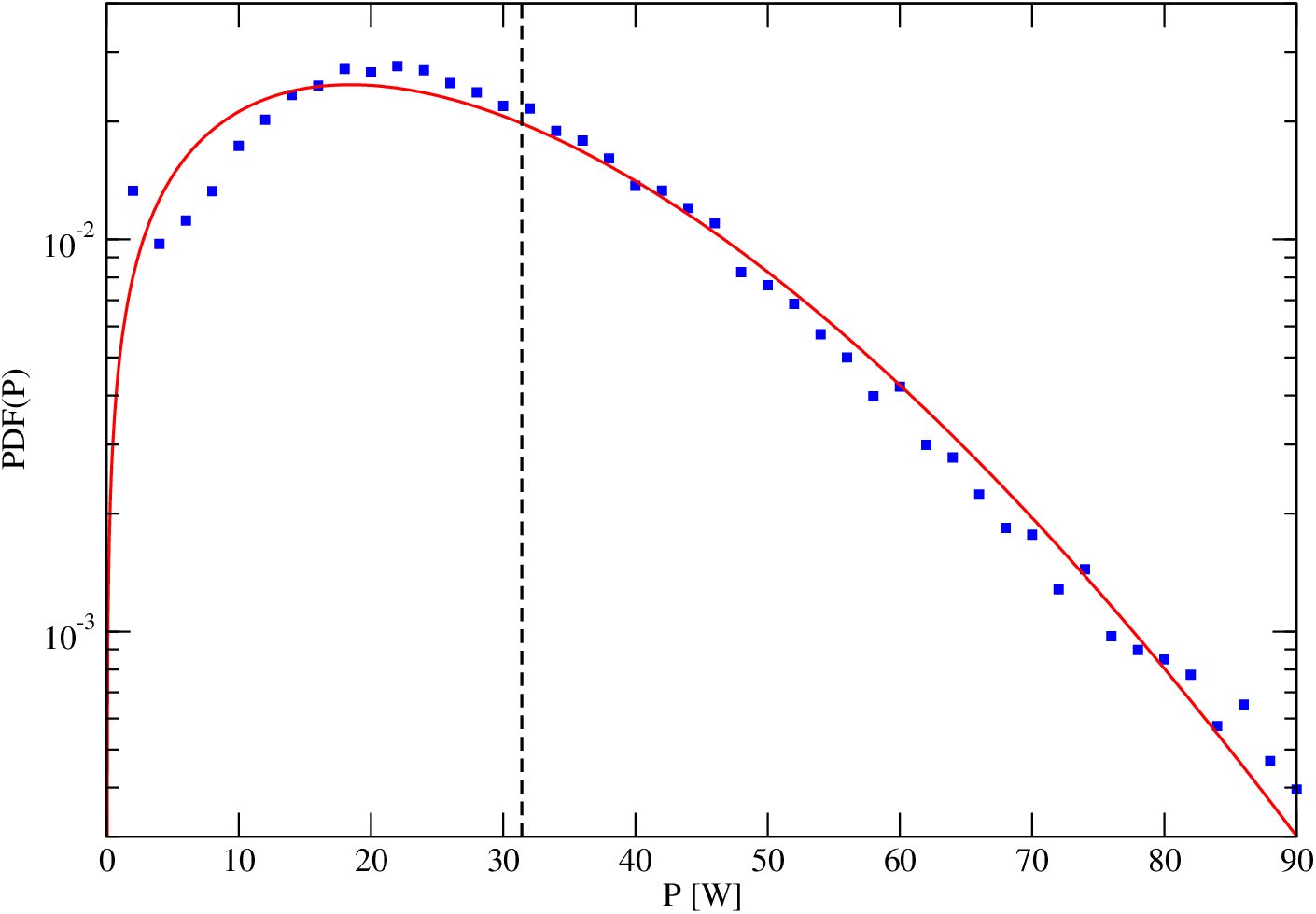}
  \caption{\label{Sfig-PDF_P} Normalized peak power distribution PDF$(P)$ for $\beta_3=0$ at $1.5$ km. 
  The data points denotes (blue squares) denote the data while the solid (red) line shows the fit with Eq.\ \eqref{Seq-PDF-P}.
  The dashed black line is at the fitted value $P_0=31.4$ W}
\end{figure}

As initial condition for the cascade model we compute the PDF of the soliton peak power, $P_q$, in the NLSE case $\beta_3=0$. We select a distance of $z=1500$m such that PDF$(|u|^2)$ has stabilized. We find that the resulting PDF$(P_q)$ can be described as
\begin{equation}
\rho(P_q)= \frac{b}{P_{0}}\left(\frac{P_q}{P_{0}}\right)^{b-1} \exp\left[-\left(\frac{P_q}{P_{0}}\right)^{b}\right],
\label{Seq-PDF-P}
\end{equation}
where $P_{0}=31.4\pm0.6$W and $b=1.69\pm0.04$. The PDF$(P_q)$ and its fit are shown in Fig.\ \ref{Sfig-PDF_P} (the fit has been performed taking the log of the data and the log of the fitting function).
The value $P_{0}$ can alternatively be estimated using energy conservation. We start with a continuous wave (CW) power of $P_\mathrm{CW}=10$W. From the autocorrelation ${\cal C}_z(\tau)$, c.p.\ Fig.\ \ref{Sfig-autocorrelation}, we measure the average time between two peaks as $\Delta T= 0.170\pm0.008$ ps. Hence the initial energy contained in the time window $\Delta T$ is $E_\mathrm{init} =  P_\mathrm{CW} \Delta T = 1.70$ pJ. 
At distances when quasi-solitons have been created the average energy contained in $\Delta T$, using \eqref{eq-period} and \eqref{Seq-PDF-P}, is 
\begin{equation}
E_\mathrm{final}= \int_{0}^{\infty} 2 P_q T_q(P_q) \rho(P_q)dP_q\simeq1.796\sqrt{\frac{|\beta_2|P_0}{\gamma}}. 
\end{equation}
From energy conservation, $E_\mathrm{final}=E_\mathrm{init}$, we find
\begin{equation}
P_0\simeq\frac{\gamma}{|\beta_2|}\left( \frac{P_\mathrm{CW} \Delta T}{1.796}\right)^2= 34 \pm 3 \mathrm{W}. 
\end{equation}

\subsection{Derivation of the effective quasi-soliton description}

In section \ref{sec-CM}, we argued that the shape of a $\beta_3 \neq 0$ quasi-soliton can be approximated by \eqref{eq-quasi-soliton}. The argument follows Ref.\ \cite{Agr13}.
The solution of \eqref{eq-gNLSE} can be approximated as a soliton-like pulse
\begin{eqnarray}
\lefteqn{
u(z,t) = \sqrt{P}\mathrm{sech}\left[\frac{t-q\left(z\right)}{T}\right] \times}\nonumber \\
 &  & \exp\left\{-i\Omega[t-q(z)]-iC\frac{[t-q(z)]^{2}}{2T^{2}}\right\}\label{Seq-quasisoliton} ,
\end{eqnarray}
where $P$, $T$ and $C$ represent the amplitude, duration and chirp.
The other two parameters are the temporal shift $q$ of the pulse
envelope and the frequency shift $\Omega$ of the pulse spectrum.
The distance-dependence of the parameters can be obtained
using the momentum method \cite{CheTE10,SanA03,Agr13}. This gives
\begin{subequations}
\begin{equation}
\frac{dT}{dz}=\left(\beta_{2}+\beta_{3}\Omega\right)\frac{C}{T},
\end{equation}
\begin{equation}
\frac{dC}{dz}=\left(\frac{4}{\pi^{2}}+C^{2}\right)\frac{\left(\beta_{2}+\beta_{3}\Omega\right)}{T^{2}}+\frac{4\gamma P}{\pi^{2}},
\end{equation}
\begin{equation}
\frac{dq}{dz}=\beta_{2}\Omega+\frac{\beta_{3}}{2}\Omega^{2}+\frac{\beta_{3}}{6T^{2}}\left(1+\frac{\pi^{2}}{4}C^{2}\right),
\end{equation}
\begin{equation}
\frac{d\Omega}{dz}=0.
\end{equation}
\end{subequations}
These equations can be solved for $C=0$, resulting in 
\begin{equation}
u(z,t)=\sqrt{P}\mathrm{sech}\left(\frac{t-v^{-1}z}{T}\right)\exp\left[-i\Omega(t-v^{-1}z)\right]\label{Seq-quasi-soliton},
\end{equation}
where $T$ and $v^{-1}$ have been given in \eqref{eq-period} and \eqref{eq-vp}.
\vspace{5mm}

\section{Calm before the storm}

\begin{figure}[tb]
\centering
(a)\includegraphics[width=0.8\columnwidth,clip]{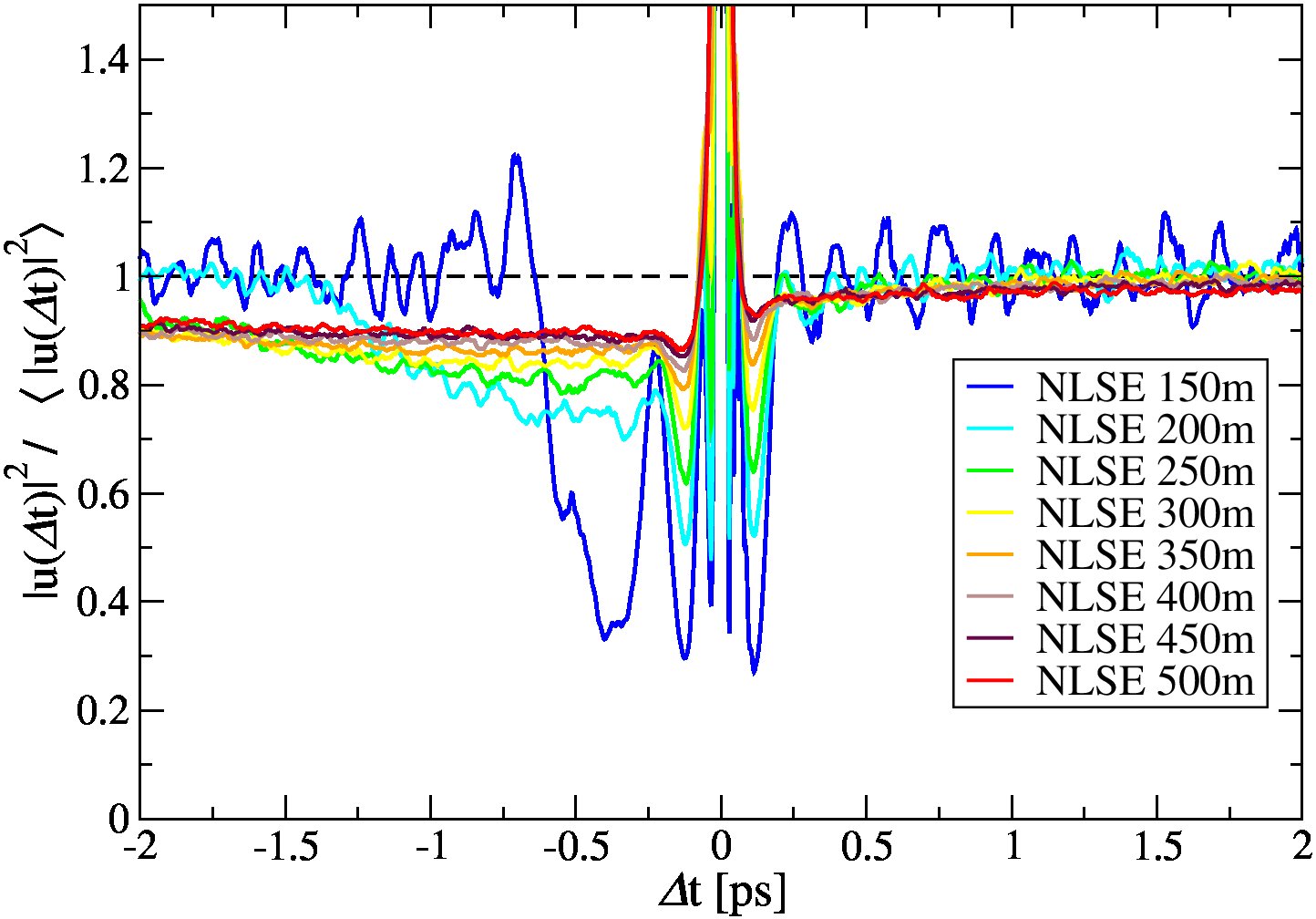}




\centering
(b)\includegraphics[width=0.8\columnwidth,clip]{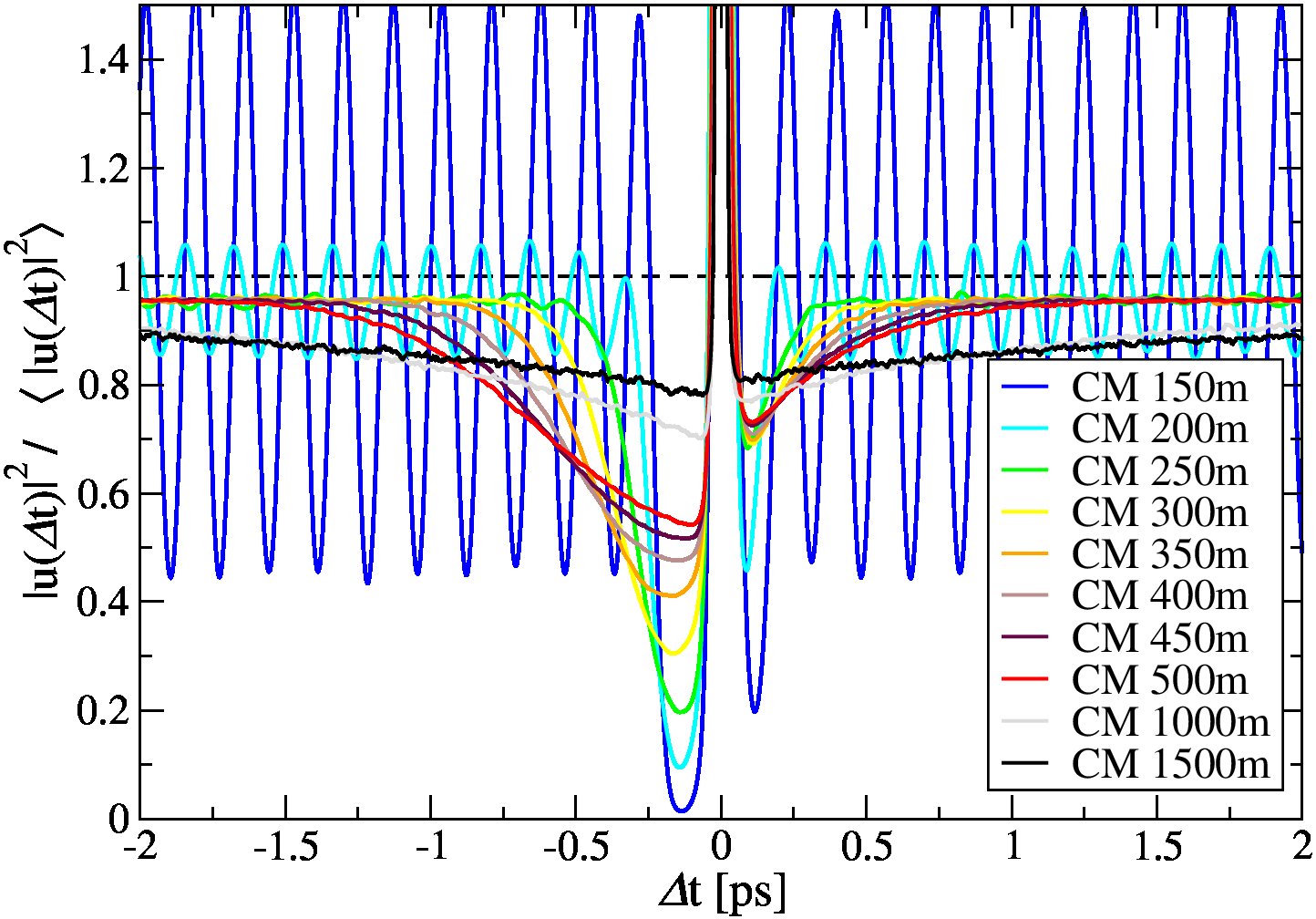}




\caption{\label{fig-Time-Series-NLSE}\label{fig-Time-Series-ECM}
Normalized averaged powers $|u(\Delta t)|^2/\langle |u(\Delta t)|^2 \rangle$ for (a) the NLSE and (b) the cascade model at times $\Delta t$ in the vicinity of a RW.  
We identify RWs as corresponding to powers equal to or larger than $150$W. Different colours indicate distances $z= 150, 200, \ldots 500$m, while the light and dark greys in denote $z=1000$ and $1500$m. The colour choice is compatible with Fig.\ \ref{fig-calmb4storm}.
}
\end{figure}

We studied the temporal vicinity of RWs and found that these are preceded by a period of reduced power values. 
We will refer to this phenomenon as "calm before the storm" \cite{BirBDS15}. 
First we consider the intensity $|u|^2$ at a certain distance $z$, then we find quasi-solitons with a peak power $P$ 
larger than a certain minimum power $P_\text{min}$. Next, we consider the time window that goes from $2$ps before to $2$ps after 
every of these events. Finally we take the average of all the events found. We choose $P_\mathrm{min}= 150$W, while the distances $z$ go from $150$m to $500$m for the gNLSE and up to $1500$m for the cascade model.
The results for the gNLSE and the cascade model are shown in Fig.\ \ref{fig-Time-Series-NLSE}. In both cases we can see a dip in the normalized power 
$|u(\Delta t)|^2/ \langle |u(\Delta t)|^2 \rangle$ before the RW event at $\Delta t =0$. 
In the cascade model, the period of calm is more pronounced, but for longer distances ($z\geq  1000$m, light gray lines in the plot) 
the intensity starts to flatten as in the gNLSE case.
For distances $z\leq 200$m, we observe strong oscillations away from $\Delta t =0$.
In the cascade model these are more regular than in the gNLSE. The cause is in the initial conditions of the cascade model, where at the beginning quasi-solitons are equally spaced, resulting in the observed correlations in the intensity.
For $z>200$m the oscillations disappear because the quasi-solitons are becoming more uncorrelated.
The number of RWs depends on the minimum power chosen, therefore the best statistics is for smaller RWs with, e.g., $P_{min}$=150W, while there are more fluctuations for $P_{min}=300$W. At large distances the statistic improves because more and more RWs are produced.

\begin{figure}[!]
  \centering
  \includegraphics[width=0.99\columnwidth,clip]{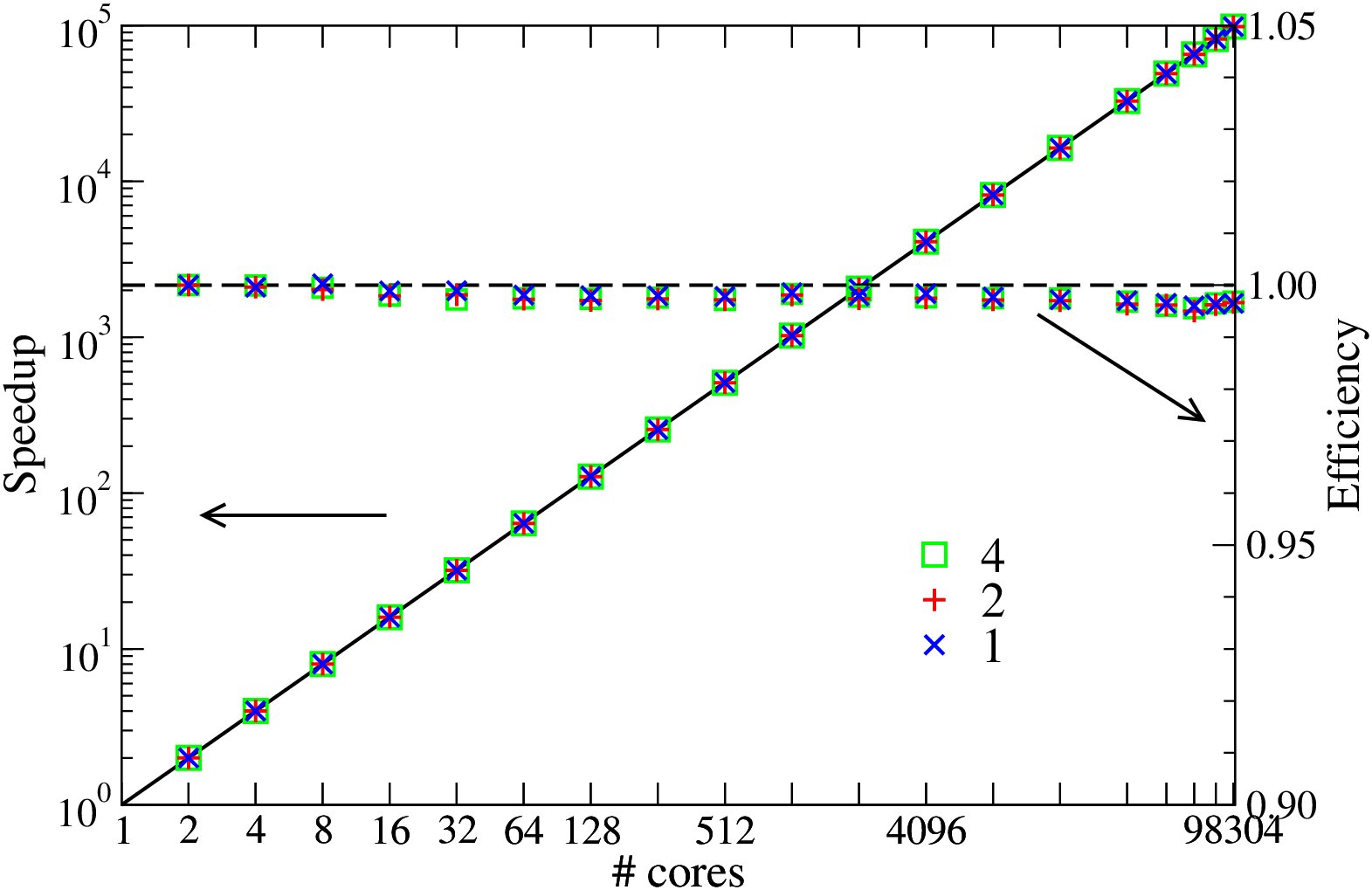}
  \caption{\label{Sfig-speedup} Parallel speedup (left axis) and efficiency (right axis) curves for the parallel implementation of the split-step Fourier method for the gNLSE. Runs for $3$ different computational loads, each double in work size, are shown by different symbols with the solid line showing perfect linear scaling while the dashed line indicating $100\%$ efficiency. The arrows indicate the corresponding axes. The computations were performed on the BlueGene/Q of the Hartree Centre. }
\end{figure}

\section{Parallel scaling}

The results presented here were run on various massively parallel high-performance computing (HPC) machines and architectures. These included HPC clusters at Warwick's Centre for Scientific Computing, the UK national facilities HECToR and ARCHER as well as the BlueGene/Q machine of the Hartree Centre. As explained in section \ref{sec-methods}, the code was MPI-parallized and can take optimal advantage of these HPC architectures. In Fig.\ \ref{Sfig-speedup}, we show the speedup and efficiency of a test run on the 98k cores at the Hartree Centre's BG/Q. 
The results show linear scaling and nearly $100\%$ efficiency.


\ifTtoD
\newpage
\clearpage
\clearpage\newpage
\setcounter{figure}{0}
\setcounter{table}{0}
\setcounter{section}{0}
\setcounter{equation}{0}
\def\thefigure{T\arabic{figure}}
\def\thetable{T\arabic{table}}
\def\thesection{T\arabic{section}}
\def\theequation{T\arabic{equation}}
\setcounter{page}{1}
\pagestyle{plain}

\section{Things to do}
\label{sec-things}
all done!
\fi

\fi\end{document}